\documentclass[aps,physrev,reprint,superscriptaddress]{revtex4-2}
\usepackage{graphicx}
\usepackage{hyperref}
\usepackage{amsmath,amssymb}
\usepackage{braket}
\usepackage{comment}
\hypersetup{
colorlinks=true,
citecolor=blue,
linkcolor=blue,
urlcolor=blue,
}
\usepackage[dvipsnames]{xcolor}	

\begin{document}

\title{Electron-electron interactions mediated by overdamped vibrations: \\ Implications for chirality-induced spin selectivity}

\author{L. Celada}
\affiliation{Dipartimento di Scienze Matematiche, Fisiche e Informatiche, Universit\`a  di Parma, Parco Area delle Scienze, 53/A, I-43124 Parma, Italy.}
\affiliation{Gruppo Collegato di Parma, INFN-Sezione Milano-Bicocca, I-43124 Parma, Italy.}
\author{R. Bellusci}
\affiliation{Dipartimento di Scienze Matematiche, Fisiche e Informatiche, Universit\`a  di Parma, Parco Area delle Scienze, 53/A, I-43124 Parma, Italy.}
\affiliation{Gruppo Collegato di Parma, INFN-Sezione Milano-Bicocca, I-43124 Parma, Italy.}
\author{D.K.A. Phan Huu}
\affiliation{Dipartimento di Scienze Matematiche, Fisiche e Informatiche, Universit\`a  di Parma, Parco Area delle Scienze, 53/A, I-43124 Parma, Italy.}
\author{A. Chiesa}
\affiliation{Dipartimento di Scienze Matematiche, Fisiche e Informatiche, Universit\`a  di Parma, Parco Area delle Scienze, 53/A, I-43124 Parma, Italy.}
\affiliation{Gruppo Collegato di Parma, INFN-Sezione Milano-Bicocca, I-43124 Parma, Italy.}
\affiliation{UdR Parma, INSTM, I-43124 Parma, Italy.}
\author{S. Carretta}
\email{stefano.carretta@unipr.it}
\affiliation{Dipartimento di Scienze Matematiche, Fisiche e Informatiche, Universit\`a  di Parma, Parco Area delle Scienze, 53/A, I-43124 Parma, Italy.}
\affiliation{Gruppo Collegato di Parma, INFN-Sezione Milano-Bicocca, I-43124 Parma, Italy.}
\affiliation{UdR Parma, INSTM, I-43124 Parma, Italy.}

\date{\today}

\begin{abstract}
Motivated by chirality-induced spin selectivity (CISS), we investigate spin polarization during molecular electron transfer through a correlated bridge in the presence of a rapidly relaxing Peierls mode.
By modulating both electronic hopping and spin--orbit coupling, the mode mediates two-body spin--spin interactions, including isotropic, anisotropic, and Dzyaloshinskii--Moriya exchange.
We derive these interactions using an exact Lang--Firsov transformation for a four-site chiral bridge.
We then include fast vibrational relaxation in the joint electron--vibration master equation and derive the reduced electronic dynamics to second order in the vibronic coupling, retaining both coherent and dissipative contributions and the finite-time buildup of the vibrational response.
We test our derivation by numerical simulations, finding spin polarization accumulated at the acceptor after unpolarized injection, with nonmonotonic dependence on damping and bridge filling.
Within the postulated local-loss model, our results show that rapid vibrational relaxation does not preclude spin polarization through a correlated bridge and provide a framework for assessing vibration-mediated interactions and dissipative dynamics as ingredients of CISS in molecular electron transfer.
\end{abstract}

\maketitle

\section{Introduction}
\label{sec:introduction}

Chirality-induced spin selectivity (CISS) refers to the correlation between molecular handedness and electron spin, whereby charge motion through chiral systems can generate spin polarization \cite{Ray1999,Naaman2019,Bloom2024}. Since its first observation, CISS has been reported in a broad range of molecular and supramolecular systems, including DNA and biomolecular assemblies, motivating interest in both its microscopic origin and its potential applications in molecular spintronics and quantum technologies \cite{Naaman2020,Naaman2020b,Latawiec2025}. More recently, time-resolved magnetic-resonance experiments on donor--chiral-bridge--acceptor molecules have probed the spin states of radical pairs generated by photoinduced electron transfer (PET) \cite{Eckvahl2023,Eckvahl2024,Latawiec2025,Chiesa2025}. These experiments provide a molecular setting for investigating the intrinsic origin of CISS \cite{Chiesa2025,Chiesa2023}, without the need to include external components such as substrates, interfaces or electrodes.

This motivated theoretical efforts to understand CISS in isolated molecules \cite{Fay2021b,Fay2023,ChiesaDMI2026}, highlighting the crucial role of spin-spin interactions between the transferred electron and the one remaining on the donor during the charge separation dynamics. In particular, we have recently shown \cite{ChiesaDMI2026} that 
simultaneous modulation of hopping and spin-orbit coupling (SOC) by torsional modes typical of these molecules \cite{phan_huu_ab_2025} mediates a Dzyaloshinskii--Moriya interaction (DMI). This interaction mixes singlet and triplet states, giving rise to a substantial spin polarization  in the spin-correlated radical pair. Moreover, the triplet component predicted by this model and its non-trivial magnetic field dependence are consistent with experimental observations  of CISS by time-resolved electron paramagnetic resonance \cite{ChiesaDMI2026}.

These models provide a good description of closed-shell molecules displaying a large HOMO-LUMO gap on the chiral bridge \cite{Eckvahl2023,Eckvahl2024,Latawiec2025}, in which the electron is transferred through the empty-LUMO channel \cite{phan_huu_ab_2025}. 
They neglect electron-electron interactions within the chiral bridge, which may be important in a different class of molecular systems. This correlated regime has been shown to be relevant to the emergence of spin polarization within the molecular bridge \cite{chiesa_many-body_2024,Fransson2019,Fransson2025,Fransson2020,Das2022}.

Also in the context of tight-binding models, low-energy molecular vibrations are emerging as a crucial ingredient for the CISS effect. 
Mechanisms involving electron--vibration scattering, polaron formation, and correlated vibronic dynamics have consequently been proposed in the context of CISS \cite{Fransson2020,Zhang2020,klein_giant_2023,vittmann2023spin,rudge2025role, NostroJPCL}. Recent work has shown that Peierls coupling enables spin-dependent interference in chiral tight-binding models even when the purely electronic system does not support spin polarization \cite{ArticoloWL}. These results motivate treating vibrations not only as a source of decoherence, but also as mediators of effective electronic interactions. 
An important question is how these vibration-mediated processes are modified when the molecular mode relaxes rapidly into its environment.
Using electronic Green functions with vibronic self-energies incorporating a finite vibrational lifetime, Fransson showed that vibration-induced exchange splitting can produce sizable spin selectivity in molecular junctions \cite{Fransson2020}.
His subsequent work identified vibrational dissipation as essential for establishing a stationary spin polarization and spin-polarized currents \cite{Fransson2025}.

Here we provide a systematic derivation of both the effective electronic Hamiltonian and the dissipative contributions that remain after eliminating the rapidly relaxing vibrational mode. 
To this end, we investigate a correlated four-site bridge coupled to a Peierls mode that modulates both spin-independent hopping and SOC. We first perform an exact Lang--Firsov transformation of the electron--vibration Hamiltonian to identify the induced one-body corrections and two-body interactions, including isotropic, anisotropic, and Dzyaloshinskii--Moriya exchange.
This transformation retains the vibrational degree of freedom.
Starting from a postulated local-loss Lindblad equation for the joint electron--vibration system, we then eliminate the rapidly relaxing mode to second order in the vibronic coupling, retaining the mixed electron--vibration correlations that generate both coherent and dissipative contributions to the reduced electronic dynamics.

We test this framework by considering unpolarized injection into a bridge prepared in a spin-balanced state, followed by spin-independent transfer to an acceptor. The dynamics produces acceptor polarization despite rapid vibrational relaxation. We investigate its dependence on damping, initial bridge filling, vibrational energy, and temperature, and identify a dominant contribution to the spin polarization that is odd under reversal of either vibronic coupling. 
Our results therefore show that rapid vibrational relaxation does not preclude spin polarization during molecular electron transfer and provide a framework for investigating vibration-mediated interactions and dissipative dynamics as ingredients of CISS, not limited to electron transfer.

\section{Microscopic model and vibrationally mediated many-body interactions}
\label{sec:microscopic_model}

We consider a four-site molecular bridge with open boundary conditions with one spinful electronic orbital per site, coupled to a single vibrational mode of energy $E_0=\hbar\omega_0$. To identify the interactions generated by the vibration, we initially retain only electronic hopping and collinear SOC. Site-energy offsets and on-site Coulomb repulsion will be introduced in the open-system electron-transfer model below. The microscopic Hamiltonian is analogous to those investigated by Fransson in Refs.~\cite{Fransson2020,Fransson2025}:
\begin{equation}
    H = H_{\rm e}
    + E_0\left(a^\dagger a+\frac{1}{2}\right)
    + (a^\dagger+a)B,
    \label{eq:H_micro_main}
\end{equation}
where
\begin{equation}
\begin{aligned}
    H_{\rm e} &= -t_{\rm hop}H_{\rm hop}
    + \lambda H_{\rm SOC}, \\
    B &= \frac{1}{\sqrt{2}}
    \left(-gH_{\rm hop}+\chi H_{\rm SOC}\right).
\end{aligned}
    \label{eq:He_B_main}
\end{equation}
Here \(t_{\rm hop}\) and \(\lambda\) are the bare hopping and SOC amplitudes, whereas \(g\) and \(\chi\) describe their Peierls modulation by the dimensionless vibrational coordinate
\((a^\dagger+a)/\sqrt{2}\). All four parameters are real and have units of energy. The electronic operators are
\begin{align}
    H_{\rm hop}
    &= \sum_{j=1}^{3}\sum_{\sigma}
    \left(
    c_{j\sigma}^\dagger c_{j+1,\sigma}
    + \mathrm{h.c.}
    \right),
    \label{eq:Hhop_main}
    \\
    H_{\rm SOC}
    &= \sum_{j=1}^{3}\sum_{\sigma\sigma'}
    \left[
    i c_{j\sigma}^\dagger
    (\sigma_z)_{\sigma\sigma'}
    c_{j+1,\sigma'}
    + \mathrm{h.c.}
    \right],
    \label{eq:HSOC_main}
\end{align}
where \(c_{j\sigma}\) annihilates an electron on site \(j\) with spin \(\sigma\), and \(\sigma_z\) is the Pauli matrix along the spin quantization axis.
For simplicity, we assume that the SOC and its vibrational modulation share a common axis on all bonds, chosen as the \(z\) direction.
Related single-axis descriptions have been adopted in previous models of spin-dependent molecular electron transfer \cite{ChiesaDMI2026,Fay2021b,Fay2023,Fransson2025}. For simplicity, we retain only nearest-neighbor couplings in both \(H_{\rm hop}\) and \(H_{\rm SOC}\).
In a purely electronic open chain with this connectivity, the spin dependence can be removed by site-local spin rotations.
A common way to overcome this constraint is to introduce additional electronic transfer pathways, for example by combining nearest-neighbor hopping with next-nearest-neighbor SOC \cite{chiesa_many-body_2024}.
Here, Peierls coupling provides additional pathways in the combined electronic--vibrational space, enabling spin-dependent interference and allowing spin polarization even with nearest-neighbor hopping and SOC \cite{ArticoloWL}.

An exact Lang--Firsov transformation removes the explicit displacement coupling in Eq.~\eqref{eq:H_micro_main}, while dressing the electronic Hamiltonian. To express this transformation, we diagonalize the single-particle matrix associated with \(B\). Denoting its eigenvalues by \(q_\alpha\) and the corresponding fermionic operators by \(d_\alpha\), with \(\alpha\) labeling the eight electronic modes, we obtain
\begin{equation}
    B = \sum_\alpha q_\alpha d_\alpha^\dagger d_\alpha.
    \label{eq:B_diag_main}
\end{equation}
The anti-Hermitian generator
\begin{equation}
    S = \frac{B}{E_0}(a^\dagger-a)
    = \frac{1}{E_0}\sum_\alpha
    q_\alpha d_\alpha^\dagger d_\alpha(a^\dagger-a)
    \label{eq:S_LF_main}
\end{equation}
defines the transformed Hamiltonian
\(\widetilde H=e^SHe^{-S}\). Since
\(e^Sae^{-S}=a-B/E_0\), the oscillator and its linear coupling combine to give
\begin{equation}
\begin{split}
    \widetilde H ={}&
    E_0\left(a^\dagger a+\frac{1}{2}\right)
    -\frac{B^2}{E_0} +\sum_{\alpha\beta}
    \widetilde h_{\alpha\beta}
    d_\alpha^\dagger d_\beta X_{\alpha\beta},
\end{split}
    \label{eq:H_LF_compact_main}
\end{equation}
where
\begin{equation}
    X_{\alpha\beta}
    =
    \exp\left[
    \frac{q_\alpha-q_\beta}{E_0}(a^\dagger-a)
    \right].
    \label{eq:Xab_main}
\end{equation}
The matrix \(\widetilde h\) is the single-particle matrix of \(H_{\rm e}\) expressed in the eigenbasis of \(B\). Equation~\eqref{eq:H_LF_compact_main} is nonperturbative in \(g\) and \(\chi\), but remains a joint electron--vibration Hamiltonian: the displacement operators \(X_{\alpha\beta}\) have not been approximated. Details of the transformation are given in Appendix~\ref{app:exact_transformation}.

The purely electronic contribution generated by the transformation is
\begin{equation}
\begin{split}
    H_{\rm ind}
    ={}&-\frac{B^2}{E_0} \\
    ={}&-\frac{g^2H_{\rm hop}^2+\chi^2H_{\rm SOC}^2}{2E_0}
    +\frac{g\chi}{2E_0}
    \{H_{\rm hop},H_{\rm SOC}\}.
\end{split}
    \label{eq:Hind_main}
\end{equation}
Normal ordering separates this contribution into one-body corrections and quartic electronic interactions. The one-body terms include site-energy shifts and next-nearest-neighbor spin-independent and spin-dependent hopping. The quartic part can be organized as
\begin{equation}
\begin{split}
    H_{\rm 4f} ={}&+\frac{g^2}{E_0}
    \left(H_{\rm Heisenberg}+H_{\rm iso}\right) +\frac{\chi^2}{E_0}
    \left(H_{\rm XXZ}+H_{\rm aniso}\right) \\
    &
    -\frac{2g\chi}{E_0}H_{\rm DM}
    +\frac{2g\chi}{E_0}H_{\rm corr}.
\end{split}
    \label{eq:H4_decomposition_main}
\end{equation}
The operators \(H_{\rm Heisenberg}\) and \(H_{\rm XXZ}\) contain nearest-neighbor isotropic and symmetric anisotropic exchange, respectively, together with density--density contributions. The residual operators \(H_{\rm iso}\) and \(H_{\rm aniso}\) collect the remaining two-body processes in the \(g^2\) and \(\chi^2\) sectors, including pair transfer and correlated hopping. Their explicit expressions are given in Appendix~\ref{app:exact_transformation}.

The interference between the two Peierls modulations produces the mixed contribution proportional to \(g\chi\), which contains the nearest-neighbor DMI,
\begin{equation}
    H_{\rm DM}
    = 2\sum_{j=1}^{3}
    \left(\mathbf{S}_j\times\mathbf{S}_{j+1}\right)_z,
    \label{eq:HDM_main}
\end{equation}
where
\(\mathbf{S}_j=\frac{1}{2}\sum_{\sigma\sigma'}
c_{j\sigma}^\dagger
\boldsymbol{\sigma}_{\sigma\sigma'}
c_{j\sigma'}\)
is the dimensionless local spin operator.
We denote the remaining mixed contribution by \(H_{\rm corr}\) and refer to it as a mixed correlated-hopping interaction. It couples electronic motion on distinct bonds, involving three sites when the bonds are adjacent and four sites when they are disjoint. 

The mixed contribution in Eq.~\eqref{eq:H4_decomposition_main} is proportional to \(g\chi\).
It therefore requires both Peierls modulations and changes sign when either coupling is reversed, while remaining unchanged when both are reversed.
This dependence on the coupling signs is distinct from the antisymmetric spin structure of the DMI, for which exchanging the two spins on a bond gives \((\mathbf{S}_{j+1}\times\mathbf{S}_j)_z=-(\mathbf{S}_j\times\mathbf{S}_{j+1})_z\).
In contrast, \(H_{\rm corr}\) includes fermionic operators on three or four sites, which cannot be recast in the form of exchange between a single pair of spins.
The DMI and these correlated-hopping processes thus belong to the same mixed \(g\chi\) sector, but have different operator structures.
The acceptor polarization depends on the complete dynamics, including the remaining electronic and vibronic terms, and need not inherit the same odd dependence on either coupling.

The DMI and mixed correlated-hopping terms can mix states with different total-spin quantum numbers while conserving the total spin projection \(S_{\rm tot}^z=\sum_j S_j^z\).
This mixing does not imply time-reversal symmetry breaking: both interactions are invariant under time reversal.
Instead, they can contribute to the redistribution of spin during the nonequilibrium electron-transfer dynamics.
In the protocol considered below, a finite acceptor polarization is accompanied by an opposite spin imbalance on the bridge, while the total spin projection remains conserved.
The accumulated polarization therefore depends on the interplay of this internal spin dynamics with vibrational relaxation and irreversible transfer, rather than on a time-reversal-symmetry breaking  in the microscopic Hamiltonian.

Finally, Eq.~\eqref{eq:H_LF_compact_main} applies to the bilinear \(H_{\rm e}\) specified above. Additional site-energy and Hubbard terms must also be transformed when working in the Lang--Firsov frame, although the contribution \(-B^2/E_0\) remains fixed by the same coupling operator. In the following section, we introduce vibrational damping -- strong in solution and frozen molecular matrices~\cite{szarka1995vibrational,chumakov2004collective,fureraj2022torsional} -- and derive the reduced electronic dynamics from the laboratory-frame master equation for the complete model.

\section{Vibrational damping and effective electronic dynamics}
\label{sec:overdamped}

We now determine how vibrational relaxation modifies the electronic dynamics generated by the microscopic model. We also introduce site-dependent orbital energies and on-site Coulomb repulsion,
so that the bare electronic Hamiltonian of the bridge is
\begin{equation}
\begin{split}
    H_{\rm el} ={}& -t_{\rm hop}H_{\rm hop}+\lambda H_{\rm SOC} \\
    &+\sum_{j=1}^{4}\sum_{\sigma}\epsilon_j n_{j\sigma}
    +U\sum_{j=1}^{4}n_{j\uparrow}n_{j\downarrow}.
\end{split}
    \label{eq:Hel_overdamped}
\end{equation}
The coupling operator \(B\) remains that of Eq.~\eqref{eq:He_B_main}.
We denote the original electron--vibration representation (before any transformation) as the laboratory frame (LAB), to distinguish it from the Lang--Firsov representation.
Here, ``frame'' refers to a unitary change of representation rather than a physical rotation.
In the LAB representation, the Hamiltonian reads

\begin{equation}
    H_{\rm LAB}=H_{\rm el}
    +E_0\left(a^\dagger a+\frac12\right)
    +B(a^\dagger+a).
    \label{eq:H_damped_start}
\end{equation}

At zero bath temperature, we model vibrational loss by
\begin{equation}
\begin{gathered}
    \hbar\dot\rho=-i[H_{\rm LAB},\rho]
    +\Gamma\mathcal D[a]\rho, \\
    \mathcal D[L]\rho=L\rho L^\dagger
    -\frac12\{L^\dagger L,\rho\},
\end{gathered}
    \label{eq:master_bare_main}
\end{equation}
where \(\rho\) is the joint electron--vibration density matrix and \(\Gamma\) has units of energy.

For \(\Gamma\gtrsim E_0\), this equation is not controlled by the usual weak-damping derivation for a microscopic vibrational reservoir, and we therefore adopt it as an idealized Markovian loss model.
More precisely, Eq.~\eqref{eq:master_bare_main} corresponds to a zero-temperature quantum-white-noise reservoir with a flat spectral density and a number-conserving system--bath coupling.
Within this idealized input--output model, the master equation is exact and does not require a further weak-coupling expansion in \(\Gamma\), although the rotating-wave and wide-band assumptions are part of the model \cite{gardiner1985input}.
The reservoir removes excitations from the vibrational mode, while the electron--vibration coupling can repopulate it during the dynamics.
Equation~\eqref{eq:master_bare_main} therefore specifies the phenomenological local-loss model adopted in this work but its correspondence to a physical regime must be tested against an accurate description of the system-bath coupling.

Under the exact Lang--Firsov transformation, the loss operator becomes
\begin{equation}
    a\longmapsto a+F,
    \qquad F\equiv-\frac{B}{E_0},
    \qquad F^\dagger=F.
    \label{eq:F_main}
\end{equation}
The Hamiltonian, initial state, and observables must undergo the same transformation.
In some analyses, the displacement operators \(X_{\alpha\beta}\) are approximated by the identity in the small-displacement regime \(|q_\alpha-q_\beta|/E_0\ll1\).
We do not impose this additional approximation.
Indeed, the vibrational-energy scan extends down to \(E_0=20~\mathrm{meV}\), where the reference couplings give \(g/E_0=0.50\) and \(\chi/E_0=0.75\), while the largest value of \(|q_\alpha-q_\beta|/E_0\) for the four-site bridge is approximately \(2.1\).
Thus, \(X_{\alpha\beta}\simeq I\) is not uniformly justified over the investigated parameter range.
We therefore derive the electronic equation directly in the LAB representation, retaining the mixed electron--vibration correlations to the required coupling order.
We use
\begin{equation}
    \gamma\equiv\frac{\Gamma}{2},
    \qquad \kappa\equiv\gamma+iE_0,
    \qquad \tau_{\rm vib}=\frac{\hbar}{\gamma},
    \label{eq:kappa_gamma_main}
\end{equation}
where \(\tau_{\rm vib}\) is the decay time of the vibrational response.

\subsection{Elimination through electron--vibration correlations}

We prescribe an initially uncorrelated vibrational vacuum,
\begin{equation}
    \rho(0)=\rho_{\rm el}^{(0)}\otimes
    |0\rangle_{\rm vib}\langle0|,
    \label{eq:initial_LAB_main}
\end{equation}
and introduce
\begin{equation}
    \rho_{\rm el}(t)=\operatorname{Tr}_{\rm vib}\rho(t),
    \qquad M(t)=\operatorname{Tr}_{\rm vib}[a\rho(t)].
    \label{eq:mixed_matrix_main}
\end{equation}
Unlike the scalar amplitude \(\langle a\rangle=\operatorname{Tr}_{\rm el}M\), the mixed matrix \(M\) retains the electronic correlations entering the exact reduced equation,
\begin{equation}
    \hbar\dot\rho_{\rm el}
    =-i[H_{\rm el},\rho_{\rm el}]
    -i[B,M+M^\dagger].
    \label{eq:rho_el_exact_main}
\end{equation}

The exact equation for \(M\) also contains
\(X=\operatorname{Tr}_{\rm vib}(a^\dagger a\rho)\) and
\(Y=\operatorname{Tr}_{\rm vib}(a^2\rho)\), through the source \(-i[B,X+Y]\).
For power counting, scale \(B\to\eta B\) while holding \(H_{\rm el}\) and the initial electronic state fixed. Starting from Eq.~\eqref{eq:initial_LAB_main}, \(M=\mathcal O(\eta)\), whereas \(X,Y=\mathcal O(\eta^2)\). Their feedback therefore changes \(M\) at third order and the electronic evolution at fourth order. Retaining the electronic generator to second order gives, after setting \(\eta=1\),
\begin{equation}
    \hbar\dot M\simeq-\kappa M-i[H_{\rm el},M]
    -iB\rho_{\rm el}.
    \label{eq:M_second_order_main}
\end{equation}
The source \(B\rho_{\rm el}\) follows from \([a,a^\dagger]=1\); vacuum fluctuations have not been discarded.

We solve Eq.~\eqref{eq:M_second_order_main} with \(M(0)=0\), without imposing \(\dot M=0\). During the response interval, electronic evolution is retained through \(U_\tau=e^{-iH_{\rm el}\tau/\hbar}\), using
\begin{equation}
    \rho_{\rm el}(t-\tau)
    \simeq U_\tau^\dagger\rho_{\rm el}(t)U_\tau.
    \label{eq:backward_density_main}
\end{equation}
Vibronic feedback in this backward propagation contributes beyond the retained coupling order. Importantly, \(H_{\rm el}\) contains the complete bare interacting Hamiltonian, not a vibration-renormalized Hamiltonian. Defining the electronic response matrix
\begin{equation}
    Q(t)=\frac{E_0^2}{\hbar}\int_0^t
    e^{-\kappa\tau/\hbar}U_\tau F U_\tau^\dagger\,d\tau,
    \label{eq:Q_finite_main}
\end{equation}
we obtain
\begin{equation}
\begin{aligned}
    M(t)&\simeq\frac{i}{E_0}Q(t)\rho_{\rm el}(t),\\
    M^\dagger(t)&\simeq-\frac{i}{E_0}
    \rho_{\rm el}(t)Q^\dagger(t).
\end{aligned}
    \label{eq:M_response_main}
\end{equation}
Substitution into Eq.~\eqref{eq:rho_el_exact_main} yields a time-local master equation for the electronic density matrix alone,
\begin{equation}
\begin{split}
    \hbar\dot\rho_{\rm el}={}&-i[H_{\rm el},\rho_{\rm el}] -[F,Q(t)\rho_{\rm el}-\rho_{\rm el}Q^\dagger(t)].
\end{split}
    \label{eq:electronic_master_main}
\end{equation}
Here \(Q(t)\) is determined by known electronic operators and has units of energy. 

Its coherent and non-Hamiltonian contributions can be separated algebraically. With
\begin{equation}
    H_{\rm eff}(t)=H_{\rm el}
    +\frac{FQ(t)-Q^\dagger(t)F}{2i},
    \label{eq:Heff_Q_main}
\end{equation}
Eq.~\eqref{eq:electronic_master_main} becomes
\begin{equation}
\begin{split}
    \hbar\dot\rho_{\rm el}={}&-i[H_{\rm eff}(t),\rho_{\rm el}] +Q\rho_{\rm el}F+F\rho_{\rm el}Q^\dagger \\
    &-\frac12\{FQ+Q^\dagger F,\rho_{\rm el}\},
\end{split}
    \label{eq:electronic_master_split_main}
\end{equation}
where \(Q=Q(t)\). Both contributions must be retained at the same coupling order.

\subsection{Finite-time and stationary response}

The response can be evaluated directly in the many-body energy eigenbasis of \(H_{\rm el}\). Writing \(H_{\rm el}|m\rangle=E_m|m\rangle\), \(\Delta_{mn}=E_m-E_n\), and \(F_{mn}=\langle m|F|n\rangle\), we find
\begin{align}
    [Q(t)]_{mn}
    &=\frac{E_0^2F_{mn}}{\kappa+i\Delta_{mn}}
    \left[1-e^{-(\kappa+i\Delta_{mn})t/\hbar}\right],
    \label{eq:Q_energy_finite_main} \\
    (Q_\infty)_{mn}
    &=\frac{E_0^2F_{mn}}{\kappa+i\Delta_{mn}}.
    \label{eq:Q_energy_stationary_main}
\end{align}
These expressions are finite for every \(\gamma>0\).

The finite-time response satisfies \(Q(0)=0\), reproducing the exact initial electronic derivative for Eq.~\eqref{eq:initial_LAB_main}. Replacing \(Q(t)\) by \(Q_\infty\) omits its initial buildup but retains electronic motion during the response. Here ``stationary'' refers only to the response coefficients, not to the electronic state. Both descriptions start from the same \(\rho_{\rm el}^{(0)}\); differences accumulated during the initial interval can persist after the response matrices have become nearly identical.

The connection with a nested-commutator expansion is
\begin{equation}
\begin{gathered}
    Q_\infty=E_0^2\sum_{n=0}^{\infty}
    \frac{(-i)^n}{\kappa^{n+1}}C_n, \\
    C_0=F,\qquad C_{n+1}=[H_{\rm el},C_n].
\end{gathered}
    \label{eq:Cn_main}
\end{equation}
Higher-order nested commutators resolve electronic motion during the response interval, but do not increase the coupling order retained in the elimination: every \(C_n\) remains linear in \(B\).
The series requires \(|\Delta_{mn}|<|\kappa|\) for every pair with \(F_{mn}\ne0\). Outside this convergence domain, Eq.~\eqref{eq:Q_energy_stationary_main} remains the direct value of the response integral, not the sum of a convergent series. The numerical comparisons below distinguish three dynamical descriptions.
The LAB curves are obtained by integrating the joint electron--vibration master equation with a converged oscillator cutoff.
The stationary and finite-time electronic curves instead follow the reduced master equation, using \(Q_\infty\) from Eq.~\eqref{eq:Q_energy_stationary_main} and \(Q(t)\) from Eq.~\eqref{eq:Q_energy_finite_main}, respectively.
The latter are labeled ``Transient electronic'' in the figures.
Both electronic responses are evaluated directly in the energy eigenbasis, without truncating the nested-commutator series.
Numerical time integration is performed for the density-matrix evolution, with acceptor transfer included as described in Sec.~\ref{sec:electron_transfer}.
At each parameter point, all descriptions use the same initial electronic state.
The subsequent coupling, vibrational-energy, and filling scans use the finite-time electronic response.

\subsection{Induced interactions and regime of validity}

To connect the stationary response with the interactions identified in Sec.~\ref{sec:microscopic_model}, write
\begin{equation}
\begin{gathered}
    R=\frac{E_0^2}{\gamma^2+E_0^2}, \\
    Q_\infty=R(\gamma-iE_0)F+\Delta Q.
\end{gathered}
    \label{eq:R_main}
\end{equation}
This is an exact decomposition of the stationary response within the electronic reduction. Equation~\eqref{eq:electronic_master_main} then reads
\begin{equation}
\begin{split}
    \hbar\dot\rho_{\rm el}={}&
    -i[H_{\rm el}-RE_0F^2,\rho_{\rm el}] +\Gamma R\mathcal D[F]\rho_{\rm el} \\
    &-[F,\Delta Q\rho_{\rm el}-\rho_{\rm el}\Delta Q^\dagger].
\end{split}
    \label{eq:stationary_master_R_main}
\end{equation}
The contribution \(-RE_0F^2=-RB^2/E_0\) contains the one-body and two-body terms derived above, multiplied by \(R\). The remainder \(\Delta Q\) contributes to both the coherent and non-Hamiltonian dynamics. Neglecting it gives the additional frozen-electronic-response approximation, with
\begin{equation}
    H_{\rm el}^{(0)}=H_{\rm el}-RE_0F^2
    \label{eq:Hel0_main}
\end{equation}
and the dissipator \(\Gamma R\mathcal D[F]\). This approximation is distinct from taking the stationary limit of the full response and is not used to generate \(U_\tau\).

A conservative working condition for the coupling expansion is
\begin{equation}
    \xi_B=\frac{\|B\|}{\gamma}\ll1,
    \label{eq:coupling_control_main}
\end{equation}
with the operator norm evaluated in the electronic sector being propagated. Neglecting electronic motion during the response would require an additional separation of time scales, for which \(|\Delta_{mn}|/\gamma\ll1\) on the coupled transitions is sufficient.

In the electron-transfer application of Sec.~\ref{sec:electron_transfer}, \(H_{\rm el}\) is extended to include the bare acceptor Hamiltonian, and the electronic equation retains the transfer term \(\Gamma_A\mathcal D[J_A]\rho_{\rm el}\). Neglecting transfer only during the short response interval requires the separate condition
\begin{equation}
    \xi_A=\frac{\Gamma_A\|J_A^\dagger J_A\|}{\gamma}\ll1,
    \label{eq:transfer_control_main}
\end{equation}
where \(\Gamma_A\) and \(J_A\) are the transfer coefficient and jump operator. The omitted transfer corrections remain second order in \(B\), with a characteristic relative scale \(\xi_A\); they are distinct from the fourth-order vibronic corrections. These scale estimates do not provide uniform error bounds at arbitrarily long times.

The reduced equation has a nonsecular, second-order (Redfield-type) structure and preserves trace and Hermiticity, but is not generally of Lindblad form with nonnegative rates \cite{Breuer,Tupkary2022}.
Positivity of \(\rho_{\rm el}(t)\) is therefore not guaranteed. We assess the reduction by monitoring its eigenvalues and comparing observables with LAB calculations converged in the vibrational Hilbert space, using identical initial preparations and evolution times. The derivation and the distinct coupling and transfer approximations are detailed in Appendix~\ref{app:adiabatic_elimination}.
Physically, the reduced dynamics combines coherent vibration-mediated interactions with electronic dissipation.
In the stationary decomposition, \(-RB^2/E_0\) retains the induced one-body terms and the complete two-body sector.
The different coupling sectors have distinct tensorial spin structures.
Here, \(k\) denotes the rank of an irreducible spherical spin tensor \(T_q^{(k)}\), while \(q=-k,\ldots,k\) labels its component with respect to the \(z\) axis; in particular, a \(q=0\) component conserves \(S_{\rm tot}^z\), since \([S_{\rm tot}^z,T_q^{(k)}]=qT_q^{(k)}\) in units with \(\hbar=1\).
Since \(H_{\rm hop}\) is a spin scalar, the \(g^2\) sector is isotropic and transforms as a rank-zero spin tensor.
The mixed \(g\chi\) sector carries rank-one, \(q=0\), spin dependence; its nearest-neighbor exchange component is the \(z\)-directed DMI, while the same sector also contains mixed one-body and correlated-hopping processes.
The \(\chi^2\) sector contains both scalar contributions and an axially anisotropic rank-two component with \(q=0\).
Only \(q=0\) components occur because the SOC and its vibrational modulation were assumed to share a common \(z\) axis.
Consequently, the coherent effective Hamiltonian, the response matrix, and the vibration-induced dissipative terms conserve \(S_{\rm tot}^z\).
The rank-one and rank-two contributions do not generally commute with \(\mathbf S_{\rm tot}^{\,2}\) and can therefore mix different total-spin multiplets while preserving their spin projection.
For noncollinear SOC axes, additional \(q\neq0\) components would generally appear and \(S_{\rm tot}^z\) would no longer be conserved.
The term \(\Gamma R\mathcal D[F]\) damps coherences between states with different eigenvalues of \(F\), whereas \(\Delta Q\) modifies both the coherent and non-Hamiltonian parts of the dynamics.
The accumulated acceptor polarization therefore reflects the interplay among these tensorial interaction channels, vibration-induced dissipation, and electron transfer.

\section{Electron-transfer protocol and spin observables}
\label{sec:electron_transfer}

We now apply the reduced electronic dynamics derived in Sec.~\ref{sec:overdamped} to electron transfer through the four-site molecular bridge. The acceptor consists of a single spinful orbital, restricted to its empty and singly occupied states, with Hamiltonian
\begin{equation}
    H_A=\epsilon_A N_A,
    \qquad
    N_A=n_{A\uparrow}+n_{A\downarrow},
    \label{eq:acceptor_H_main}
\end{equation}
where \(n_{A\sigma}=c_{A\sigma}^\dagger c_{A\sigma}\).
No coherent hopping connects the bridge and the acceptor. In this section, we denote the bridge Hamiltonian of Eq.~\eqref{eq:Hel_overdamped} by \(H_{\rm B}\) and write \(H_{\rm el}=H_{\rm B}+H_A\) for the complete bare electronic Hamiltonian.

As a reference configuration before electron injection, we consider a half-filled bridge containing four electrons and an empty acceptor,
\begin{equation}
    N_{\rm bridge}=4,
    \qquad N_A=0,
    \label{eq:initial_sector_main}
\end{equation}
where \(N_{\rm bridge}=\sum_{j,\sigma}n_{j\sigma}\). This will be the starting point for injecting an additional electron and studying the subsequent electron transfer and possible spin polarization.
This configuration illustrates
the development of acceptor spin polarization during electron transfer;
the dependence on the initial bridge filling is examined separately below.
We prescribe the static preparation Hamiltonian
\begin{equation}
    H_{\rm prep}=H_{\rm el}-\frac{R}{E_0}B^2,
    \label{eq:Hprep_main}
\end{equation}
with \(R\) defined in Eq.~\eqref{eq:R_main}.
Its restriction to the sector in Eq.~\eqref{eq:initial_sector_main}
acts on \(\binom{8}{4}=70\) states.
At zero temperature, we choose
a spin-balanced ground-state density matrix, without assuming that the state is a total-spin singlet.
Specifically,
\begin{equation}
    \rho_4=
    \frac{\Pi_{\rm gs}^{(4)}}
    {\operatorname{Tr}\Pi_{\rm gs}^{(4)}},
    \label{eq:rho4_main}
\end{equation}
where \(\Pi_{\rm gs}^{(4)}\) projects onto the complete ground eigenspace of the restricted \(H_{\rm prep}\).
For a nondegenerate ground state, this reduces to its pure-state projector.
For a degenerate ground level, it avoids selecting an arbitrary spin-polarized component.
The preparation is time-reversal invariant and has equal local spin-up and spin-down occupations.
It is not assumed to be the equilibrium state of the interacting electron--vibration Lindblad model.

An additional electron is injected instantaneously at bridge site \(1\),
with equal prior weights for spin up and spin down, and no coherence between the two injection channels.
After injection, the complete dynamical sector is
\begin{equation}
\begin{gathered}
    N_{\rm tot}=N_{\rm bridge}+N_A=5,
    \qquad N_A\leq1,\\
    \dim\mathcal H_5=
    \binom{8}{5}+2\binom{8}{4}=196.
\end{gathered}
    \label{eq:dynamic_sector_main}
\end{equation}
The two contributions correspond to an empty acceptor with five bridge electrons and a singly occupied acceptor with four bridge electrons, respectively.
Let \(\Pi_4\) and \(\Pi_5\) project onto the preparation and dynamical sectors.
Using the injection maps
\(C_{1\sigma}^\dagger=\Pi_5c_{1\sigma}^\dagger\Pi_4\),
we construct
\begin{equation}
    \rho_I=
    \frac{
    \displaystyle\sum_\sigma
    C_{1\sigma}^\dagger\rho_4C_{1\sigma}
    }{
    \displaystyle\operatorname{Tr}
    \left[
    \sum_\sigma C_{1\sigma}^\dagger\rho_4C_{1\sigma}
    \right]
    }.
    \label{eq:injected_state_main}
\end{equation}
The two contributions are summed before normalization.
Pauli exclusion is enforced by the creation operators, and the spin-balanced preparation gives equal traces for the two injection channels.
The resulting \(\rho_I\) has no initial spin imbalance.
All occupation states in the specified sectors are retained: no additional energy-window projection is applied to injection or propagation.
The donor is represented only through this preparation step and is not retained as a dynamical degree of freedom.

Transfer from bridge site \(4\) to the acceptor is described by
\begin{equation}
    J_A=\sum_{\sigma=\uparrow,\downarrow}
    c_{A\sigma}^\dagger
    \left(1-n_{A\bar\sigma}\right)c_{4\sigma},
    \label{eq:acceptor_jump_main}
\end{equation}
where \(\bar\sigma\) denotes the spin opposite to \(\sigma\).
The factor \(1-n_{A\bar\sigma}\) enforces the absence of double occupation in full Fock-space notation. Both spin components enter with the same amplitude and the same transfer coefficient \(\Gamma_A\), so the extraction channel conserves spin without imposing a preferred spin orientation.
The sum is taken at the operator level and used in the single dissipator
\(\Gamma_A\mathcal D[J_A]\), rather than in two independent spin-resolved dissipators.
The jump transfers one electron to an empty acceptor and vanishes when the acceptor is occupied.
We treat the acceptor as an irreversible trap, omit reverse transfer, and take the forward-transfer coefficient \(\Gamma_A\) to be energy independent. 
The reduced electronic equation is
\begin{equation}
\begin{split}
    \hbar\dot\rho_{\rm el}={}&
    -i[H_{\rm el},\rho_{\rm el}]
    +\Gamma_A\mathcal D[J_A]\rho_{\rm el}\\
    &-[F,Q(t)\rho_{\rm el}
    -\rho_{\rm el}Q^\dagger(t)].
\end{split}
    \label{eq:electron_transfer_master_main}
\end{equation}
Here \(F=-B/E_0\), and the response is evaluated from the complete bare \(H_{\rm el}\) in the dynamical sector using Eq.~\eqref{eq:Q_energy_finite_main}.
For the stationary-response calculation, \(Q(t)\) is replaced by \(Q_\infty\) from Eq.~\eqref{eq:Q_energy_stationary_main}.
The last commutator contains both coherent and non-Hamiltonian vibronic contributions.
The preparation correction in Eq.~\eqref{eq:Hprep_main} is therefore not added separately to the propagation Hamiltonian.
Acceptor transfer is retained throughout the evolution; its omission during the short vibrational response interval is the separate approximation discussed in Sec.~\ref{sec:overdamped}.

For each parameter set, \(\rho_4\) and \(\rho_I\) are constructed once and used in all dynamical descriptions.
The initial conditions are
\begin{equation}
\begin{aligned}
    \rho_{\rm LAB}(0)
    &=\rho_I\otimes|0\rangle_{\rm vib}\langle0|, \qquad \rho_{\rm el}(0)=\rho_I.
\end{aligned}
    \label{eq:transfer_initial_states_main}
\end{equation}
The LAB calculation propagates the joint master equation
\eqref{eq:master_bare_main}, with the complete bridge--acceptor Hamiltonian and the additional term
\(\Gamma_A\mathcal D[J_A]\rho_{\rm LAB}\)
on the right-hand side.
This factorized initial state is prescribed rather than obtained by equilibrating the coupled electron--vibration system.
Electronic observables are evaluated from
\(\operatorname{Tr}_{\rm vib}\rho_{\rm LAB}(t)\)
or from the reduced electronic calculation using identical operator matrices.

We monitor the spin-resolved acceptor populations and their sum,
\begin{equation}
\begin{aligned}
    n_{A\sigma}(t)
    &=\operatorname{Tr}[\rho_{\rm el}(t)n_{A\sigma}],\\
    n_A(t)&=n_{A\uparrow}(t)+n_{A\downarrow}(t),
\end{aligned}
    \label{eq:acceptor_populations_main}
\end{equation}
together with the spin imbalance
\(m_A(t)=n_{A\uparrow}(t)-n_{A\downarrow}(t)\).
The acceptor spin polarization, expressed in percent, is
\begin{equation}
    P_A(t)=100\,\frac{m_A(t)}{n_A(t)},
    \qquad n_A(t)>0.
    \label{eq:PA_main}
\end{equation}
Since the acceptor can contain at most one electron, \(n_A(t)\) is its occupation probability. Dividing by this probability gives the spin polarization of the occupied-acceptor subensemble.
The polarization is undefined at \(t=0\), when the acceptor is empty, and is reported only when its occupation is numerically resolved.
Recording \(n_A\) and \(m_A\) alongside \(P_A\) distinguishes the transferred population from its spin composition.

The dynamics conserves the total electron number and, for the collinear SOC considered here, the total spin projection.
Consequently,
\begin{equation}
\begin{aligned}
    \langle N_{\rm bridge}\rangle_t+n_A(t)&=5,\\
    m_A(t)+\sum_{j=1}^{4}m_j(t)&=0,
\end{aligned}
    \label{eq:transfer_balance_main}
\end{equation}
where \(m_j(t)=2\operatorname{Tr}[\rho_{\rm el}(t)S_j^z]\)
is evaluated from the full electronic density matrix, including both acceptor-occupation sectors.
An acceptor spin imbalance is therefore accompanied by an opposite total imbalance on the bridge.
The protocol probes polarization developing during nonequilibrium electron transfer, without imposing a spin-dependent injection or extraction coefficient.

\section{Results}
\label{sec:results}

As a benchmark of the introduced framework, we now investigate the electron-transfer (ET) dynamics and the resulting acceptor spin polarization.
In all calculations, the response is evaluated directly in the energy eigenbasis of the bare interacting electronic Hamiltonian.
All results in this section refer to zero temperature.

Unless otherwise specified, the parameters are those in Table~\ref{tab:baseline_parameters}.
The parameter set defines a representative model test bed for correlated molecular systems, rather than a quantitative parametrization of a specific molecule.
The reference bridge has \(t_{\rm hop}/U=0.15\), placing it in a strongly correlated regime comparable to that discussed for organic charge-transfer systems \cite{fortunelli1997ab,kandpal2009revision}.
Its nonuniform site energies follow a rescaled profile inspired by the charge-transfer model of Ref.~\cite{Ratner2013}.
The reduced energy separations are used to define a model electronic landscape.
The chosen SOC, \(\lambda=1~\mathrm{meV}\), lies in the meV range commonly considered in theoretical descriptions of light-atom chiral molecules \cite{Medina2015,geyer_chirality-induced_2019}.
The Peierls modulations of hopping and SOC are deliberately chosen to be large in order to probe a regime of strong vibronic effects.
The reference vibrational energy \(E_0=\hbar\omega_0=100~\mathrm{meV}\) is chosen partly for numerical convenience in benchmarking the reduced electronic dynamics against LAB calculations.
For the adopted couplings, \(g/(\hbar\omega_0)=0.10\) and \(\chi/(\hbar\omega_0)=0.15\); together with rapid vibrational relaxation, this choice permits converged LAB calculations with a modest oscillator cutoff, as documented in Appendix~\ref{app:additional_results}.
Lower vibrational energies are explored separately in Sec.~\ref{subsec:physical_controls} using the reduced electronic description.

\begin{table}[t]
    \caption{Reference parameters.
    The damping and transfer coefficients have units of energy.
    The coupling, vibrational-energy, and filling scans use \(\Gamma=1~\mathrm{eV}\), as specified in the text.}
    \label{tab:baseline_parameters}
    \begin{ruledtabular}
    \begin{tabular}{lc}
        Parameter & Value \\
        \hline
        \(t_{\rm hop}\) & \(30~\mathrm{meV}\) \\
        \(\lambda\) & \(1~\mathrm{meV}\) \\
        \(U\) & \(200~\mathrm{meV}\) \\
        \(E_0\) & \(100~\mathrm{meV}\) \\
        \(g\) & \(10~\mathrm{meV}\) \\
        \(\chi\) & \(15~\mathrm{meV}\) \\
        \(\epsilon_{1,\ldots,4}\) & \(20,100,142,165~\mathrm{meV}\) \\
        \(\epsilon_A\) & \(-1~\mathrm{eV}\) \\
        \(\Gamma\) & \(2~\mathrm{eV}\) \\
        \(\Gamma_A\) & \(10~\mathrm{meV}\) \\
    \end{tabular}
    \end{ruledtabular}
\end{table}

We first compare the LAB dynamics with the stationary and finite-time electronic responses at \(\Gamma=2~\mathrm{eV}\).
This comparatively large damping is deliberately chosen as a stringent benchmark for the electronic reduction, rather than as a realistic molecular vibrational linewidth.
At this value, the response time is very short and the control parameters of the elimination are small, allowing a direct comparison with converged LAB simulations in a regime where the approximations underlying the reduced dynamics are expected to be most reliable.
For this reference point,
\begin{equation}
    \xi_B\simeq0.057,
    \qquad
    \xi_A=0.020,
    \label{eq:reference_control_parameters}
\end{equation}
with norms evaluated over the complete allowed five-electron sector, and the response-envelope decay time is \(\tau_{\rm vib}=0.658~\mathrm{fs}\).
The accuracy of the reduction is assessed by comparison with LAB simulations retaining the vibrational mode explicitly and converged with respect to its occupation cutoff.
The damping dependence and the choice of \(\Gamma=1~\mathrm{eV}\) for the subsequent parameter scans are discussed below in Sec.~\ref{subsec:representative_dynamics}.

\subsection{Acceptor spin polarization and validation}
\label{subsec:representative_dynamics}

\begin{figure*}[t]
    \centering
    \includegraphics[width=\textwidth] {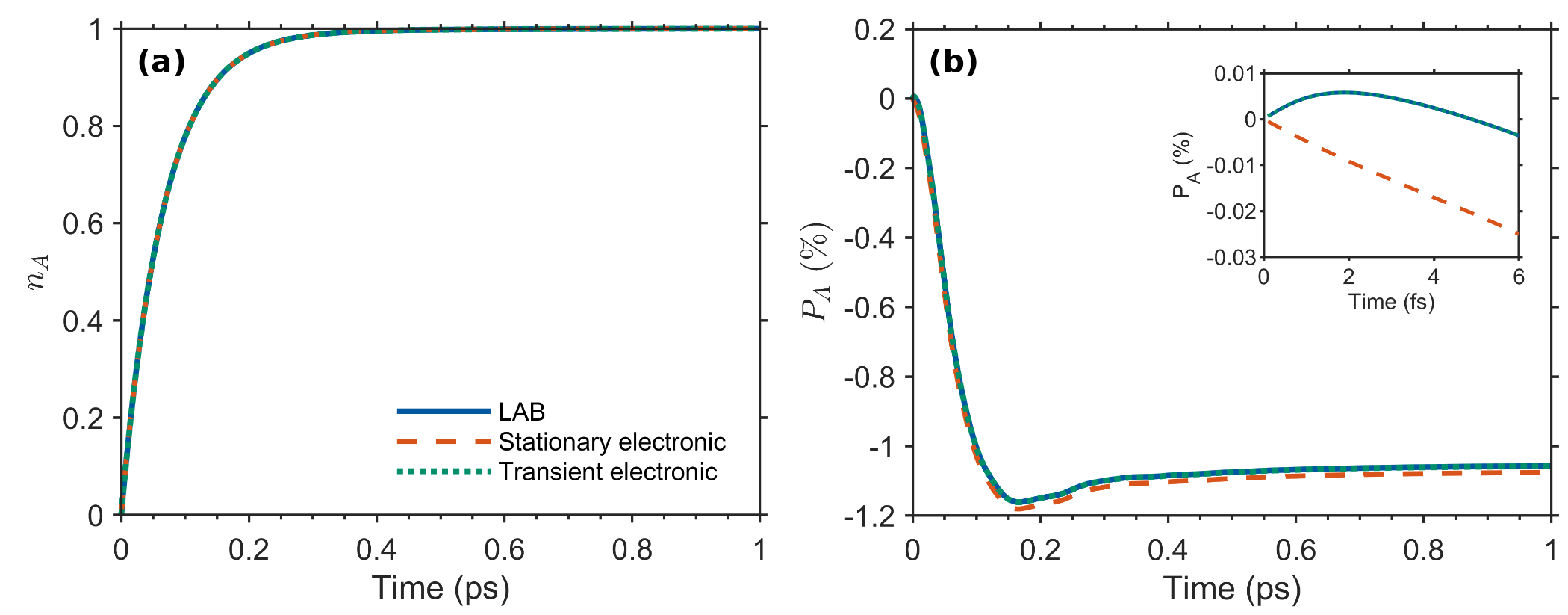}
    \caption{Reference ET dynamics for Table~\ref{tab:baseline_parameters}: (a) acceptor occupation \(n_A(t)\) and (b) acceptor spin polarization \(P_A(t)\).
    The LAB calculation with \(n_{\max}=4\) is compared with the stationary and finite-time electronic responses; the latter is labeled ``Transient electronic'' in the legend.
    All descriptions start from the same injected electronic state.
    The inset resolves the first \(6~\mathrm{fs}\); the main time axes are in picoseconds.
    Bridge occupation follows from \(\langle N_{\rm bridge}\rangle=5-n_A\).
    The polarization is defined only for nonzero acceptor occupation.}
    \label{fig:reference_dynamics}
\end{figure*}

Figure~\ref{fig:reference_dynamics} shows the dynamics following unpolarized injection into the half-filled bridge.
The acceptor population approaches unity, while particle-number conservation gives \(\langle N_{\rm bridge}\rangle=5-n_A\).
The acceptor-transfer coefficient \(\Gamma_A=10~\mathrm{meV}\) is chosen as a phenomenological ultrafast trapping rate, allowing the acceptor population to approach its late-time value within the \(1~\mathrm{ps}\) simulation window.
At the same time, transfer remains much slower than the vibrational response, and the corresponding control parameter is small, \(\xi_A=0.020\).
At \(T_{\rm final}=1~\mathrm{ps}\), the LAB calculation gives
\[
n_A\simeq1.00,
\qquad
\langle N_{\rm bridge}\rangle\simeq4.00,
\]
indicating nearly complete transfer.
The spin-resolved acceptor populations are
\[
n_{A\uparrow}\simeq0.495,
\qquad
n_{A\downarrow}\simeq0.505,
\]
corresponding to
\begin{equation}
    P_A^{\rm LAB}(T_{\rm final})=-1.06\%,
    \qquad T_{\rm final}=1~\mathrm{ps}.
    \label{eq:PA_reference}
\end{equation}
This is the polarization of the occupied-acceptor subensemble, as defined in Eq.~\eqref{eq:PA_main}, rather than the transferred population itself.
Its negative sign denotes an excess of spin-down electrons on the acceptor.
The opposite total spin imbalance remains on the bridge, consistently with Eq.~\eqref{eq:transfer_balance_main}.

The polarization develops nonmonotonically.
After a small positive initial excursion \cite{chiesa_many-body_2024}, the LAB result reaches approximately \(-1.16\%\) at \(0.17~\mathrm{ps}\) before approaching its final value.
The finite-time response reproduces this initial behavior, whereas the stationary response starts with a negative signal.
At \(T_{\rm final}=1~\mathrm{ps}\), the stationary and finite-time calculations give approximately \(-1.08\%\) and \(-1.06\%\), respectively.
Thus, resolving the initial buildup of the vibrational response improves the quantitative agreement with LAB.
Further convergence and positivity checks are reported in Appendix~\ref{app:additional_results}.

Figure~\ref{fig:Gamma_scan_main} tests the reduced description over $\Gamma=0.2$--$2~\mathrm{eV}$, extending substantially beyond the deeply damped reference benchmark. The LAB polarization depends nonmonotonically on damping, reaching approximately $-1.52\%$ at $\Gamma=1~\mathrm{eV}$ and $+4.20\%$ at $0.2~\mathrm{eV}$, with a sign reversal at intermediate damping. This behavior is consistent with a changing balance between dissipative and coherent vibronic contributions, whose leading large-$\Gamma$ scalings at fixed electronic parameters and vibronic couplings are $\Gamma^{-1}$ and $\Gamma^{-2}$, respectively. These different damping dependences provide a qualitative interpretation of the sign reversal, while the calculations retain the full electronic response beyond these asymptotic forms.

The finite-time electronic response reproduces the LAB damping dependence very well, including the sign reversal. Near the crossing, at $\Gamma=0.4~\mathrm{eV}$, it retains the LAB sign, whereas the stationary response predicts the opposite sign, highlighting the importance of resolving the initial buildup of the vibrational response. The agreement remains very good even where the conservative small-parameter conditions are no longer well satisfied, demonstrating the practical accuracy of the finite-time description for the investigated dynamics beyond the perturbative reference regime. The subsequent coupling, vibrational-energy, and filling scans use $\Gamma=1~\mathrm{eV}$, which provides a compromise between perturbative control and a more pronounced spin response.

\begin{figure}[t]
    \centering
    \includegraphics[width=\columnwidth]{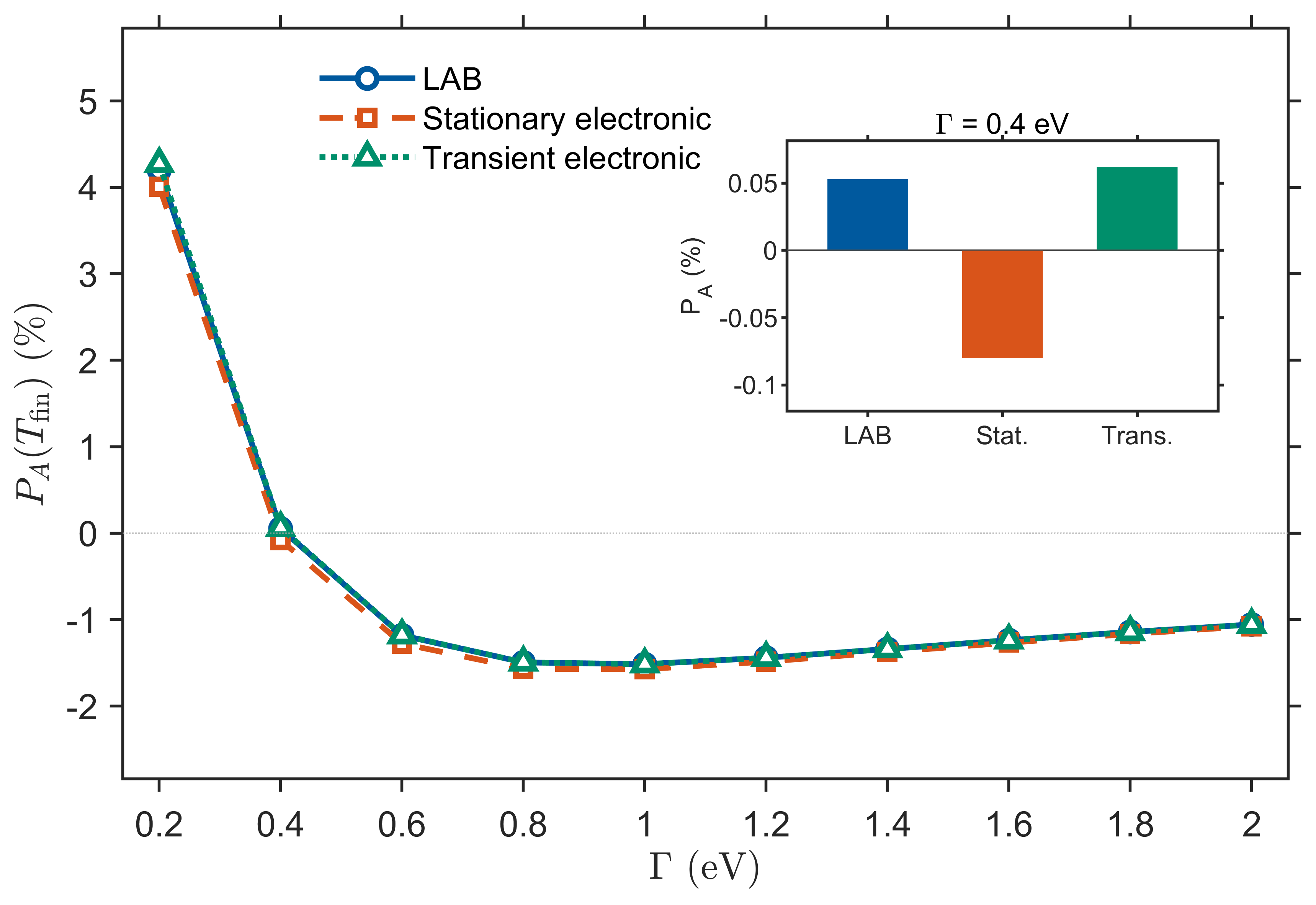}
    \caption{Final acceptor polarization at \(T_{\rm final}=1~\mathrm{ps}\) as a function of the vibrational damping.
    LAB, stationary electronic, and finite-time electronic calculations use the same electronic preparation at each \(\Gamma\), while the preparation changes between damping values through \(R(\Gamma)\).
    The finite-time response tracks the LAB result across the scan and retains its sign near the zero crossing at \(\Gamma=0.4~\mathrm{eV}\), where the stationary response predicts the opposite sign.
    The LAB calculations use \(n_{\max}=4\), checked against \(n_{\max}=5\) at every sampled value of \(\Gamma\).}
    \label{fig:Gamma_scan_main}
\end{figure}

\subsection{Vibronic couplings and sign-reversal symmetry}
\label{subsec:vibronic_symmetry}

We next vary the hopping and SOC modulation strengths over \(g,\chi=0,1,\ldots,20~\mathrm{meV}\), using the finite-time response at \(\Gamma=1~\mathrm{eV}\) and \(T_{\rm final}=1~\mathrm{ps}\).
The reference ratio is \(\chi/\lambda=15\).
Here \(\chi\) measures the variation of the SOC with a normalized vibrational coordinate, whereas \(\lambda\) is its equilibrium value.
Peierls coupling in torsionally flexible molecules can produce a modulation exceeding the static SOC, as discussed in Ref.~\cite{ChiesaDMI2026}.
The present values explore this regime without constituting a quantitative parametrization of a specific molecule.

\begin{figure*}[t]
    \centering
    \begin{minipage}[t]{0.49\textwidth}
        \centering
        \textbf{(a)}\par\smallskip
        \includegraphics[width=\linewidth] {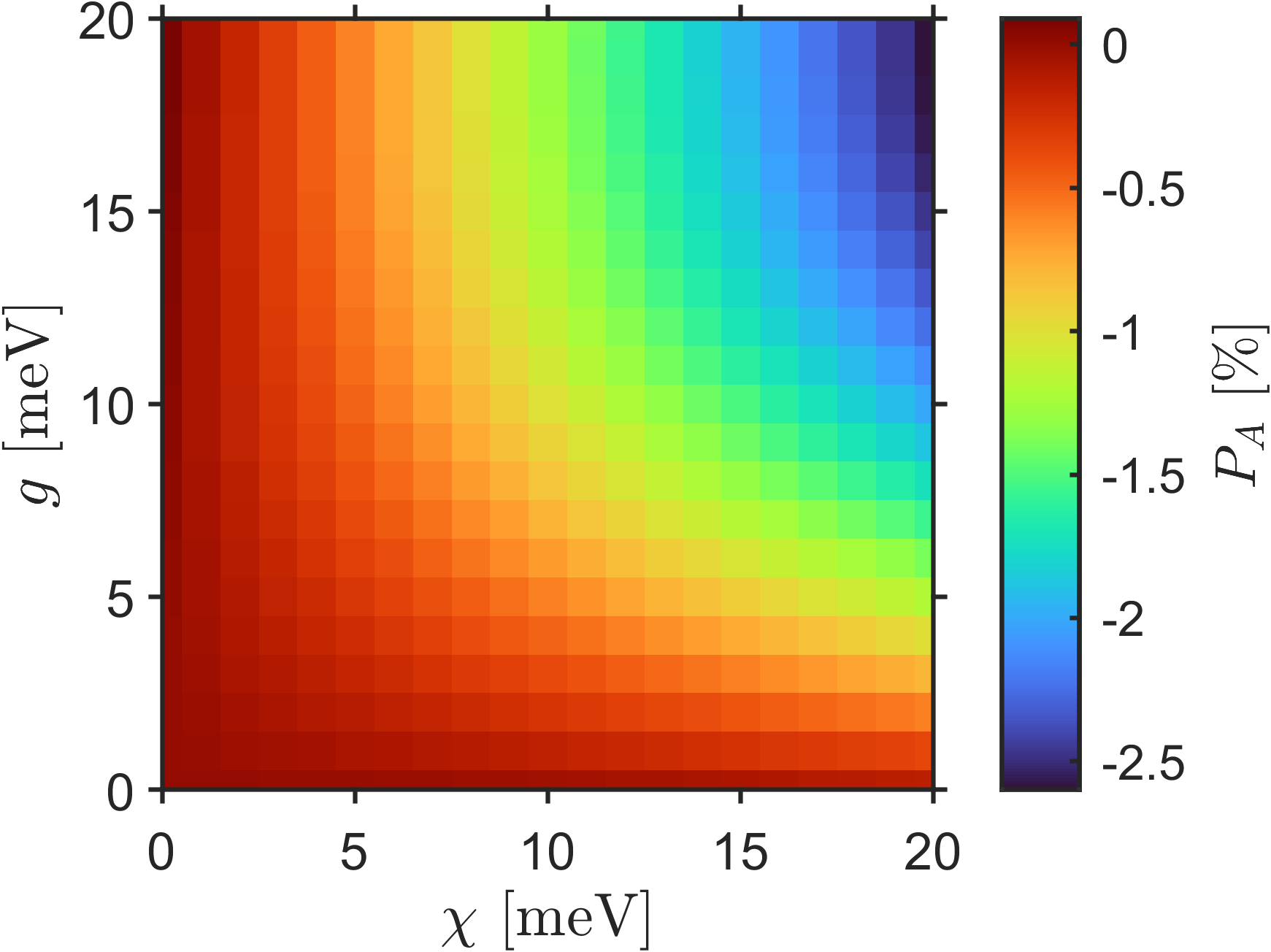}
    \end{minipage}
    \hfill
    \begin{minipage}[t]{0.49\textwidth}
        \centering
        \textbf{(b)}\par\smallskip
        \includegraphics[width=\linewidth] {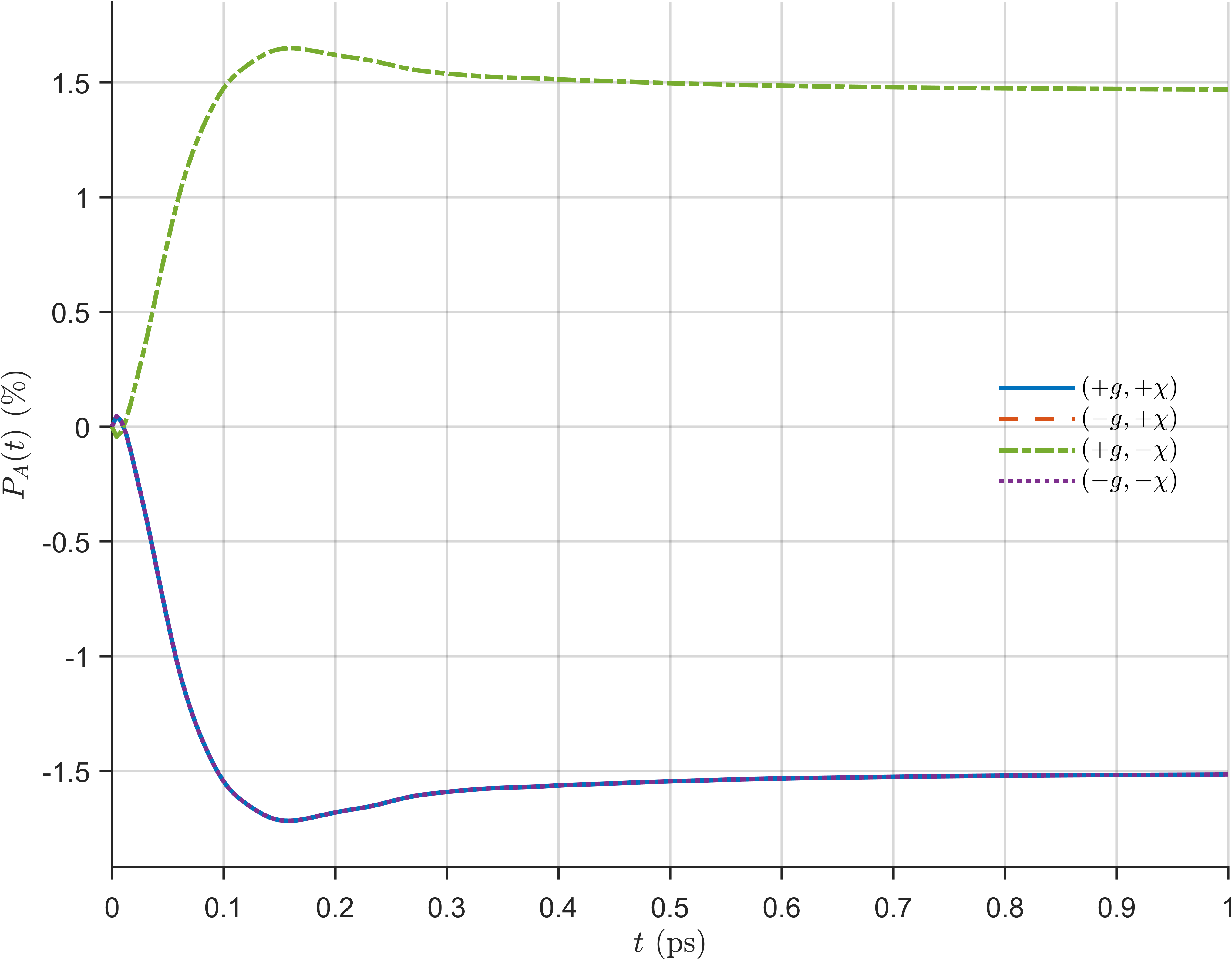}
    \end{minipage}
    \caption{Coupling dependence in the finite-time electronic model at \(\Gamma=1~\mathrm{eV}\).
    (a) Final polarization at \(T_{\rm final}=1~\mathrm{ps}\) over the \(21\times21\) coupling grid.
    Small nonzero edge signals are quantified in the text.
    (b) Time-dependent polarization for the four sign combinations at \(|g|=10~\mathrm{meV}\) and \(|\chi|=15~\mathrm{meV}\).
    Simultaneous reversal of both couplings leaves the curves unchanged; reversing either one changes the final sign but not with exactly equal magnitude.
    The preparation is recomputed for every coupling configuration.
    The zero-time point in (b) is a plotting convention; the polarization of an empty acceptor is undefined.}
    \label{fig:vibronic_symmetry}
\end{figure*}

Figure~\ref{fig:vibronic_symmetry}(a) shows an enhancement of the negative polarization when both modulations are increased, reaching \(-2.60\%\) at \(g=\chi=20~\mathrm{meV}\).
However, the signal does not vanish identically on the coordinate axes: at \((g,\chi)=(20,0)~\mathrm{meV}\), it is \(+0.09\%\), whereas at \((0,20)~\mathrm{meV}\) it is \(-0.15\%\).
Thus, the map cannot be described by an exact law proportional to \(g\chi\). 

The sign-reversal test in Fig.~\ref{fig:vibronic_symmetry}(b) resolves the symmetry content more directly.
For \(|g|=10~\mathrm{meV}\) and \(|\chi|=15~\mathrm{meV}\), define \(P_{s_gs_\chi}=P_A(s_g|g|,s_\chi|\chi|;T_{\rm final})\).
The final values are
\begin{equation}
    \begin{aligned}
        P_{++}=P_{--}&\simeq-1.52\%,\\
        P_{-+}=P_{+-}&\simeq+1.47\%.
    \end{aligned}
    \label{eq:sign_flip_values}
\end{equation}
Simultaneous reversal sends \(B\to-B\) and \(Q(t)\to-Q(t)\), leaving both the reduced generator and the preparation Hamiltonian unchanged.
Within this protocol,
\begin{equation}
    P_A(-g,-\chi;t)=P_A(g,\chi;t).
    \label{eq:simultaneous_reversal_main}
\end{equation}
The odd--even and even--odd sectors therefore vanish.
The remaining components are
\begin{equation}
    \begin{aligned}
        P_{\rm ee}
        &=\frac{P_{++}+P_{-+}+P_{+-}+P_{--}}{4},\\
        P_{\rm oo}
        &=\frac{P_{++}-P_{-+}-P_{+-}+P_{--}}{4},
    \end{aligned}
    \label{eq:Poo_main}
\end{equation}
which give \(P_{\rm ee}=-0.023\%\) and \(P_{\rm oo}=-1.49\%\).
The odd--odd component accounts for approximately \(98.48\%\) of \(|P_{++}|\), but does not exhaust it.
The parity decomposition does not uniquely identify the microscopic origin of the even--even remainder.
In particular, it does not isolate a contribution proportional only to \(t_{\rm hop}\lambda\): bare SOC, the even vibronic sectors, and the coupling dependence of the preparation are all retained.
For fixed \(t_{\rm hop}\), \(\lambda\), and all other electronic parameters, the coupling-parity symmetry allows
\begin{equation}
    P_A(g,\chi)=\Psi(g^2,\chi^2)
    +g\chi\,\Phi(g^2,\chi^2),
    \label{eq:PA_gchi_symmetry}
\end{equation}
where the observation time is also held fixed.
Equation~\eqref{eq:PA_gchi_symmetry} characterizes the dependence on the two Peierls couplings. Even though all Hamiltonian terms are included in the analysis, the sign-reversal shows that the dependence on the varied vibronic couplings is overwhelmingly odd--odd, so that its leading contribution is proportional to \(g\chi\).
This is the same mixed Peierls sector that generates the DMI and the mixed correlated-hopping interaction in Eq.~\eqref{eq:H4_decomposition_main}, together with the corresponding mixed one-body contribution.
The numerical symmetry therefore identifies the mixed hopping--SOC modulation as the dominant contribution associated with the \(g,\chi\) dependence, with minor corrections due for instance to contributions involving \(g\lambda\) and \(t_{\rm hop}\chi\).

\subsection{Vibrational energy and electronic filling}
\label{subsec:physical_controls}

\begin{figure*}[t]
    \centering
    \begin{minipage}[t]{0.49\textwidth}
        \centering
        \textbf{(a)}\par\smallskip
        \includegraphics[width=\linewidth] {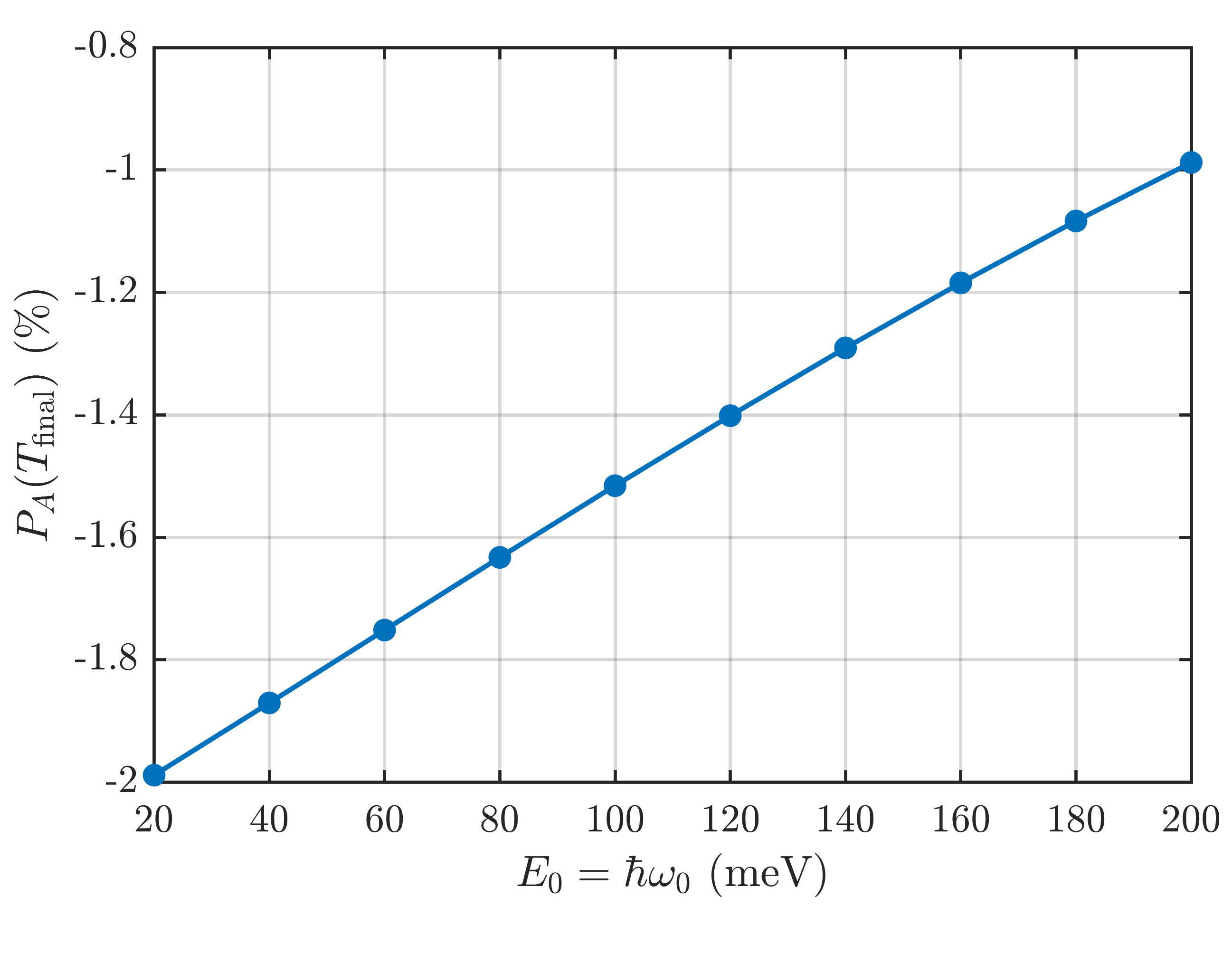}
    \end{minipage}
    \hfill
    \begin{minipage}[t]{0.49\textwidth}
        \centering
        \textbf{(b)}\par\smallskip
        \includegraphics[width=\linewidth] {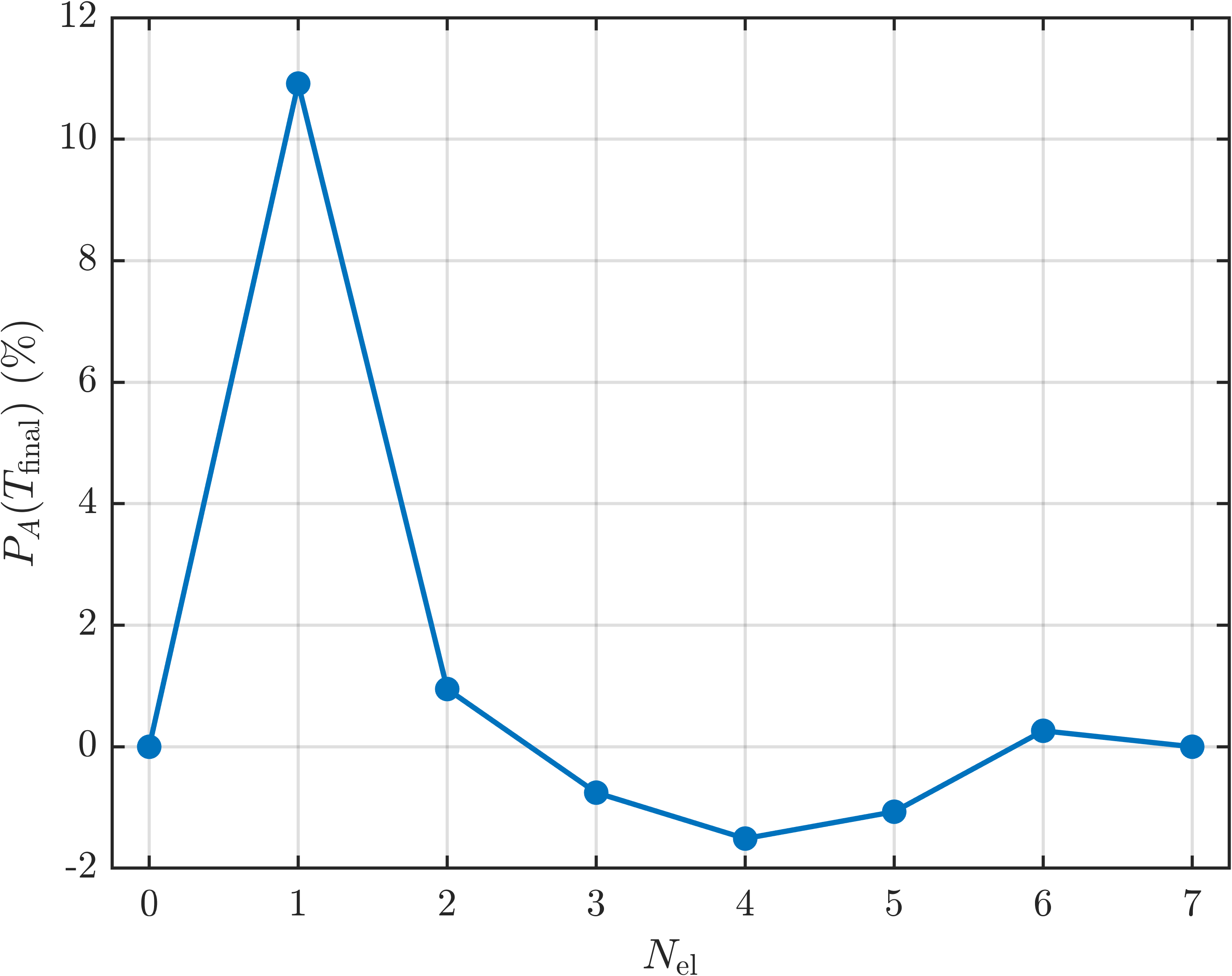}
    \end{minipage}
    \caption{Vibrational-energy and filling dependence at zero temperature, calculated with the finite-time electronic response and \(\Gamma=1~\mathrm{eV}\).
    (a) Final polarization versus \(E_0\) for a bridge initially containing four electrons, at \(T_{\rm final}=1~\mathrm{ps}\).
    The magnitude increases as the vibrational energy is reduced.
    (b) Final polarization versus the number \(N_{\rm el}\) of bridge electrons before injection, with \(E_0=100~\mathrm{meV}\) and \(T_{\rm final}=10~\mathrm{ps}\).
    The largest magnitude occurs at \(N_{\rm el}=1\), whereas the most negative value occurs at \(N_{\rm el}=4\).
    Other parameters follow Table~\ref{tab:baseline_parameters}.
    The spin-balanced preparation is recomputed at every point.
    Lines connecting the integer fillings are guides to the eye.}
    \label{fig:energy_filling}
\end{figure*}

The vibrational energy provides a further control of the polarization.
Figure~\ref{fig:energy_filling}(a) shows the finite-time electronic result for \(E_0=20\)--\(200~\mathrm{meV}\), with \(g=10~\mathrm{meV}\), \(\chi=15~\mathrm{meV}\), and \(\Gamma=1~\mathrm{eV}\).
Lowering \(E_0\) monotonically increases the polarization magnitude, from approximately \(0.99\%\) at \(200~\mathrm{meV}\) to \(1.99\%\) at \(20~\mathrm{meV}\).
Thus, over the explored range, reducing the vibrational energy approximately doubles \(|P_A|\).

The inverse-energy coefficients of the exact Lang--Firsov interaction, \(-B^2/E_0\), motivate examining low-energy modes.
However, the observed enhancement is not a direct measurement of a \(1/E_0\) interaction law in the damped system.
The leading stationary coherent contribution is instead
\begin{equation}
    -\frac{R}{E_0}B^2
    =-\frac{E_0}{\gamma^2+E_0^2}B^2.
    \label{eq:E0_damped_interaction_main}
\end{equation}
Since \(E_0<\gamma=0.5~\mathrm{eV}\) throughout this scan, its prefactor decreases in magnitude as the mode energy is lowered.
The increase in \(|P_A|\) therefore cannot be attributed to growth of this term alone.
The full response retains both coherent and dissipative contributions, with energy dependence through \(\gamma+i(E_0+\Delta_{mn})\), and the preparation also changes with \(E_0\).
Moreover, the explicit inverse power in \(F=-B/E_0\) cancels the prefactor proportional to \(E_0\) in \(Q(t)\) in the products entering the generator.
The result identifies an enhancement for lower-energy modes in the sampled interval, not a divergent low-energy law or an interior resonance.
Because \(B\), \(J_A\), and both dissipative coefficients are fixed, \(\xi_B=0.114\) and \(\xi_A=0.040\) remain unchanged along this scan.
This trend is relevant to the donor--bridge--acceptor molecule that motivates the present work.
The ab initio analysis of Ref.~\cite{phan_huu_ab_2025} identifies low-energy torsional modes at approximately \(1.2\), \(1.8\), and \(3.7~\mathrm{meV}\), associated with rotations between the molecular fragments and capable of modulating their electronic conjugation.
Related CISS models consider torsional energies in the \(1\)--\(10~\mathrm{meV}\) range, including the \(2\)--\(4~\mathrm{meV}\) interval \cite{ChiesaDMI2026}.
Such low-energy modes are therefore a natural target for future studies of vibration-mediated spin polarization.

Figure~\ref{fig:energy_filling}(b) shows the dependence on the initial bridge filling \(N_{\rm el}=0,\ldots,7\).
For this scan, \(E_0=100~\mathrm{meV}\) and the evolution is extended to \(T_{\rm final}=10~\mathrm{ps}\) to compare the different fillings in their late-time transfer regime.
All allowed states are retained, with no filling-dependent energy cutoff.
For odd \(N_{\rm el}\), the preparation is an equal incoherent mixture of the two time-reversed lowest-energy states with \(S^z_{\rm bridge}=\pm1/2\), so that both even and odd fillings start from spin-balanced ensembles.

The largest polarization magnitude is approximately \(+10.92\%\) at \(N_{\rm el}=1\), corresponding to two electrons after injection, whereas at half filling (\(N_{\rm el}=4\)) the polarization is about \(-1.51\%\).
The signal changes sign with filling and vanishes at the nearly empty and completely filled limits, where spin conservation and Pauli blocking strongly constrain the dynamics.

The control with uniform bridge energies, \(\epsilon_j=0\), is reported in Appendix~\ref{app:additional_results}.
In that case, the large positive response at \(N_{\rm el}=1\) is replaced by approximately \(-0.79\%\), while the largest magnitude shifts to \(N_{\rm el}=2\).
The pronounced \(N_{\rm el}=1\) peak is therefore not a generic consequence of low filling, but depends strongly on the electronic energy landscape and the associated initial preparation.
Its disappearance for uniform site energies points to a specific interplay between the sparse two-electron dynamics after injection and the nonuniform bridge spectrum, rather than to a universal low-filling optimum.
The filling scan therefore establishes a strong occupation dependence of the polarization without assigning it to a single microscopic interaction channel.
The finite-temperature extension, including thermal electronic preparation and vibrational absorption, is discussed separately in Appendix~\ref{app:additional_results}.

\section{Conclusions and outlook}
\label{sec:conclusions}

We have developed an open-system framework for electronic-electron dynamics mediated by a rapidly relaxing molecular vibration. The reduced dynamics retains both coherent and dissipative vibronic contributions, whose relative importance depends on damping. In the stationary decomposition, the explicit induced interaction $-RB^2/E_0$ is strongly suppressed; the full response also retains additional coherent and non-Hamiltonian contributions through $\Delta Q$. At fixed electronic parameters and vibronic couplings, the coherent and dissipative corrections scale as $\Gamma^{-2}$ and $\Gamma^{-1}$, respectively, in the large-$\Gamma$ limit. The leading vibronic correction is therefore dissipative in this limit. The polarization reported here is obtained from the combined coherent and dissipative dynamics.
The exact Lang--Firsov transformation of the microscopic electron--vibration Hamiltonian identifies the induced electronic contribution \(-B^2/(\hbar\omega_0)\), including one-body corrections and two-body isotropic, anisotropic, Dzyaloshinskii--Moriya, and correlated-hopping interactions.
Starting from the joint Lindblad equation, we have then eliminated the mixed electron--vibration correlations systematically to second order in the vibronic coupling.
The resulting purely electronic dynamics retains both the coherent vibration-mediated interactions and the dissipative terms generated by relaxation of the same microscopic mode, together with the finite-time buildup of its response.
Postulating the Lindbald equation for the system-bath dynamics in the present strongly damped regime is a critical assumption which needs to be double checked against microscopic reservoir models.
This would establish how the balance between coherent and dissipative contributions depends on the reservoir.

The physical relevance of vibration-induced exchange and dissipation for CISS has been emphasized in earlier work by Fransson.
In particular, vibration-induced exchange splitting was shown to produce sizable spin selectivity in molecular junctions \cite{Fransson2020}, while more recent work identified dissipation as an essential ingredient for establishing a stationary spin polarization \cite{Fransson2025}.
Our approach is complementary: rather than starting from the stationary transport response, it makes explicit the many-electron operators generated by the vibration and derives their coherent and incoherent evolution from the same damping parameter.
This also shows how the damping rate enters the transient electronic response before any specific transport or electron-transfer protocol is imposed.

Numerical simulations of electron-transfer provide a benchmark and a concrete illustration of this framework.
Unpolarized injection followed by spin-independent transfer produces percent-level acceptor polarization despite vibrational relaxation occurring on a much shorter timescale than the electronic dynamics.
Resolving the finite-time buildup of the vibrational response improves agreement with simulations retaining the vibrational mode explicitly.
The analysis further shows that the dependence of the polarization  on the two Peierls modulations is dominated by  contributions  proportional to \(g\chi\).
This identifies the mixed hopping--SOC modulation as the dominant coupling-dependent contribution in the present scan.
The strong dependence on damping, vibrational energy, and bridge filling likewise shows that the final polarization results from the complete nonequilibrium dynamics rather than from a single effective term.

A central physical conclusion is therefore that rapid relaxation and vibration-mediated spin--spin interactions are not mutually exclusive.
Even a short-lived vibrational mode can leave coherent interaction terms in the reduced electronic dynamics, while its coupling to the environment simultaneously generates dissipative electronic contributions.
Dissipation should consequently not be regarded only as a mechanism that destroys the vibronically mediated response.
In the present model, changing the damping can substantially modify and even reverse the polarization, and decreasing the very large reference damping can enhance \(|P_A|\) over parts of the explored parameter range.
The dependence is nevertheless nonmonotonic, so our results do not support a universal statement that weaker or stronger relaxation always enhances spin polarization.
Rather, damping controls the balance between coherent interaction, dissipative dynamics, and the finite response of the eliminated vibration.

This framework is not restricted to CISS or to molecular electron transfer.
More generally, whenever a molecular or collective coordinate modulates electronic operators, eliminating a relaxing mode can leave both effective electron--electron interactions and dissipative electronic terms whose relative importance depends on the mode frequency, damping, and electronic spectrum.
Extensions to multiple modes, structured or microscopic vibrational environments, and slower relaxation would therefore provide a route toward more realistic molecular applications.
For CISS specifically, incorporating molecule-dependent parameters and noncollinear SOC axes would allow closer comparison with individual systems.
Retaining the donor spin and the correlated photoexcited-pair preparation would also connect the present framework directly to the vibration-mediated DMI mechanism of Ref.~\cite{ChiesaDMI2026}, allowing the effect of vibrational relaxation on singlet--triplet mixing and coherence to be investigated explicitly.

\section*{Acknowledgments}
\noindent
The work was funded by the Horizon Europe Programme within the ERC-Synergy project CASTLE (proj. n. 101071533).
Views and opinions expressed are however those of the author(s) only and do not necessarily reflect those of the European Union or the European Commission.
Neither the European Union nor the granting authority can be held responsible for them.\\
AI tools (ChatGPT and Codex) were used to assist with checking analytical calculations, implementing numerical code for the applications presented in this work and improving the language of the manuscript.
The authors reviewed all AI-assisted output and take full responsibility for the content of the manuscript and the accompanying code.

\section*{Data Availability}
\noindent
The numerical data and analysis scripts that support the findings of this study are available from the corresponding author upon reasonable request.

\onecolumngrid
\appendix

\section{Exact Lang--Firsov transformation and interaction decomposition}
\label{app:exact_transformation}

In this Appendix we provide the derivation of the exact transformation used in Sec.~\ref{sec:microscopic_model}.
We consider the microscopic Hamiltonian
\begin{equation}
    H = H_{\rm e} + E_0\left(a^\dagger a+\frac{1}{2}\right) + (a^\dagger+a)B ,
    \label{eq:app_H_start}
\end{equation}
with \(H_{\rm e}\) and \(B\) defined in Eq.~\eqref{eq:He_B_main}.

\subsection{Diagonalization of the electron--vibration coupling}
Introducing the spinor
\begin{equation}
    \mathbf{c} = \left(c_{1\uparrow},c_{1\downarrow},\ldots,c_{4\uparrow},c_{4\downarrow}\right)^T ,
\end{equation}
the Hamiltonian can be expressed as
\begin{equation}
    H = \mathbf{c}^\dagger h\,\mathbf{c}
    + E_0\left(a^\dagger a+\frac{1}{2}\right)
    + \mathbf{c}^\dagger v\,\mathbf{c}\,(a^\dagger+a),
    \label{eq:app_matrix_form}
\end{equation}
where \(h\) and \(v\) are the single-particle matrices associated with \(H_{\rm e}\) and \(B\), respectively.
Since \(v=v^\dagger\), we introduce a unitary matrix \(P\) such that
\begin{equation}
    v = P Q_{\rm int} P^\dagger,
    \qquad
    Q_{\rm int} = \operatorname{diag}(q_1,q_1,q_2,q_2,q_3,q_3,q_4,q_4).
    \label{eq:app_diag_v}
\end{equation}
For the four-site open chain, the four spin-degenerate eigenvalues can be written compactly as
\begin{equation}
    (q_1,q_2,q_3,q_4)
    = \frac{\Delta}{\sqrt{2}}\left(-\phi,-\phi^{-1},\phi^{-1},\phi\right),
    \label{eq:app_q}
\end{equation}
where
\begin{equation}
    \Delta=\sqrt{g^2+\chi^2},
    \qquad
    \phi=\frac{1+\sqrt{5}}{2}.
\end{equation}
The appearance of \(\phi\) follows from the standing-wave eigenvectors of a four-site chain with open boundaries.
Defining
\begin{equation}
    \mathbf{d}=P^\dagger\mathbf{c},
    \qquad
    \widetilde h=P^\dagger hP,
    \label{eq:app_d_def}
\end{equation}
the Hamiltonian becomes
\begin{equation}
    H = \sum_{\alpha\beta}\widetilde h_{\alpha\beta}d_\alpha^\dagger d_\beta
    + E_0\left(a^\dagger a+\frac{1}{2}\right)
    + \sum_\alpha q_\alpha d_\alpha^\dagger d_\alpha(a^\dagger+a).
    \label{eq:app_H_d}
\end{equation}
In sums over \(\alpha\), \(q_\alpha\) denotes the diagonal entry of \(Q_{\rm int}\) associated with mode \(d_\alpha\), including both members of each degenerate pair.

\subsection{Lang--Firsov transformation}
We introduce the anti-Hermitian generator
\begin{equation}
    S = \frac{1}{E_0}\sum_\alpha q_\alpha d_\alpha^\dagger d_\alpha(a^\dagger-a),
    \qquad
    S^\dagger=-S,
    \label{eq:app_S}
\end{equation}
and transform the Hamiltonian according to
\begin{equation}
    \widetilde H=e^SHe^{-S}.
\end{equation}
The bosonic operators transform as
\begin{equation}
    e^Sae^{-S} = a-\frac{1}{E_0}\sum_\alpha q_\alpha d_\alpha^\dagger d_\alpha ,
    \label{eq:app_a_transform}
\end{equation}
and analogously for \(a^\dagger\), while
\begin{equation}
    e^Sd_\alpha e^{-S}
    = d_\alpha\exp\left[-\frac{q_\alpha}{E_0}(a^\dagger-a)\right].
    \label{eq:app_d_transform}
\end{equation}
Consequently,
\begin{equation}
    e^Sd_\alpha^\dagger d_\beta e^{-S}
    = d_\alpha^\dagger d_\beta
    \exp\left[\frac{q_\alpha-q_\beta}{E_0}(a^\dagger-a)\right].
    \label{eq:app_bilinear_transform}
\end{equation}
Collecting the bosonic and electron--vibration contributions, the terms linear in \(a^\dagger+a\) cancel exactly, yielding
\begin{equation}
\begin{split}
    \widetilde H ={}&
    \sum_{\alpha\beta}\widetilde h_{\alpha\beta}d_\alpha^\dagger d_\beta
    \exp\left[\frac{q_\alpha-q_\beta}{E_0}(a^\dagger-a)\right]+E_0\left(a^\dagger a+\frac{1}{2}\right)
    -\frac{1}{E_0}\left(\sum_\alpha q_\alpha d_\alpha^\dagger d_\alpha\right)^2 .
\end{split}
\label{eq:app_H_exact}
\end{equation}
No expansion in the electron--vibration couplings has been performed in Eq.~\eqref{eq:app_H_exact}.
Since the transformation \(P\) diagonalizes the operator \(B\),
\begin{equation}
    \sum_\alpha q_\alpha d_\alpha^\dagger d_\alpha = B,
\end{equation}
the purely electronic contribution generated by the transformation can equivalently be written directly in the original site basis as
\begin{equation}
    H_{\rm ind} = -\frac{B^2}{E_0}.
    \label{eq:app_Hind_B2}
\end{equation}
Using Eq.~\eqref{eq:He_B_main}, this becomes
\begin{equation}
    H_{\rm ind}
    = -\frac{1}{2E_0}
    \left[
        g^2H_{\rm hop}^2
        +\chi^2H_{\rm SOC}^2
        -g\chi\left\{H_{\rm hop},H_{\rm SOC}\right\}
    \right].
    \label{eq:app_Hind_compact}
\end{equation}
Equation~\eqref{eq:app_Hind_compact} makes explicit that the mixed interaction channel originates from the simultaneous vibrational modulation of hopping and SOC.

\subsection{One-body and quartic contributions}
The operators \(H_{\rm hop}\) and \(H_{\rm SOC}\) are fermionic bilinears.
Their products therefore contain terms involving either two or four fermionic operators.
Expanding Eq.~\eqref{eq:app_Hind_compact} in the original site basis and repeatedly using the canonical anticommutation relations,
\begin{equation}
    \{c_{i\sigma},c_{j\sigma'}^\dagger\}=\delta_{ij}\delta_{\sigma\sigma'},
    \qquad
    \{c_{i\sigma},c_{j\sigma'}\}=0,
    \label{eq:app_anticommutation}
\end{equation}
the induced interaction can be separated as
\begin{equation}
    H_{\rm ind}=H_{\rm ind}^{(2)}+H_{\rm ind}^{(4)},
    \label{eq:app_split}
\end{equation}
where \(H_{\rm ind}^{(2)}\) contains the effective one-body terms and \(H_{\rm ind}^{(4)}\) contains the two-body interactions, with two creation and two annihilation operators.
The induced one-body contribution is
\begin{equation}
\begin{split}
    H_{\rm ind}^{(2)} ={}&
    -\frac{g^2+\chi^2}{2E_0}
    \sum_\sigma\left[n_{1\sigma}+n_{4\sigma}+2(n_{2\sigma}+n_{3\sigma})\right] \\
    &-\frac{g^2-\chi^2}{2E_0}
    \sum_{j=1}^{2}\sum_\sigma
    \left(c_{j\sigma}^\dagger c_{j+2,\sigma}+\mathrm{h.c.}\right) \\
    &+\frac{g\chi}{E_0}
    \sum_{j=1}^{2}\sum_{\sigma\sigma'}
    \left[i c_{j\sigma}^\dagger(\sigma_z)_{\sigma\sigma'}c_{j+2,\sigma'}+\mathrm{h.c.}\right].
\end{split}
\label{eq:app_Hind_onebody}
\end{equation}
Thus, even though the microscopic electron--vibration coupling acts only on nearest-neighbor bonds, the transformation generates next-nearest-neighbor hopping and spin-dependent hopping processes.\\
The genuine quartic contribution can be grouped according to its dependence on the two Peierls couplings:
\begin{equation}
\begin{split}
    H_{\rm ind}^{(4)} ={}&
    \frac{g^2}{E_0}\left(H_{\rm Heisenberg}+H_{\rm iso}\right)
    +\frac{\chi^2}{E_0}\left(H_{\rm XXZ}+H_{\rm aniso}\right) -\frac{2g\chi}{E_0}\left(H_{\rm DM}-H_{\rm corr}\right).
\end{split}
\label{eq:app_quartic_decomposition}
\end{equation}
Introducing
\begin{equation}
    S_j^\mu = \frac{1}{2}\sum_{\sigma\sigma'}c_{j\sigma}^\dagger(\sigma_\mu)_{\sigma\sigma'}c_{j\sigma'},
    \qquad
    n_j=\sum_\sigma n_{j\sigma},
    \label{eq:app_spin_def}
\end{equation}
the nearest-neighbor contributions are
\begin{equation}
    H_{\rm Heisenberg}
    = \sum_{j=1}^{3}
    \left[
        2S_j^zS_{j+1}^z
        +\frac{1}{2}n_jn_{j+1}
        +S_j^-S_{j+1}^+
        +S_j^+S_{j+1}^-
    \right],
    \label{eq:app_HHeis}
\end{equation}
\begin{equation}
    H_{\rm XXZ}
    = \sum_{j=1}^{3}
    \left[
        2S_j^zS_{j+1}^z
        +\frac{1}{2}n_jn_{j+1}
        -S_j^-S_{j+1}^+
        -S_j^+S_{j+1}^-
    \right],
    \label{eq:app_HXXZ}
\end{equation}
and
\begin{equation}
    H_{\rm DM}
    = 2\sum_{j=1}^{3}\left(\mathbf{S}_j\times\mathbf{S}_{j+1}\right)_z.
    \label{eq:app_HDM}
\end{equation}
For completeness, it is useful to introduce a compact notation for the remaining quartic terms without explicitly listing all the individual fermionic products.
We denote by \([O]_{\rm 4f}\) the part of an operator \(O\) containing two creation and two annihilation operators after normal ordering with respect to the fermionic vacuum.
Equation~\eqref{eq:app_Hind_compact} then gives
\begin{align}
    H_{\rm Heisenberg}+H_{\rm iso}
    &= -\frac{1}{2}\left[H_{\rm hop}^2\right]_{\rm 4f},
    \label{eq:app_iso_compact}
    \\
    H_{\rm XXZ}+H_{\rm aniso}
    &= -\frac{1}{2}\left[H_{\rm SOC}^2\right]_{\rm 4f},
    \label{eq:app_aniso_compact}
    \\
    H_{\rm DM}-H_{\rm corr}
    &= -\frac{1}{4}\left[\left\{H_{\rm hop},H_{\rm SOC}\right\}\right]_{\rm 4f}.
    \label{eq:app_chiral_compact}
\end{align}
These relations provide an exact and compact definition of the residual operators \(H_{\rm iso}\), \(H_{\rm aniso}\), and \(H_{\rm corr}\), where \(H_{\rm corr}\) is the mixed correlated-hopping interaction introduced in the main text.
These residual operators collect the quartic processes that extend beyond the nearest-neighbor exchange terms explicitly displayed above, including correlated hopping, pair-transfer, and bond--bond processes.
The structure of Eqs.~\eqref{eq:app_iso_compact}--\eqref{eq:app_chiral_compact} also makes the different symmetry sectors transparent.
The \(g^2\) and \(\chi^2\) contributions are even under sign reversal of the corresponding Peierls couplings, whereas the mixed contribution is proportional to \(g\chi\) and is therefore odd under either \(g\rightarrow-g\) or \(\chi\rightarrow-\chi\).
This coupling parity characterizes the interaction decomposition, but does not by itself impose exact oddness of the full acceptor polarization under reversal of either coupling.

\subsection{Explicit form in the original site basis and consistency check}
\label{app:explicit_H}
For completeness, we finally report the exact transformed Hamiltonian fully re-expressed in the original site basis.
This expression is algebraically equivalent to the compact form in Eq.~\eqref{eq:H_LF_compact_main}, but makes explicit all vibration-induced one-body, many-body, and vibrationally dressed electronic processes.
Defining
\begin{equation}
    \hat p=\frac{i}{\sqrt{2}}(a^\dagger-a),
    \qquad
    \zeta=\frac{\sqrt{g^2+\chi^2}}{E_0},
    \qquad
    \phi=\frac{1+\sqrt{5}}{2},
    \label{eq:recap_definitions}
\end{equation}
the transformed Hamiltonian for \(g^2+\chi^2>0\) reads
\allowdisplaybreaks[4]
\begin{align}
\widetilde H ={}&
E_0\left(a^\dagger a+\frac{1}{2}\right)
-\frac{g^2+\chi^2}{2E_0}
\sum_{\sigma}
\Big[n_{1,\sigma}+n_{4,\sigma}+2(n_{2,\sigma}+n_{3,\sigma})\Big]
\notag\\
&-\frac{g^2-\chi^2}{2E_0}
\sum_{j=1}^{2}\sum_{\sigma}
\Big[c^\dagger_{j,\sigma}c_{j+2,\sigma}+\mathrm{h.c.}\Big]
+\frac{g\chi}{E_0}
\sum_{j=1}^{2}\sum_{\sigma}
\Big[i\,c^\dagger_{j,\sigma}\sigma_z c_{j+2,\sigma}+\mathrm{h.c.}\Big]
\notag\\
&-\frac{gt_{\rm hop}+\lambda\chi}{g^2+\chi^2}
\sum_{j=1}^{3}\sum_{\sigma}
\Big[
g\left(c^\dagger_{j,\sigma}c_{j+1,\sigma}+\mathrm{h.c.}\right)
-\chi\left(i\,c^\dagger_{j,\sigma}\sigma_z c_{j+1,\sigma}+\mathrm{h.c.}\right)
\Big]
\notag\\
&+\frac{g^2}{E_0}\left(H_{\mathrm{Heisenberg}}+H_{\mathrm{iso}}\right)
+\frac{\chi^2}{E_0}\left(H_{\mathrm{XXZ}}+H_{\mathrm{aniso}}\right)
-\frac{2g\chi}{E_0}\left(H_{\mathrm{DM}}-H_{\mathrm{corr}}\right)
\notag\\
&+\frac{1}{5}\frac{g\lambda-t_{\rm hop}\chi}{g^2+\chi^2}
\sum_{\sigma}\sum_{j=1}^{3}
\Big[
\chi\,c_{j,\sigma}^{\dagger}c_{j+1,\sigma}
+ig\,c_{j,\sigma}^{\dagger}\sigma_z c_{j+1,\sigma}
+\mathrm{h.c.}
\Big]
\notag\\
&\qquad\times
\Big[
(2+2\delta_{j,2})\cos(\zeta\hat p)
+(\phi^2-\delta_{j,2})\cos(2\phi^{-1}\zeta\hat p)
+(\phi^{-2}-\delta_{j,2})\cos(2\phi\zeta\hat p)
\Big]
\notag\\
&+\frac{1}{5}\frac{g\lambda-t_{\rm hop}\chi}{(g^2+\chi^2)^{3/2}}
\sum_{\sigma}
\Bigg\{
2g\chi
\Big[
i\left(c_{1,\sigma}^{\dagger}c_{3,\sigma}
-c_{2,\sigma}^{\dagger}c_{4,\sigma}\right)
+\mathrm{h.c.}
\Big]
\notag\\
&\qquad\qquad
-(g^2-\chi^2)
\Big[
\left(
c_{1,\sigma}^{\dagger}\sigma_z c_{3,\sigma}
-c_{2,\sigma}^{\dagger}\sigma_z c_{4,\sigma}
\right)
+\mathrm{h.c.}
\Big]
\Bigg\}
\notag\\
&\qquad\times
\Big[
-2\sin(\zeta\hat p)
+\phi^2\sin(2\phi^{-1}\zeta\hat p)
-\phi^{-2}\sin(2\phi\zeta\hat p)
\Big]
\notag\\
&+\frac{1}{5}\frac{g\lambda-t_{\rm hop}\chi}{(g^2+\chi^2)^2}
\sum_{\sigma}
\Big[
(\chi^3-3g^2\chi)c_{1,\sigma}^{\dagger}c_{4,\sigma}
-i(g^3-3g\chi^2)c_{1,\sigma}^{\dagger}\sigma_z c_{4,\sigma}
+\mathrm{h.c.}
\Big]
\notag\\
&\qquad\times
\Big[
-4\cos(\zeta\hat p)
+\phi^3\cos(2\phi^{-1}\zeta\hat p)
-\phi^{-3}\cos(2\phi\zeta\hat p)
\Big]
\notag\\
&+\frac{2}{5}\frac{g\lambda-t_{\rm hop}\chi}{\sqrt{g^2+\chi^2}}
\Bigg[
4\left(S_1^z-S_4^z+S_2^z-S_3^z\right)\sin(\zeta\hat p)
\notag\\
&\qquad
+\left[
\phi^3(S_1^z-S_4^z)-\phi(S_2^z-S_3^z)
\right]\sin(2\phi^{-1}\zeta\hat p)
\notag\\
&\qquad
+\left[
\phi^{-3}(S_1^z-S_4^z)-\phi^{-1}(S_2^z-S_3^z)
\right]\sin(2\phi\zeta\hat p)
\Bigg].
\label{eq:H_final_explicit}
\end{align}
In the terms containing a single sum over the spin index, \(\sigma_z\) denotes the corresponding eigenvalue of the Pauli matrix in the chosen spin basis, i.e., \(+1\) for \(\sigma=\uparrow\) and \(-1\) for \(\sigma=\downarrow\).
Equation~\eqref{eq:H_final_explicit} is the explicit representation in the original \(c_{j\sigma}\) basis of the compact transformed Hamiltonian in Eq.~\eqref{eq:H_LF_compact_main}.
No approximation has been introduced in passing from one representation to the other.
At \(g=\chi=0\), the apparent singularities in this representation are avoided by using the compact form, for which \(B=S=0\) and \(\widetilde H=H\).
The analytical relation between the microscopic and transformed Hamiltonians is
\begin{equation}
    \widetilde H_{\rm explicit}=U_{\rm LF}HU_{\rm LF}^\dagger,
    \qquad U_{\rm LF}=e^S.
\end{equation}
As an independent consistency check, we numerically compared the explicit Hamiltonian in Eq.~\eqref{eq:H_final_explicit} with a direct implementation of the Lang--Firsov transformation of the microscopic Hamiltonian, expressing both in the same basis.
This comparison was used to check the coefficients, relative signs, and operator structure of the explicit expression.

\section{Damped vibrational response and reduced electronic dynamics}
\label{app:adiabatic_elimination}

This Appendix derives the electronic equation used in Sec.~\ref{sec:overdamped}, including the irreversible acceptor transfer introduced in Sec.~\ref{sec:electron_transfer}.
We first relate the laboratory and Lang--Firsov descriptions exactly, then construct the finite-time vibrational response and its stationary limit.
The electronic evolution is obtained separately by retaining the leading mixed electron--vibration correlations.

\subsection{Dissipative model and exact Lang--Firsov transformation}
\label{appB:frames}
Let \(H_{\rm B}\) denote the complete bare bridge Hamiltonian, including the site energies and on-site Coulomb repulsion, and let \(H_A\) describe the acceptor:
\begin{equation}
\begin{aligned}
    H_{\rm B}&=-t_{\rm hop}H_{\rm hop}+\lambda H_{\rm SOC}
    +\sum_{j=1}^{4}\sum_\sigma\epsilon_j n_{j\sigma}
    +U\sum_{j=1}^{4}n_{j\uparrow}n_{j\downarrow},\\
    H_A&=\epsilon_A N_A,
    \qquad H_{\rm el}=H_{\rm B}+H_A.
\end{aligned}
\label{eq:appB_Hel}
\end{equation}
The electronic Hamiltonian \(H_{\rm el}\) does not contain a vibration-induced correction.
The laboratory-frame Hamiltonian is
\begin{equation}
    H_{\rm LAB}=H_{\rm el}+E_0\left(a^\dagger a+\frac12\right)+B(a^\dagger+a),
    \qquad B=\frac{-gH_{\rm hop}+\chi H_{\rm SOC}}{\sqrt{2}},
    \qquad E_0=\hbar\omega_0,
    \label{eq:appB_HLAB}
\end{equation}
where \(H_{\rm hop}\) and \(H_{\rm SOC}\) are defined in Sec.~\ref{sec:microscopic_model}.
The acceptor is restricted to its empty and singly occupied states, \(N_A\leq1\), and no coherent bridge--acceptor hopping is included.
The irreversible transfer channel is the single operator
\begin{equation}
    J_A=\sum_{\sigma=\uparrow,\downarrow}J_{A\sigma},
    \qquad J_{A\sigma}=c_{A\sigma}^\dagger(1-n_{A\bar\sigma})c_{4\sigma},
    \label{eq:appB_JA}
\end{equation}
with no reverse-transfer jump.
The sum over spin is taken before constructing the dissipator.\\
At zero bath temperature, the joint density matrix \(\rho\) obeys
\begin{equation}
    \hbar\dot\rho=-i[H_{\rm LAB},\rho]+\Gamma\mathcal D[a]\rho+\Gamma_A\mathcal D[J_A]\rho,
    \qquad \mathcal D[L]Z=LZL^\dagger-\frac12\{L^\dagger L,Z\}.
    \label{eq:appB_LAB_master}
\end{equation}
The coefficients \(\Gamma\) and \(\Gamma_A\) have units of energy; the corresponding rates contain an additional factor \(1/\hbar\).
Equation~\eqref{eq:appB_LAB_master} specifies the phenomenological Markovian model used throughout the derivation.
The change of frame is exact within this model and does not require strong damping.\\
As in Appendix~\ref{app:exact_transformation}, define
\begin{equation}
    S=\frac{B}{E_0}(a^\dagger-a),
    \qquad U_{\rm LF}=e^S,
    \qquad F=-\frac{B}{E_0}.
    \label{eq:appB_LF_definitions}
\end{equation}
Since \(B^\dagger=B\) and the electronic and vibrational operators commute, \(S^\dagger=-S\) and \(F^\dagger=F\).
The transformed density matrix and operators are
\begin{equation}
    \widetilde\rho=U_{\rm LF}\rho U_{\rm LF}^\dagger,
    \qquad \widetilde O=U_{\rm LF}OU_{\rm LF}^\dagger.
    \label{eq:appB_LF_state_operators}
\end{equation}
A dissipator transforms covariantly under this time-independent unitary transformation:
\begin{equation}
\begin{aligned}
    U_{\rm LF}\bigl(\mathcal D[L]\rho\bigr)U_{\rm LF}^\dagger
    &=\widetilde L\widetilde\rho\widetilde L^\dagger
    -\frac12\{\widetilde L^\dagger\widetilde L,\widetilde\rho\}=\mathcal D[\widetilde L]\widetilde\rho.
\end{aligned}
\label{eq:appB_dissipator_covariance}
\end{equation}
Consequently, both loss and transfer operators must be transformed along with the Hamiltonian.\\
For the annihilation operator,
\begin{equation}
    [S,a]=-\frac{B}{E_0},
    \qquad [S,B]=0,
\end{equation}
so the Baker--Campbell--Hausdorff series terminates after the first commutator:
\begin{equation}
    \widetilde a=a+F,
    \qquad \widetilde a^\dagger=a^\dagger+F,
    \qquad \widetilde B=B,
    \qquad \widetilde F=F.
    \label{eq:appB_transformed_a}
\end{equation}
The transformed Hamiltonian is
\begin{equation}
    \widetilde H=\widetilde H_{\rm B}+H_A+E_0\left(a^\dagger a+\frac12\right)-\frac{B^2}{E_0},
    \qquad \widetilde H_{\rm B}=U_{\rm LF}H_{\rm B}U_{\rm LF}^\dagger.
    \label{eq:appB_HLF}
\end{equation}
Here \([S,H_A]=0\), because \(B\) acts only on the bridge.
All terms in \(H_{\rm B}\), including the site-energy and Hubbard contributions, are included in \(\widetilde H_{\rm B}\).
Their transformed displacement dependence has not been discarded.\\
Similarly, \([B,c_{A\sigma}^\dagger]=[B,n_{A\sigma}]=0\), giving
\begin{equation}
    \widetilde J_A=\sum_\sigma c_{A\sigma}^\dagger(1-n_{A\bar\sigma})\widetilde c_{4\sigma},
    \qquad \widetilde c_{4\sigma}=U_{\rm LF}c_{4\sigma}U_{\rm LF}^\dagger.
    \label{eq:appB_transformed_JA}
\end{equation}
The exact transformed master equation is therefore
\begin{equation}
    \hbar\dot{\widetilde\rho}
    =-i[\widetilde H,\widetilde\rho]
    +\Gamma\mathcal D[a+F]\widetilde\rho
    +\Gamma_A\mathcal D[\widetilde J_A]\widetilde\rho.
    \label{eq:appB_LF_master}
\end{equation}
The coefficients of the dissipators are unchanged.
The initial state and observables obey
\begin{equation}
    \widetilde\rho(0)=U_{\rm LF}\rho(0)U_{\rm LF}^\dagger,
    \qquad \operatorname{Tr}(\rho O)=\operatorname{Tr}(\widetilde\rho\widetilde O).
    \label{eq:appB_initial_observables}
\end{equation}
In particular, \(\widetilde n_{A\sigma}=n_{A\sigma}\), whereas a factorized LAB initial state need not remain factorized in the LF representation.

\subsection{Vibrational fluctuations, response time, and initial preparation}
\label{appB:response_time}
Strong damping shortens the response time, but does not suppress the vibrational quadrature as an operator.
For
\begin{equation}
    K=a^\dagger-a,
    \qquad \hat p=\frac{iK}{\sqrt{2}},
\end{equation}
the vacuum has
\begin{equation}
    \langle K\rangle_0=0,
    \qquad \langle K^2\rangle_0=-1,
    \qquad \langle\hat p^2\rangle_0-\langle\hat p\rangle_0^2=\frac12.
    \label{eq:appB_vacuum_fluctuations}
\end{equation}
The uncoupled oscillator vacuum is stationary under \(\mathcal D[a]\) for every \(\Gamma\), but retains these fluctuations.
The replacement \(K\simeq0\) therefore does not follow from increasing the loss coefficient.\\
The same distinction appears in the LF displacement factors.
For real \(z\), the commutator \([za^\dagger,-za]=z^2 I\) is a scalar, so that
\begin{equation}
    e^{zK}=e^{-z^2/2}e^{za^\dagger}e^{-za},
    \qquad \langle0|e^{zK}|0\rangle=e^{-z^2/2}.
    \label{eq:appB_displacement_average}
\end{equation}
For the factors multiplying electronic bilinears in Appendix~\ref{app:exact_transformation}, this gives
\begin{equation}
    X_{\alpha\beta}=\exp\left[\frac{q_\alpha-q_\beta}{E_0}K\right],
    \qquad \langle0|X_{\alpha\beta}|0\rangle
    =\exp\left[-\frac{(q_\alpha-q_\beta)^2}{2E_0^2}\right].
    \label{eq:appB_Xab_average}
\end{equation}
The response time can be obtained exactly from the physical LAB amplitude
\begin{equation}
    \alpha_{\rm vib}(t)=\operatorname{Tr}[\rho(t)a].
    \label{eq:appB_alpha_definition}
\end{equation}
The required Hamiltonian commutator is \([a,H_{\rm LAB}]=E_0a+B\).
Cyclicity of the full trace and \(aa^\dagger a=a^\dagger a^2+a\) give
\begin{equation}
\begin{aligned}
    \operatorname{Tr}[a\mathcal D[a]\rho]
    &=\frac12\operatorname{Tr}[(a^\dagger a^2-aa^\dagger a)\rho]
    =-\frac12\alpha_{\rm vib},\\
    \operatorname{Tr}[a\mathcal D[J_A]\rho]
    &=\operatorname{Tr}\left[\left(J_A^\dagger aJ_A-\frac12\{J_A^\dagger J_A,a\}\right)\rho\right]
    =0.
\end{aligned}
\label{eq:appB_alpha_dissipators}
\end{equation}
The second equality follows from \([a,J_A]=[a,J_A^\dagger]=0\).
Acceptor transfer thus has no direct contribution to the amplitude equation, although it changes the evolving electronic state and hence the forcing \(\langle B\rangle\).
With
\begin{equation}
    \gamma=\frac{\Gamma}{2},
    \qquad \kappa=\gamma+iE_0,
    \label{eq:appB_gamma_kappa}
\end{equation}
the exact equation and its integrating-factor solution are
\begin{align}
    \hbar\dot\alpha_{\rm vib}(t)
    &=-\kappa\alpha_{\rm vib}(t)-i\langle B\rangle_{\rm LAB}(t),
    \label{eq:appB_alpha_eom}\\
    \alpha_{\rm vib}(t)
    &=e^{-\kappa t/\hbar}\alpha_{\rm vib}(0)
    -\frac{i}{\hbar}\int_0^t e^{-\kappa(t-s)/\hbar}\langle B\rangle_{\rm LAB}(s)\,ds.
    \label{eq:appB_alpha_integral}
\end{align}
Since
\begin{equation}
    e^{-\kappa\tau/\hbar}=e^{-\gamma\tau/\hbar}e^{-iE_0\tau/\hbar},
    \qquad \tau_{\rm vib}=\frac{\hbar}{\gamma}=\frac{2\hbar}{\Gamma},
    \label{eq:appB_response_time}
\end{equation}
\(E_0/\hbar\) determines the oscillatory phase and \(\gamma/\hbar\) determines the decay of the response envelope.
The electronic forcing can maintain a nonzero amplitude after the homogeneous transient has decayed.\\
To distinguish the amplitudes in the two frames, define \(\langle O\rangle_{\rm LF}=\operatorname{Tr}(\widetilde\rho O)\).
The exact transformation gives
\begin{equation}
    \alpha_{\rm vib}=\langle a\rangle_{\rm LAB}=\langle a+F\rangle_{\rm LF},
    \qquad \langle F\rangle_{\rm LAB}=\langle F\rangle_{\rm LF}.
    \label{eq:appB_amplitude_frames}
\end{equation}
Thus \(\alpha_{\rm vib}\) is not generally the expectation value of the canonical symbol \(a\) in the LF state.
Using \(B=-E_0F\) and differentiating the complete relation in Eq.~\eqref{eq:appB_amplitude_frames}, one obtains
\begin{equation}
    \hbar\frac{d\langle a\rangle_{\rm LF}}{dt}
    =-\kappa\langle a\rangle_{\rm LF}-\gamma\langle F\rangle_{\rm LF}
    -\hbar\frac{d\langle F\rangle_{\rm LF}}{dt}.
    \label{eq:appB_canonical_LF_eom}
\end{equation}
The last term is required by the change of frame.
It cannot be omitted while retaining corrections associated with the electronic time dependence of \(F\).\\
We prescribe the initial state in the laboratory frame as
\begin{equation}
    \rho(0)=\rho_{\rm el}^{(0)}\otimes|0\rangle_{\rm vib}\langle0|,
    \qquad \operatorname{Tr}_{\rm el}\rho_{\rm el}^{(0)}=1.
    \label{eq:appB_initial_state}
\end{equation}
This assumption describes an initially unexcited mode and does not require a particular electronic pure state or filling.
It implies \(\alpha_{\rm vib}(0)=0\), but not \(\langle a\rangle_{\rm LF}(0)=0\).
Combining Eqs.~\eqref{eq:appB_alpha_integral} and \eqref{eq:appB_amplitude_frames}, and writing \(\tau=t-s\), gives the exact relation
\begin{equation}
    \langle a\rangle_{\rm LF}(t)
    =-\langle F\rangle_{\rm LF}(t)
    +\frac{iE_0}{\hbar}\int_0^t e^{-\kappa\tau/\hbar}\langle F\rangle_{\rm LF}(t-\tau)\,d\tau.
    \label{eq:appB_LF_amplitude_integral}
\end{equation}
In particular, \(\langle a\rangle_{\rm LF}(0)=-\langle F\rangle_{\rm LF}(0)\).
The corresponding LF initial state is obtained from Eq.~\eqref{eq:appB_initial_observables}, not by separately assigning a vacuum to the transformed canonical oscillator.\\
The amplitude solution identifies the response time but still depends on the evolving electronic expectation value.
It does not determine the mixed electron--vibration correlations needed for the electronic density-matrix equation.
We first construct the response associated with electronic motion during this interval, and subsequently derive its role in that equation.

\subsection{Electronic response, stationary limit, and commutator expansion}
\label{appB:response_matrix}
Define the LAB reduced density matrix and the bare electronic propagator by
\begin{equation}
    \rho_{\rm el}(t)=\operatorname{Tr}_{\rm vib}\rho(t),
    \qquad U_\tau=e^{-iH_{\rm el}\tau/\hbar}.
    \label{eq:appB_electronic_propagator}
\end{equation}
During the short response interval, we approximate electronic propagation by the coherent evolution generated by the complete bare \(H_{\rm el}\):
\begin{equation}
    \rho_{\rm el}(t)\simeq U_\tau\rho_{\rm el}(t-\tau)U_\tau^\dagger,
    \qquad \rho_{\rm el}(t-\tau)\simeq U_\tau^\dagger\rho_{\rm el}(t)U_\tau.
    \label{eq:appB_backward_density}
\end{equation}
This neglects acceptor transfer and vibronic feedback only during the response interval, not during the complete evolution.
The coupling-order justification and the separate transfer condition are derived below.\\
Because \(F\) is invariant under the LF transformation, cyclicity of the electronic trace gives
\begin{equation}
\begin{aligned}
    \langle F\rangle_{\rm LF}(t-\tau)
    &=\operatorname{Tr}_{\rm el}[\rho_{\rm el}(t-\tau)F]\simeq\operatorname{Tr}_{\rm el}[\rho_{\rm el}(t)F(-\tau)],
    \qquad F(-\tau)\equiv U_\tau F U_\tau^\dagger.
\end{aligned}
\label{eq:appB_delayed_F}
\end{equation}
Substitution into Eq.~\eqref{eq:appB_LF_amplitude_integral} yields
\begin{equation}
    \langle a\rangle_{\rm LF}(t)\simeq\operatorname{Tr}_{\rm el}[\rho_{\rm el}(t)\mathcal A(t)],
    \qquad \mathcal A(t)=-F+\frac{iE_0}{\hbar}\int_0^t e^{-\kappa\tau/\hbar}F(-\tau)\,d\tau.
    \label{eq:appB_A_finite}
\end{equation}
It is useful to collect the integral into a response matrix with units of energy:
\begin{equation}
    Q(t)=\frac{E_0^2}{\hbar}\int_0^t e^{-\kappa\tau/\hbar}F(-\tau)\,d\tau,
    \qquad \mathcal A(t)=-F+\frac{i}{E_0}Q(t).
    \label{eq:appB_Q_finite}
\end{equation}
Here \(\mathcal A(t)\) is dimensionless and acts only on the electronic Hilbert space.
It represents an amplitude within the coherent-propagation approximation, not a canonical bosonic annihilation operator.\\
For \(\gamma>0\), the stationary response is defined by extending the upper integration limit:
\begin{equation}
    Q_\infty=\frac{E_0^2}{\hbar}\int_0^\infty e^{-\kappa\tau/\hbar}F(-\tau)\,d\tau,
    \qquad \mathcal A_\infty=-F+\frac{i}{E_0}Q_\infty.
    \label{eq:appB_Q_stationary}
\end{equation}
The word ``stationary'' refers to the response matrices, not to the electronic state, which continues to evolve.
Subtracting the finite-time response gives
\begin{equation}
    Q_\infty-Q(t)=\frac{E_0^2}{\hbar}\int_t^\infty e^{-\kappa\tau/\hbar}F(-\tau)\,d\tau.
    \label{eq:appB_Q_tail}
\end{equation}
Unitary invariance of the operator norm implies \(\|F(-\tau)\|=\|F\|\).
The triangle inequality therefore gives
\begin{equation}
\begin{aligned}
    \|Q_\infty-Q(t)\|
    &\leq\frac{E_0^2\|F\|}{\hbar}\int_t^\infty e^{-\gamma\tau/\hbar}\,d\tau
    =\frac{E_0^2\|F\|}{\gamma}e^{-\gamma t/\hbar},\\
    \|\mathcal A_\infty-\mathcal A(t)\|
    &\leq\frac{E_0\|F\|}{\gamma}e^{-\gamma t/\hbar}.
\end{aligned}
\label{eq:appB_response_tail_bound}
\end{equation}
Thus the response tail becomes small for \(t\gg\tau_{\rm vib}\).
At the initial time, however, \(Q(0)=0\) and \(\mathcal A(0)=-F\); using the stationary coefficients from \(t=0\) omits their buildup.
The bound in Eq.~\eqref{eq:appB_response_tail_bound} concerns response matrices, not the electronic states generated by different responses.
State differences accumulated during the initial interval can persist after the response coefficients have become nearly identical.\\
The relation to a nested-commutator expansion follows by differentiating the propagated coupling:
\begin{equation}
    \frac{dF(-\tau)}{d\tau}=-\frac{i}{\hbar}[H_{\rm el},F(-\tau)],
    \qquad C_0=F,
    \qquad C_{n+1}=[H_{\rm el},C_n].
    \label{eq:appB_commutator_definition}
\end{equation}
For example, \(C_1=H_{\rm el}F-FH_{\rm el}\) and \(C_2=H_{\rm el}^2F-2H_{\rm el}FH_{\rm el}+FH_{\rm el}^2\).
Repeated differentiation at \(\tau=0\) gives
\begin{equation}
    F(-\tau)=\sum_{n=0}^{\infty}\frac{(-i\tau/\hbar)^n}{n!}C_n
    =F-\frac{i\tau}{\hbar}C_1-\frac{\tau^2}{2\hbar^2}C_2+\cdots.
    \label{eq:appB_F_Taylor}
\end{equation}
The response moments are
\begin{equation}
    I_n=\int_0^\infty\tau^n e^{-\kappa\tau/\hbar}\,d\tau,
    \qquad I_0=\frac{\hbar}{\kappa},
    \qquad I_n=\frac{n\hbar}{\kappa}I_{n-1}
    =n!\left(\frac{\hbar}{\kappa}\right)^{n+1},
    \label{eq:appB_response_moments}
\end{equation}
where the recurrence follows by integration by parts and \(\operatorname{Re}\kappa>0\) removes the boundary term at infinity.
Where termwise integration is justified, Eqs.~\eqref{eq:appB_Q_stationary} and \eqref{eq:appB_F_Taylor} give
\begin{equation}
    Q_\infty=E_0^2\sum_{n=0}^{\infty}\frac{(-i)^n}{\kappa^{n+1}}C_n,
    \qquad \mathcal A_\infty=-F+iE_0\sum_{n=0}^{\infty}\frac{(-i)^n}{\kappa^{n+1}}C_n.
    \label{eq:appB_response_series}
\end{equation}
Combining the two terms proportional to \(F\), the LF amplitude response becomes
\begin{equation}
    \mathcal A_\infty=-\frac{\gamma}{\kappa}F
    +\frac{E_0}{\kappa^2}C_1-\frac{iE_0}{\kappa^3}C_2
    -\frac{E_0}{\kappa^4}C_3+\cdots.
    \label{eq:appB_A_series_explicit}
\end{equation}
The coefficients beyond the first term depend on retaining the derivative of \(\langle F\rangle_{\rm LF}\) in Eq.~\eqref{eq:appB_canonical_LF_eom}.
The zeroth-order amplitude \(-\gamma F/\kappa\) corresponds to neglecting electronic motion during the response, but does not by itself specify a Hamiltonian or a dissipator.\\
The convergence domain is determined by the electronic transitions coupled by \(F\).
Let
\begin{equation}
    H_{\rm el}|m\rangle=E_m|m\rangle,
    \qquad \Delta_{mn}=E_m-E_n,
    \qquad F_{mn}=\langle m|F|n\rangle.
    \label{eq:appB_energy_basis}
\end{equation}
Each additional commutator introduces a factor of \(\Delta_{mn}\), giving
\begin{equation}
    (C_k)_{mn}=\Delta_{mn}^{\,k}F_{mn},
    \qquad (Q_\infty)_{mn}
    =\frac{E_0^2F_{mn}}{\kappa}\sum_{k=0}^{\infty}
    \left(-\frac{i\Delta_{mn}}{\kappa}\right)^k.
    \label{eq:appB_geometric_series}
\end{equation}
For \(F_{mn}\neq0\), this geometric series converges precisely when
\begin{equation}
    |\Delta_{mn}|<|\kappa|=\sqrt{\gamma^2+E_0^2}.
    \label{eq:appB_series_convergence}
\end{equation}
The condition must hold for every coupled pair if the full matrix series is used.
Within that domain, summation yields \(E_0^2F_{mn}/(\kappa+i\Delta_{mn})\).
The relevant scales are the actual many-body transition energies, not the Hubbard parameter alone. Direct evaluation of the integral avoids any commutator truncation.
In the same eigenbasis,
\begin{equation}
    [F(-\tau)]_{mn}=e^{-i\Delta_{mn}\tau/\hbar}F_{mn},
\end{equation}
so that
\begin{equation}
    [Q(t)]_{mn}=\frac{E_0^2F_{mn}}{\hbar}
    \int_0^t e^{-[\kappa+i\Delta_{mn}]\tau/\hbar}\,d\tau
    =\frac{E_0^2F_{mn}}{\kappa+i\Delta_{mn}}
    \left[1-e^{-[\kappa+i\Delta_{mn}]t/\hbar}\right].
    \label{eq:appB_Q_spectral_finite}
\end{equation}
Since \(\operatorname{Re}(\kappa+i\Delta_{mn})=\gamma>0\), its stationary limit is
\begin{equation}
    (Q_\infty)_{mn}=\frac{E_0^2F_{mn}}{\kappa+i\Delta_{mn}}.
    \label{eq:appB_Q_spectral_stationary}
\end{equation}
This integral exists even when Eq.~\eqref{eq:appB_series_convergence} fails.
Outside the convergence domain, Eq.~\eqref{eq:appB_Q_spectral_stationary} is the value of the integral, not the sum of a convergent geometric series.
The matrices are subsequently transformed back to the electronic occupation basis for propagation.\\
Every \(C_n\) is linear in \(F\), irrespective of its commutator order.
Resolving additional commutators therefore refines the electronic motion within the response; it does not increase the coupling order retained in the vibrational elimination.
Similarly, evaluating the integral directly removes commutator-cutoff and quadrature errors, but does not restore the transfer or vibronic feedback omitted from the short-time propagator.
A sufficient condition for the additional approximation of freezing electronic motion would be \(E_{\rm el}/\gamma\ll1\), with \(E_{\rm el}\) a relevant transition-energy scale.
That condition is not imposed in Eqs.~\eqref{eq:appB_Q_spectral_finite} and \eqref{eq:appB_Q_spectral_stationary}, which retain this motion explicitly.

\subsection{Mixed correlations and second-order electronic elimination}
\label{appB:mixed_correlations}
The amplitude construction must now be supplemented by the electronic information carried by the joint state.
Define the electronic matrices
\begin{equation}
    M(t)=\operatorname{Tr}_{\rm vib}[a\rho(t)],
    \qquad X(t)=\operatorname{Tr}_{\rm vib}[a^\dagger a\rho(t)],
    \qquad Y(t)=\operatorname{Tr}_{\rm vib}[a^2\rho(t)].
    \label{eq:appB_mixed_matrices}
\end{equation}
Only vibrational indices are traced out; in particular, \(M\) is not the scalar amplitude.
Rather,
\begin{equation}
    \operatorname{Tr}_{\rm el}M=\alpha_{\rm vib},
    \qquad M^\dagger=\operatorname{Tr}_{\rm vib}(\rho a^\dagger)
    =\operatorname{Tr}_{\rm vib}(a^\dagger\rho).
    \label{eq:appB_M_trace_adjoint}
\end{equation}
A vanishing \(\alpha_{\rm vib}\) does not imply \(M=0\).
The partial trace is cyclic for vibrational factors only: electronic matrices must retain their order.\\
Partial tracing Eq.~\eqref{eq:appB_LAB_master} gives
\begin{equation}
\begin{aligned}
    \operatorname{Tr}_{\rm vib}[a^\dagger a,\rho]&=0,
    \qquad \operatorname{Tr}_{\rm vib}\mathcal D[a]\rho=0,\\
    \operatorname{Tr}_{\rm vib}[H_{\rm el},\rho]&=[H_{\rm el},\rho_{\rm el}],
    \qquad \operatorname{Tr}_{\rm vib}\mathcal D[J_A]\rho
    =\mathcal D[J_A]\rho_{\rm el}.
\end{aligned}
\label{eq:appB_partial_trace_identities}
\end{equation}
For example, cyclicity over the vibrational space makes the three partial traces in \(\mathcal D[a]\rho\) equal to \(\operatorname{Tr}_{\rm vib}(a^\dagger a\rho)\); their coefficients sum to zero.
These identities hold also for correlated electron--vibration states.
For the interaction term,
\begin{equation}
    \operatorname{Tr}_{\rm vib}[B(a^\dagger+a),\rho]
    =B(M+M^\dagger)-(M+M^\dagger)B
    =[B,M+M^\dagger].
\end{equation}
The exact reduced equation is thus
\begin{equation}
    \hbar\dot\rho_{\rm el}=-i[H_{\rm el},\rho_{\rm el}]
    +\Gamma_A\mathcal D[J_A]\rho_{\rm el}
    -i[B,M+M^\dagger].
    \label{eq:appB_rho_exact}
\end{equation}
Vibrational loss does not appear directly in this partial trace, but affects the electronic dynamics through \(M\).\\
To obtain the exact equation for \(M\), multiply the joint master equation by \(a\) before taking the vibrational trace.
The electronic commutator gives \([H_{\rm el},M]\), while
\begin{equation}
\begin{aligned}
    \operatorname{Tr}_{\rm vib}\bigl(a[a^\dagger a,\rho]\bigr)&=M,\\
    \operatorname{Tr}_{\rm vib}\bigl(a\mathcal D[a]\rho\bigr)
    &=\frac12\operatorname{Tr}_{\rm vib}[(a^\dagger a^2-aa^\dagger a)\rho]
    =-\frac12M,\\
    \operatorname{Tr}_{\rm vib}\bigl(a\mathcal D[J_A]\rho\bigr)
    &=J_AMJ_A^\dagger-\frac12\{J_A^\dagger J_A,M\}
    =\mathcal D[J_A]M.
\end{aligned}
\label{eq:appB_M_free_terms}
\end{equation}
The last expression is a linear operation on a matrix; it does not assume that \(M\) is a density matrix.
Unlike the corresponding contribution to \(\dot\alpha_{\rm vib}\), the transfer term in \(\dot M\) does not generally vanish.\\
For the interaction contribution, \(aa^\dagger=a^\dagger a+I\) yields
\begin{equation}
\begin{aligned}
    \operatorname{Tr}_{\rm vib}[aB(a^\dagger+a)\rho]
    &=B(X+\rho_{\rm el}+Y),\\
    \operatorname{Tr}_{\rm vib}[a\rho B(a^\dagger+a)]
    &=(X+Y)B.
\end{aligned}
\label{eq:appB_M_coupling_products}
\end{equation}
Only the vibrational factors have been moved through the partial trace in the second line.
Their difference is
\begin{equation}
    \operatorname{Tr}_{\rm vib}\bigl(a[B(a^\dagger+a),\rho]\bigr)
    =B\rho_{\rm el}+[B,X+Y].
    \label{eq:appB_M_vacuum_source}
\end{equation}
The source \(B\rho_{\rm el}\) comes from the identity in \(aa^\dagger=a^\dagger a+I\).
It remains present for an initially unoccupied mode, retaining the vacuum fluctuation responsible for the response.
Collecting the terms gives
\begin{equation}
    \hbar\dot M=-\kappa M-i[H_{\rm el},M]
    +\Gamma_A\mathcal D[J_A]M
    -iB\rho_{\rm el}-i[B,X+Y].
    \label{eq:appB_M_exact}
\end{equation}
Equations~\eqref{eq:appB_rho_exact} and \eqref{eq:appB_M_exact} are exact within the specified LAB model.
For the factorized initial vacuum in Eq.~\eqref{eq:appB_initial_state},
\begin{equation}
    M(0)=X(0)=Y(0)=0.
    \label{eq:appB_initial_moments}
\end{equation}
To determine the retained coupling order, introduce an auxiliary dimensionless parameter \(\eta_{\rm c}\) through
\begin{equation}
    B\longrightarrow\eta_{\rm c}B,
    \qquad g\longrightarrow\eta_{\rm c}g,
    \qquad \chi\longrightarrow\eta_{\rm c}\chi,
    \label{eq:appB_coupling_counting}
\end{equation}
while holding \(H_{\rm el}\) and the initial electronic density matrix fixed.
This parameter counts insertions of the interaction Hamiltonian; it is set to unity after the retained orders have been identified.
For the initial vacuum, the interaction contribution over a short interval is
\begin{equation}
    \delta\rho_{\rm int}
    =-\frac{i\eta_{\rm c}\delta t}{\hbar}
    \left[B\rho_{\rm el}^{(0)}\otimes|1\rangle\langle0|
    -\rho_{\rm el}^{(0)}B\otimes|0\rangle\langle1|\right].
    \label{eq:appB_first_order_coherence}
\end{equation}
The number states in this expression refer to the vibration.
A single interaction insertion creates zero--one coherences, not a one-boson population.
Their partial trace vanishes, so the electronic density matrix has no interaction correction at first order.
However, \(M\) detects one of these coherences, giving
\begin{equation}
    \delta M=-\frac{i\eta_{\rm c}\delta t}{\hbar}B\rho_{\rm el}^{(0)}.
\end{equation}
The matrix \(X\) first receives a contribution from a population such as \(|1\rangle\langle1|\), requiring two interaction insertions.
The matrix \(Y\) first detects a zero--two coherence, since \(a^2|2\rangle=\sqrt{2}|0\rangle\), which also requires two insertions.
Electronic propagation and zero-temperature loss change the coefficients of these contributions but not the minimum number of coupling insertions required to generate them.
At fixed time,
\begin{equation}
    M=\mathcal O(\eta_{\rm c}),
    \qquad X=\mathcal O(\eta_{\rm c}^2),
    \qquad Y=\mathcal O(\eta_{\rm c}^2).
    \label{eq:appB_moment_orders}
\end{equation}
The higher-moment source in Eq.~\eqref{eq:appB_M_exact} therefore begins at third order:
\begin{equation}
    [\eta_{\rm c}B,X+Y]=\mathcal O(\eta_{\rm c}^3),
    \qquad \delta M_{\rm higher}=\mathcal O(\eta_{\rm c}^3).
\end{equation}
Its feedback into Eq.~\eqref{eq:appB_rho_exact} contains a further coupling factor and begins at fourth order,
\begin{equation}
    -i[\eta_{\rm c}B,\delta M_{\rm higher}+\delta M_{\rm higher}^\dagger]
    =\mathcal O(\eta_{\rm c}^4).
    \label{eq:appB_omitted_feedback_order}
\end{equation}
Retaining the leading \(M\) thus retains the second-order electronic contribution.
Neglecting \([B,X+Y]\) does not assert that vibrational populations and higher coherences are physically zero; it omits their additional feedback beyond the retained order.
Nor is the joint state assumed to remain factorized during the evolution.
Replacing it by \(\rho_{\rm el}(t)\otimes|0\rangle\langle0|\) at every time would instead set \(M=0\) and remove the second-order contribution itself.\\
The response envelope supplies a characteristic scale for this counting.
For one coupling insertion, unitary electronic propagation and exponential damping give
\begin{equation}
    \frac{\eta_{\rm c}\|B\|}{\hbar}\int_0^\infty e^{-\gamma\tau/\hbar}\,d\tau
    =\frac{\eta_{\rm c}\|B\|}{\gamma}.
\end{equation}
After setting \(\eta_{\rm c}=1\), a conservative working condition is
\begin{equation}
    \xi_B=\frac{\|B\|}{\gamma}=\frac{2\|B\|}{\Gamma}\ll1,
    \label{eq:appB_epsilon_B}
\end{equation}
with the operator norm evaluated in the electronic sector being propagated.
The characteristic sizes are \(\|M\|=\mathcal O(\xi_B)\) and \(\|X\|,\|Y\|=\mathcal O(\xi_B^2)\).
Accordingly, the retained and omitted vibronic contributions to the right-hand side of \(\hbar\dot\rho_{\rm el}\) have characteristic scales \(\|B\|^2/\gamma\) and \(\|B\|^4/\gamma^3\), respectively, with ratio \(\xi_B^2\).
These are perturbative scale estimates, not relative-error bounds for a particular observable or guarantees of uniform accuracy at arbitrarily long times.\\
The coherent response also neglects acceptor transfer while solving for \(M\) over the short vibrational response interval.
Writing \(G=J_A^\dagger J_A\), this requires the separate condition
\begin{equation}
    \xi_A=\frac{\Gamma_A\|G\|}{\gamma}\ll1.
    \label{eq:appB_xi_A}
\end{equation}
Transfer remains present in the electronic equation at all times.
Its omission from the response neglects corrections to the second-order vibronic contribution with characteristic relative scale \(\xi_A\).
These corrections remain second order in \(B\), unlike the fourth-order corrections associated with the higher moments.
The two approximations must therefore be assessed separately.
With both stated explicitly, the mixed-matrix equation to be solved is
\begin{equation}
    \hbar\dot M\simeq-\kappa M-i[H_{\rm el},M]-iB\rho_{\rm el}.
    \label{eq:appB_M_approximate}
\end{equation}
For the time-independent bare Hamiltonian, the derivative of the interaction-picture matrix is
\begin{equation}
    \frac{d}{dt}(U_t^\dagger M U_t)
    =U_t^\dagger\left(\dot M+\frac{i}{\hbar}[H_{\rm el},M]\right)U_t.
\end{equation}
Multiplication by the scalar integrating factor removes the remaining homogeneous term:
\begin{equation}
    \frac{d}{dt}\left(e^{\kappa t/\hbar}U_t^\dagger M(t)U_t\right)
    \simeq-\frac{i}{\hbar}e^{\kappa t/\hbar}U_t^\dagger B\rho_{\rm el}(t)U_t.
    \label{eq:appB_M_integrating_factor}
\end{equation}
Integrating from zero with \(M(0)=0\), and using \(U_tU_s^\dagger=U_{t-s}\), gives
\begin{equation}
\begin{aligned}
    M(t)&\simeq-\frac{i}{\hbar}\int_0^t e^{-\kappa(t-s)/\hbar}
    U_{t-s}B\rho_{\rm el}(s)U_{t-s}^\dagger\,ds\\
    &=-\frac{i}{\hbar}\int_0^t e^{-\kappa\tau/\hbar}
    U_\tau B\rho_{\rm el}(t-\tau)U_\tau^\dagger\,d\tau.
\end{aligned}
\label{eq:appB_M_memory}
\end{equation}
No commutation of \(B\) or \(\rho_{\rm el}\) with the propagators has been assumed.
This integral solves the approximate equation without imposing \(\dot M=0\).\\
We now apply the same coherent backward propagation as in Eq.~\eqref{eq:appB_backward_density}.
Vibronic feedback changes this backward electronic propagation from second order in \(B\); its inclusion in Eq.~\eqref{eq:appB_M_memory} would change \(M\) from third order and the electronic equation from fourth order.
The omitted transfer contribution instead belongs to the separate approximation controlled by \(\xi_A\).
Preserving the order of all electronic factors,
\begin{equation}
    U_\tau B\left(U_\tau^\dagger\rho_{\rm el}(t)U_\tau\right)U_\tau^\dagger
    =\left(U_\tau B U_\tau^\dagger\right)\rho_{\rm el}(t).
    \label{eq:appB_memory_product_order}
\end{equation}
The density matrix can therefore be taken outside the integral on its right:
\begin{equation}
    M(t)\simeq-\frac{i}{\hbar}
    \left[\int_0^t e^{-\kappa\tau/\hbar}U_\tau B U_\tau^\dagger\,d\tau\right]\rho_{\rm el}(t)
    =\frac{i}{E_0}Q(t)\rho_{\rm el}(t).
    \label{eq:appB_M_closed}
\end{equation}
Taking the adjoint reverses the matrix order:
\begin{equation}
    M^\dagger(t)\simeq-\frac{i}{E_0}\rho_{\rm el}(t)Q^\dagger(t),
    \qquad M+M^\dagger\simeq\frac{i}{E_0}
    \left[Q(t)\rho_{\rm el}-\rho_{\rm el}Q^\dagger(t)\right].
    \label{eq:appB_M_adjoint_closed}
\end{equation}
The same \(Q(t)\) found from the amplitude response now enters through the mixed correlations, with its position relative to \(\rho_{\rm el}\) fixed by the derivation.\\
Using \(B=-E_0F\), the coupling term becomes
\begin{equation}
    -i[B,M+M^\dagger]
    \simeq-i\left[-E_0F,\frac{i}{E_0}(Q\rho_{\rm el}-\rho_{\rm el}Q^\dagger)\right]
    =-[F,Q\rho_{\rm el}-\rho_{\rm el}Q^\dagger].
\end{equation}
The resulting reduced electronic model is
\begin{equation}
    \hbar\dot\rho_{\rm el}=-i[H_{\rm el},\rho_{\rm el}]
    +\Gamma_A\mathcal D[J_A]\rho_{\rm el}
    -[F,Q(t)\rho_{\rm el}-\rho_{\rm el}Q^\dagger(t)].
    \label{eq:appB_electronic_master}
\end{equation}
The equality defines the approximate electronic evolution under the coupling and transfer assumptions described above.
Only \(\rho_{\rm el}\) remains an unknown dynamical matrix; \(Q(t)\) is calculated from the known electronic operators.
Neither the bosonic amplitude nor \(M\), \(X\), and \(Y\) must be propagated alongside it.

\subsection{Coherent and dissipative contributions, stationary dynamics, and validity}
\label{appB:effective_generator}
The coherent correction follows from the matrix products in Eq.~\eqref{eq:appB_electronic_master}.
Suppressing only the explicit time argument of \(Q\), expand
\begin{equation}
    -[F,Q\rho_{\rm el}-\rho_{\rm el}Q^\dagger]
    =-FQ\rho_{\rm el}-\rho_{\rm el}Q^\dagger F
    +Q\rho_{\rm el}F+F\rho_{\rm el}Q^\dagger.
    \label{eq:appB_generator_products}
\end{equation}
The first two products can be separated into the Hermitian and anti-Hermitian parts of \(FQ\):
\begin{equation}
    -FQ\rho_{\rm el}-\rho_{\rm el}Q^\dagger F
    =-\frac12\{FQ+Q^\dagger F,\rho_{\rm el}\}
    -\frac12[FQ-Q^\dagger F,\rho_{\rm el}].
    \label{eq:appB_commutator_anticommutator_split}
\end{equation}
Since \((FQ-Q^\dagger F)^\dagger=-(FQ-Q^\dagger F)\), the second term has Hamiltonian form:
\begin{equation}
    \delta H(t)=\frac{FQ(t)-Q^\dagger(t)F}{2i},
    \qquad \delta H^\dagger(t)=\delta H(t),
    \qquad H_{\rm eff}(t)=H_{\rm el}+\delta H(t).
    \label{eq:appB_effective_Hamiltonian}
\end{equation}
Inserting Eq.~\eqref{eq:appB_commutator_anticommutator_split} into Eq.~\eqref{eq:appB_generator_products} gives the equivalent electronic equation
\begin{equation}
\begin{split}
    \hbar\dot\rho_{\rm el}={}&-i[H_{\rm eff}(t),\rho_{\rm el}]
    +\Gamma_A\mathcal D[J_A]\rho_{\rm el}+Q(t)\rho_{\rm el}F+F\rho_{\rm el}Q^\dagger(t)
    -\frac12\{FQ(t)+Q^\dagger(t)F,\rho_{\rm el}\}.
\end{split}
\label{eq:appB_electronic_master_split}
\end{equation}
This is an algebraic rearrangement, with no further approximation.
Both the coherent correction and the non-Hamiltonian terms are second order in the vibronic coupling and must be retained together.
The non-Hamiltonian contribution is not, in general, a single Lindblad dissipator.\\
For the stationary response, introduce
\begin{equation}
    R=\frac{E_0^2}{\gamma^2+E_0^2},
    \qquad \frac{E_0^2}{\kappa}=R(\gamma-iE_0),
    \qquad Q_\infty=R(\gamma-iE_0)F+\Delta Q.
    \label{eq:appB_R_decomposition}
\end{equation}
The remainder is defined by subtraction,
\begin{equation}
    \Delta Q\equiv Q_\infty-\frac{E_0^2}{\kappa}F,
    \label{eq:appB_delta_Q}
\end{equation}
which is meaningful also outside the convergence domain of the commutator series.
Within that domain,
\begin{equation}
\begin{aligned}
    \Delta Q&=E_0^2\sum_{n=1}^{\infty}\frac{(-i)^n}{\kappa^{n+1}}C_n,\\
    Q_\infty&=R(\gamma-iE_0)
    \left[F-\frac{i}{\kappa}C_1-\frac{1}{\kappa^2}C_2
    +\frac{i}{\kappa^3}C_3+\cdots\right].
\end{aligned}
\label{eq:appB_Q_R_series}
\end{equation}
Displaying \(R\) does not truncate the response to its first term.
The corresponding leading coefficient in \(\mathcal A_\infty\) satisfies
\begin{equation}
    -\frac{\gamma}{\kappa}=-(1-R)+i\frac{\gamma}{E_0}R.
\end{equation}
For the coherent contribution,
\begin{equation}
    FQ_\infty-Q_\infty^\dagger F
    =-2iRE_0F^2+F\Delta Q-\Delta Q^\dagger F,
\end{equation}
and hence
\begin{equation}
    H_{{\rm eff},\infty}=H_{\rm el}-RE_0F^2
    +\frac{F\Delta Q-\Delta Q^\dagger F}{2i}.
    \label{eq:appB_H_stationary}
\end{equation}
The term directly proportional to \(F^2\) is
\begin{equation}
    -RE_0F^2=-\frac{R}{E_0}B^2.
    \label{eq:appB_static_induced_term}
\end{equation}
Its one-body and two-body components are those identified in Appendix~\ref{app:exact_transformation}, multiplied by \(R\).
The remainder \(\Delta Q\) supplies further coherent contributions through the last term of Eq.~\eqref{eq:appB_H_stationary}.\\
The non-Hamiltonian part can be separated with the same decomposition:
\begin{equation}
\begin{aligned}
    Q_\infty\rho_{\rm el}F+F\rho_{\rm el}Q_\infty^\dagger
    &=2\gamma R F\rho_{\rm el}F
    +\Delta Q\rho_{\rm el}F+F\rho_{\rm el}\Delta Q^\dagger,\\
    FQ_\infty+Q_\infty^\dagger F
    &=2\gamma R F^2+F\Delta Q+\Delta Q^\dagger F.
\end{aligned}
\label{eq:appB_nonHamiltonian_R_products}
\end{equation}
The imaginary scalar coefficients cancel in each sum.
Using \(2\gamma=\Gamma\) and \(F^\dagger=F\), the terms proportional to \(R\) combine into
\begin{equation}
    2\gamma R F\rho_{\rm el}F-\gamma R\{F^2,\rho_{\rm el}\}
    =\Gamma R\mathcal D[F]\rho_{\rm el}.
    \label{eq:appB_R_dissipator}
\end{equation}
The full stationary-response equation is therefore
\begin{equation}
\begin{split}
    \hbar\dot\rho_{\rm el}={}&-i[H_{{\rm eff},\infty},\rho_{\rm el}]
    +\Gamma_A\mathcal D[J_A]\rho_{\rm el}
    +\Gamma R\mathcal D[F]\rho_{\rm el}+\Delta Q\rho_{\rm el}F+F\rho_{\rm el}\Delta Q^\dagger
    -\frac12\{F\Delta Q+\Delta Q^\dagger F,\rho_{\rm el}\}.
\end{split}
\label{eq:appB_stationary_split}
\end{equation}
Both parts of \(\Delta Q\) are present: its coherent contribution is contained in \(H_{{\rm eff},\infty}\), while its non-Hamiltonian contribution is displayed explicitly.
Combining them reverses the product separation in Eq.~\eqref{eq:appB_commutator_anticommutator_split}, giving the compact equivalent form
\begin{equation}
\begin{split}
    \hbar\dot\rho_{\rm el}={}&-i[H_{\rm el}-RE_0F^2,\rho_{\rm el}]
    +\Gamma_A\mathcal D[J_A]\rho_{\rm el}
    +\Gamma R\mathcal D[F]\rho_{\rm el}-[F,\Delta Q\rho_{\rm el}-\rho_{\rm el}\Delta Q^\dagger].
\end{split}
\label{eq:appB_stationary_full}
\end{equation}
Only if electronic motion during the response is additionally neglected, \(F(-\tau)\simeq F\), does one obtain \(Q_\infty\simeq E_0^2F/\kappa\) and \(\Delta Q\simeq0\).
The resulting frozen-electronic-response equation is
\begin{equation}
    \hbar\dot\rho_{\rm el}=-i[H_{\rm el}-RE_0F^2,\rho_{\rm el}]
    +\Gamma_A\mathcal D[J_A]\rho_{\rm el}
    +\Gamma R\mathcal D[F]\rho_{\rm el}.
    \label{eq:appB_frozen_response}
\end{equation}
This extra approximation is distinct from taking the stationary limit of the full response.
In particular, the Hamiltonian \(H_{\rm el}-RE_0F^2\) obtained here is not the Hamiltonian used to generate the propagator \(U_\tau\) in the derivation.\\
The initial conditions remain the same in all descriptions:
\begin{equation}
    \rho_{\rm el}(0)=\rho_{\rm el}^{(0)}.
\end{equation}
For the finite-time response,
\begin{equation}
    Q(0)=0,
    \qquad \delta H(0)=0,
    \qquad H_{\rm eff}(0)=H_{\rm el},
\end{equation}
so every vibronic contribution vanishes at the initial time and
\begin{equation}
    \hbar\dot\rho_{\rm el}(0)
    =-i[H_{\rm el},\rho_{\rm el}^{(0)}]
    +\Gamma_A\mathcal D[J_A]\rho_{\rm el}^{(0)}.
    \label{eq:appB_initial_derivative}
\end{equation}
This is the exact initial reduced LAB derivative for the preparation in Eq.~\eqref{eq:appB_initial_state}.
The stationary-response calculation starts from the same electronic density matrix but generally has a different initial derivative, because it inserts \(Q_\infty\) immediately.
The exponentially small response tail at late times does not remove the state difference generated during the initial interval.\\
The approximation levels can now be distinguished explicitly.
The coupling expansion omits fourth-order vibronic feedback at fixed time, with \(\xi_B\ll1\) providing a conservative working regime.
Neglecting transfer during the response omits additional second-order terms with characteristic relative scale \(\xi_A\).
Taking \(Q(t)\to Q_\infty\) omits the response buildup, while taking \(\Delta Q\to0\) additionally freezes coherent electronic motion during the response.
Finally, a truncation of the commutator series would introduce an evaluation error of its own, separate from all these physical approximations.
If a finite \(Q_\infty^{(N_c)}\) is used, the same matrix must enter every term, including \(\Delta Q^{(N_c)}=Q_\infty^{(N_c)}-E_0^2F/\kappa\).
Increasing \(N_c\) does not raise the coupling order of the reduced equation.
The calculations in the main text instead use the direct spectral response.\\
Trace and Hermiticity preservation follow directly from Eq.~\eqref{eq:appB_electronic_master}.
For the vibronic term,
\begin{equation}
    \operatorname{Tr}_{\rm el}[F,Q\rho_{\rm el}-\rho_{\rm el}Q^\dagger]=0.
\end{equation}
For Hermitian \(\rho_{\rm el}\),
\begin{equation}
    (Q\rho_{\rm el}-\rho_{\rm el}Q^\dagger)^\dagger
    =-(Q\rho_{\rm el}-\rho_{\rm el}Q^\dagger),
\end{equation}
so its commutator with Hermitian \(F\) is Hermitian.
These properties do not guarantee positivity: the full response equation is not generally of Lindblad form with nonnegative rates.
The eigenvalues of \(\rho_{\rm el}(t)\) and the observables of interest must therefore be checked against LAB calculations converged in the vibrational Hilbert space, using the same electronic preparation and evolution time.
Neither convergence of the response evaluation nor small working ratios alone provides a uniform bound on observable errors at arbitrarily long times.

\section{Numerical implementation of the electron-transfer protocol}
\label{app:numerics}

This Appendix specifies the numerical implementation of the electron-transfer protocol introduced in Sec.~\ref{sec:electron_transfer}.
We compare the laboratory-frame dynamics, retaining the vibrational mode explicitly, with the stationary and finite-time electronic responses derived in Appendix~\ref{app:adiabatic_elimination}.
The stationary calculation uses \(Q_\infty\), whereas the finite-time calculation retains \(Q(t)\) and resolves the initial response buildup.
All descriptions use the same electronic parameters, acceptor-transfer operator, and injected electronic density matrix.
We describe the reference preparation with four bridge electrons before injection; numerical results and convergence data are reported separately with the corresponding benchmarks.

\subsection{Electronic Hamiltonian and particle-number sectors}
\label{appC:electronic_sectors}
Following the notation of the main text, the bare bridge Hamiltonian is
\begin{equation}
    H_{\rm B}=-t_{\rm hop}H_{\rm hop}+\lambda H_{\rm SOC}
    +\sum_{j=1}^{4}\sum_\sigma\epsilon_j n_{j\sigma}
    +U\sum_{j=1}^{4}n_{j\uparrow}n_{j\downarrow}.
    \label{eq:appC_HB}
\end{equation}
The complete bare electronic Hamiltonian and the vibrational coupling are
\begin{equation}
    H_{\rm el}=H_{\rm B}+\epsilon_A N_A,
    \qquad B=\frac{-gH_{\rm hop}+\chi H_{\rm SOC}}{\sqrt{2}},
    \qquad F=-\frac{B}{E_0}.
    \label{eq:appC_Hel_B_F}
\end{equation}
The site-energy and Hubbard terms are included both in the bare electronic propagation and in the Hamiltonian used to evaluate the response.
No coherent hopping connects the bridge and the acceptor.
We retain the conventions
\begin{equation}
    E_0=\hbar\omega_0,
    \qquad \gamma=\frac{\Gamma}{2},
    \qquad \kappa=\gamma+iE_0,
    \qquad R=\frac{E_0^2}{\gamma^2+E_0^2}.
    \label{eq:appC_response_parameters}
\end{equation}
The coefficients \(\Gamma\) and \(\Gamma_A\) have units of energy, and their conversion to inverse-time rates requires division by \(\hbar\).
The hopping amplitude is denoted by \(t_{\rm hop}\) to distinguish it from the evolution time \(t\).\\
The particle-number operators are
\begin{equation}
    N_{\rm bridge}=\sum_{j=1}^{4}\sum_\sigma n_{j\sigma},
    \qquad N_A=n_{A\uparrow}+n_{A\downarrow},
    \qquad N_{\rm tot}=N_{\rm bridge}+N_A.
    \label{eq:appC_number_operators}
\end{equation}
The acceptor is restricted to its empty and singly occupied states.
Electronic matrices are constructed in an occupation basis with one fixed ordering of the eight bridge spin orbitals and the two acceptor spin orbitals.
The same fermionic sign convention is used for the Hamiltonian, injection maps, transfer operator, and observables.

The reference preparation belongs to the sector
\begin{equation}
    N_{\rm bridge}=4,
    \qquad N_A=0,
    \qquad \dim\mathcal H_4=\binom{8}{4}=70.
    \label{eq:appC_preparation_sector}
\end{equation}
After injection, the electronic evolution takes place in the complete five-electron sector with at most one electron on the acceptor:
\begin{equation}
    N_{\rm tot}=5,
    \qquad N_A\leq1,
    \qquad \dim\mathcal H_5=\binom{8}{5}+2\binom{8}{4}=56+140=196.
    \label{eq:appC_dynamical_sector}
\end{equation}
The 56 states in the first contribution have an empty acceptor and five bridge electrons.
The 140 states in the second contribution have four bridge electrons and a singly occupied acceptor with either spin orientation.
All occupation states in these sectors are retained.

\subsection{Ground-state preparation and incoherent injection}
\label{appC:preparation_injection}
The reference electronic state is prepared using
\begin{equation}
    H_{\rm prep}=H_{\rm el}-RE_0F^2
    =H_{\rm el}-\frac{R}{E_0}B^2.
    \label{eq:appC_Hprep}
\end{equation}
This prescribed Hamiltonian contains the bare electronic terms and the leading stationary vibration-induced correction.
It is distinct from the complete effective Hamiltonian, which also contains the coherent contribution involving \(\Delta Q\).
Its ground state is not assumed to be the equilibrium state of the dissipative LAB model.\\
Let \(\Pi_4\) project onto \(\mathcal H_4\).
We diagonalize the restricted preparation Hamiltonian
\begin{equation}
    H_4=\left.\Pi_4H_{\rm prep}\Pi_4\right|_{\mathcal H_4}.
    \label{eq:appC_H4}
\end{equation}
The acceptor-energy term vanishes in this sector because \(N_A=0\).
For a nondegenerate ground state, the preparation is
\begin{equation}
    H_4|\psi_0^{(4)}\rangle=E_{\rm gs}^{(4)}|\psi_0^{(4)}\rangle,
    \qquad \rho_4=|\psi_0^{(4)}\rangle\langle\psi_0^{(4)}|,
    \label{eq:appC_pure_preparation}
\end{equation}
where \(E_{\rm gs}^{(4)}\) is the lowest eigenvalue of \(H_4\).
If the ground level is degenerate, the preparation is the normalized projector onto the complete ground eigenspace:
\begin{equation}
    \rho_4=\frac{\Pi_{\rm gs}^{(4)}}{\operatorname{Tr}_{\mathcal H_4}\Pi_{\rm gs}^{(4)}}.
    \label{eq:appC_degenerate_preparation}
\end{equation}
This avoids choosing an arbitrary spin-polarized vector within a degenerate eigenspace.
For the time-reversal-invariant preparation Hamiltonian, the resulting density matrix is also time-reversal invariant and has equal local spin-up and spin-down occupations.
For each parameter set, including the value of \(\Gamma\) entering \(R\), the state \(\rho_4\) is constructed once and used in every dynamical description being compared.
Changes of the preparation between parameter points are therefore common to the LAB and electronic calculations.\\
Injection is an instantaneous preparation step at bridge site \(1\), after which the donor is not retained as a dynamical degree of freedom.
With \(\Pi_5\) the projector onto \(\mathcal H_5\), the injection maps are
\begin{equation}
    C_{1\sigma}^\dagger=\Pi_5c_{1\sigma}^\dagger\Pi_4,
    \qquad C_{1\sigma}=\Pi_4c_{1\sigma}\Pi_5.
    \label{eq:appC_injection_maps}
\end{equation}
Here \(C_{1\sigma}^\dagger\) is a \(196\times70\) matrix mapping the preparation space into the dynamical space.
The maps connect different particle-number sectors; projecting a creation operator within one fixed-number sector would instead give zero.
For a pure preparation, the possible injected vectors are
\begin{equation}
    |\phi_\uparrow\rangle=C_{1\uparrow}^\dagger|\psi_0^{(4)}\rangle,
    \qquad |\phi_\downarrow\rangle=C_{1\downarrow}^\dagger|\psi_0^{(4)}\rangle.
    \label{eq:appC_injected_vectors}
\end{equation}
They need not have unit norm, because Pauli exclusion removes the component in which the injected spin orbital is already occupied.\\
The unpolarized injector assigns equal prior weights to the two spin orientations.
The channels are added incoherently at the density-matrix level, giving, for a general preparation,
\begin{equation}
    \rho_I^{\rm unnorm}
    =\frac12C_{1\uparrow}^\dagger\rho_4C_{1\uparrow}
    +\frac12C_{1\downarrow}^\dagger\rho_4C_{1\downarrow}.
    \label{eq:appC_injection_unnormalized}
\end{equation}
No cross terms between the injected spin orientations are included.
In particular, this is not the density matrix of the coherent superposition \(|\phi_\uparrow\rangle+|\phi_\downarrow\rangle\).
The normalized post-injection state is
\begin{equation}
    \rho_I=
    \frac{C_{1\uparrow}^\dagger\rho_4C_{1\uparrow}+C_{1\downarrow}^\dagger\rho_4C_{1\downarrow}}
    {\operatorname{Tr}_{\mathcal H_5}\left[C_{1\uparrow}^\dagger\rho_4C_{1\uparrow}+C_{1\downarrow}^\dagger\rho_4C_{1\downarrow}\right]}.
    \label{eq:appC_rhoI}
\end{equation}
The two contributions are summed before normalization, rather than normalized separately.
The common prior factor \(1/2\) cancels between numerator and denominator, and the denominator must be nonzero for injection to be possible.
Pauli exclusion is enforced by the creation operators themselves.
Since injection changes only the bridge occupation, the initial-state checks are
\begin{equation}
    \operatorname{Tr}\rho_I=1,
    \qquad \operatorname{Tr}(\rho_I N_{\rm bridge})=5,
    \qquad \operatorname{Tr}(\rho_I N_A)=0.
    \label{eq:appC_injection_checks}
\end{equation}

\subsection{Irreversible transfer to the acceptor}
\label{appC:acceptor_transfer}

The spin components of transfer from bridge site \(4\) are
\begin{equation}
    J_{A\uparrow}=c_{A\uparrow}^\dagger(1-n_{A\downarrow})c_{4\uparrow},
    \qquad J_{A\downarrow}=c_{A\downarrow}^\dagger(1-n_{A\uparrow})c_{4\downarrow}.
    \label{eq:appC_transfer_components}
\end{equation}
The occupation factors exclude double occupation of the acceptor.
The transfer channel is constructed by adding the two components at the operator level:
\begin{equation}
    J_A=J_{A\uparrow}+J_{A\downarrow}.
    \label{eq:appC_single_jump}
\end{equation}
It enters the electronic equation through the single dissipator
\begin{equation}
    \Gamma_A\mathcal D[J_A]\rho_{\rm el}
    =\Gamma_A\left(J_A\rho_{\rm el}J_A^\dagger-\frac12\{J_A^\dagger J_A,\rho_{\rm el}\}\right).
    \label{eq:appC_transfer_dissipator}
\end{equation}
In particular, its gain term is
\begin{equation}
\begin{split}
    J_A\rho_{\rm el}J_A^\dagger={}&
    J_{A\uparrow}\rho_{\rm el}J_{A\uparrow}^\dagger
    +J_{A\downarrow}\rho_{\rm el}J_{A\downarrow}^\dagger+J_{A\uparrow}\rho_{\rm el}J_{A\downarrow}^\dagger
    +J_{A\downarrow}\rho_{\rm el}J_{A\uparrow}^\dagger.
\end{split}
\label{eq:appC_transfer_gain}
\end{equation}
The cross terms are retained, so this superoperator is generally different from the sum of two independent spin-resolved dissipators.
Incoherent addition during injection and addition at the operator level during acceptor transfer describe two distinct preparation and dynamical processes.
The transfer jump preserves the electron spin and the total particle number.
It transfers one electron to an empty acceptor and vanishes when the acceptor is already occupied.
No reverse-transfer channel is included.

\subsection{LAB and electronic propagation from the same initial state}
\label{appC:propagation}

For the LAB calculation, the oscillator is initially in its vacuum and uncorrelated with the injected electronic state:
\begin{equation}
    \rho_{\rm LAB}(0)=\rho_I\otimes|0\rangle_{\rm vib}\langle0|.
    \label{eq:appC_LAB_initial}
\end{equation}
This factorized state is prescribed at \(t=0\), rather than obtained by equilibrating the interacting electron--vibration system.
The vibrational Hilbert space is represented by the number states \(|0\rangle,\ldots,|n_{\max}\rangle\), giving a joint Hilbert-space dimension \(196(n_{\max}+1)\).
With the tensor factors displayed explicitly, the Hamiltonian is
\begin{equation}
    H_{\rm LAB}=H_{\rm el}\otimes I_{\rm vib}
    +I_{\rm el}\otimes E_0\left(a^\dagger a+\frac12I_{\rm vib}\right)
    +B\otimes(a^\dagger+a).
    \label{eq:appC_LAB_H}
\end{equation}
The joint density matrix obeys
\begin{equation}
    \hbar\dot\rho_{\rm LAB}
    =-i[H_{\rm LAB},\rho_{\rm LAB}]
    +\Gamma\mathcal D[I_{\rm el}\otimes a]\rho_{\rm LAB}
    +\Gamma_A\mathcal D[J_A\otimes I_{\rm vib}]\rho_{\rm LAB}.
    \label{eq:appC_LAB_evolution}
\end{equation}
The electronic density matrix used for observable evaluation is
\begin{equation}
    \rho_{\rm el}^{\rm LAB}(t)=\operatorname{Tr}_{\rm vib}\rho_{\rm LAB}(t).
    \label{eq:appC_LAB_partial_trace}
\end{equation}
Every reduced electronic calculation starts from precisely the same matrix:
\begin{equation}
    \rho_{\rm el}^{\rm eff}(0)=\rho_{\rm el}^{\rm LAB}(0)=\rho_I.
    \label{eq:appC_common_initial}
\end{equation}
The electronic evolution is
\begin{equation}
\begin{split}
    \hbar\dot\rho_{\rm el}^{\rm eff}={}&
    -i[H_{\rm el},\rho_{\rm el}^{\rm eff}]
    +\Gamma_A\mathcal D[J_A]\rho_{\rm el}^{\rm eff}-[F,Q(t)\rho_{\rm el}^{\rm eff}-\rho_{\rm el}^{\rm eff}Q^\dagger(t)].
\end{split}
\label{eq:appC_effective_evolution}
\end{equation}
The stationary-response calculation replaces \(Q(t)\) by \(Q_\infty\), while the finite-time calculation retains its initial buildup.
Both evolve \(196\times196\) electronic density matrices in the same occupation basis.
The last commutator contains the complete coherent and non-Hamiltonian vibronic contributions retained in the reduction.
The static correction used in \(H_{\rm prep}\) is not added separately to \(H_{\rm el}\) during propagation.\\
The response is calculated using the complete bare \(H_{\rm el}\) in \(\mathcal H_5\), not the preparation Hamiltonian or the vibration-renormalized effective Hamiltonian.
As derived in Appendix~\ref{app:adiabatic_elimination}, the stationary integral is equivalently the solution of
\begin{equation}
    \kappa Q_\infty+i[H_{\rm el},Q_\infty]=E_0^2F.
    \label{eq:appC_Q_matrix_equation}
\end{equation}
For \(\gamma>0\), this linear matrix equation has a unique solution.
To evaluate it, let
\begin{equation}
    V^\dagger H_{\rm el}V=\operatorname{diag}(E_1,\ldots,E_{196}),
    \qquad F^{E}=V^\dagger F V,
    \qquad \Delta_{mn}=E_m-E_n.
    \label{eq:appC_energy_basis}
\end{equation}
Equation~\eqref{eq:appC_Q_matrix_equation} then reduces element by element to
\begin{equation}
    (\kappa+i\Delta_{mn})(Q_\infty^{E})_{mn}=E_0^2F^{E}_{mn},
    \qquad (Q_\infty^{E})_{mn}=\frac{E_0^2F^{E}_{mn}}{\kappa+i\Delta_{mn}},
    \label{eq:appC_Q_stationary_spectral}
\end{equation}
followed by the basis transformation \(Q_\infty=VQ_\infty^{E}V^\dagger\).
Here the superscript \(E\) denotes the electronic energy basis, and \(m,n\) label many-body eigenstates rather than bridge sites.
The denominator has strictly positive real part \(\gamma\), including at electronic degeneracies.
All states in \(\mathcal H_5\) enter this evaluation.\\
The finite-time response can be obtained from the same stationary matrix.
Writing \(U_t=e^{-iH_{\rm el}t/\hbar}\), the response tail is
\begin{equation}
\begin{split}
    Q_\infty-Q(t)
    &=\frac{E_0^2}{\hbar}\int_t^\infty e^{-\kappa\tau/\hbar}U_\tau F U_\tau^\dagger\,d\tau=e^{-\kappa t/\hbar}U_tQ_\infty U_t^\dagger.
\end{split}
\label{eq:appC_Q_tail}
\end{equation}
The last equality follows by shifting the integration variable and using the time independence of \(H_{\rm el}\).
Consequently,
\begin{equation}
    Q(t)=Q_\infty-e^{-\kappa t/\hbar}U_tQ_\infty U_t^\dagger.
    \label{eq:appC_Q_finite}
\end{equation}
In the energy basis, the same expression is
\begin{equation}
    [Q^{E}(t)]_{mn}=(Q_\infty^{E})_{mn}
    \left[1-e^{-[\kappa+i\Delta_{mn}]t/\hbar}\right],
    \qquad Q(t)=VQ^{E}(t)V^\dagger.
    \label{eq:appC_Q_finite_spectral}
\end{equation}
This construction gives \(Q(0)=0\) and approaches \(Q_\infty\) as the response envelope decays.
It supplies the response at the required times without numerical quadrature of its memory integral or a separate dynamical equation for \(Q\).
It introduces no additional physical approximation beyond the electronic reduction specified in Appendix~\ref{app:adiabatic_elimination}.

\subsection{Observables, consistency checks, and numerical convergence}
\label{appC:observables_checks}
All electronic observables are evaluated from \(\rho_{\rm el}^{\rm LAB}(t)\) or \(\rho_{\rm el}^{\rm eff}(t)\) using identical operator matrices.
In the following definitions, \(\rho_{\rm el}\) denotes either state.
The spin-resolved acceptor populations and their sum are
\begin{equation}
    n_{A\sigma}(t)=\operatorname{Tr}[\rho_{\rm el}(t)n_{A\sigma}],
    \qquad n_A(t)=n_{A\uparrow}(t)+n_{A\downarrow}(t).
    \label{eq:appC_acceptor_populations}
\end{equation}
Because the acceptor is empty or singly occupied, \(n_A(t)\) is its occupation probability for a physical density matrix.
Particle-number conservation requires
\begin{equation}
    \langle N_{\rm bridge}\rangle_t+n_A(t)=5.
    \label{eq:appC_number_balance}
\end{equation}
We record the acceptor spin imbalance separately from its polarization, which is expressed in percent:
\begin{equation}
    m_A(t)=n_{A\uparrow}(t)-n_{A\downarrow}(t),
    \qquad P_A(t)=100\,\frac{m_A(t)}{n_A(t)},
    \qquad n_A(t)>0.
    \label{eq:appC_acceptor_polarization}
\end{equation}
The polarization is undefined at \(t=0\), when the acceptor is empty.
It is reported only where the occupation is resolved above the numerical error.
Recording \(n_A\) and \(m_A\) alongside \(P_A\) distinguishes the transferred population from its spin composition.\\
For each bridge site, the local charge occupation and spin imbalance are
\begin{equation}
\begin{aligned}
    n_j(t)&=\operatorname{Tr}[\rho_{\rm el}(t)(n_{j\uparrow}+n_{j\downarrow})],\\
    m_j(t)&=2\operatorname{Tr}[\rho_{\rm el}(t)S_j^z],
    \qquad S_j^z=\frac12(n_{j\uparrow}-n_{j\downarrow}).
\end{aligned}
\label{eq:appC_bridge_observables}
\end{equation}
The quantities \(m_j\) are spin imbalances, not percentages normalized by the local charge occupation.
For the collinear SOC considered here, \(H_{\rm el}\), \(B\), and \(J_A\) conserve the total spin projection.
The unpolarized initial preparation therefore gives
\begin{equation}
    m_A(t)+\sum_{j=1}^{4}m_j(t)=0.
    \label{eq:appC_spin_balance}
\end{equation}
An acceptor spin imbalance is accompanied by the opposite total imbalance on the bridge.
This identity uses the full electronic density matrix, including both acceptor-occupation sectors.\\
To resolve the bridge contribution before transfer, we also select the rows and columns corresponding to an empty acceptor:
\begin{equation}
    \rho_{\rm el}^{(A=0)}(t)=\Pi_{A=0}\rho_{\rm el}(t)\Pi_{A=0},
    \label{eq:appC_empty_acceptor_block}
\end{equation}
where \(\Pi_{A=0}\) projects onto the five-electron bridge block.
Within the allowed acceptor space, this is also the block with acceptor spin quantum number \(S_A^z=0\); singly occupied acceptor states have \(S_A^z=\pm1/2\).
The projected local observable is
\begin{equation}
    m_j^{(A=0)}(t)
    =2\operatorname{Tr}[\Pi_{A=0}\rho_{\rm el}(t)\Pi_{A=0}S_j^z].
    \label{eq:appC_projected_bridge_spin}
\end{equation}
It measures the difference between the two local spin occupations contributed by the empty-acceptor block.
The selected block retains its original statistical weight, with
\begin{equation}
    \operatorname{Tr}\rho_{\rm el}^{(A=0)}(t)=1-n_A(t),
    \label{eq:appC_empty_acceptor_weight}
\end{equation}
and is not normalized separately.
Thus \(m_j^{(A=0)}\) includes the probability that transfer has not yet occurred and is not the polarization conditioned on an empty acceptor.
It must not replace the full \(m_j\) in the spin balance of Eq.~\eqref{eq:appC_spin_balance}.\\
Before propagation, the electronic matrices are checked for the required symmetries, including
\begin{equation}
    [N_{\rm tot},H_{\rm el}]=[N_{\rm tot},B]=[N_{\rm tot},J_A]=0.
    \label{eq:appC_number_commutators}
\end{equation}
During propagation, normalization, Hermiticity, and the smallest eigenvalue of the electronic density matrix are monitored, together with the particle-number balance.
Negative eigenvalues that persist beyond numerical error under refinement are reported as a limitation of the effective approximation.
For physical states, \(0\leq n_A(t)\leq1\), but complete transfer is not assumed.
The acceptor occupation is evaluated explicitly rather than set to its limiting value.\\
For LAB convergence, the oscillator cutoff is increased until the observables entering the comparison are stable over the selected evolution interval.
Occupation near the upper vibrational boundary is monitored as an additional diagnostic.
Time-integration convergence is checked separately from vibrational-cutoff convergence before a discrepancy between LAB and effective dynamics is interpreted physically.
These tests concern numerical accuracy, whereas the size of the residual effective-model discrepancy also reflects the approximations of the electronic reduction.\\
The observables are compared on a common time grid.
We evaluate
\begin{equation}
\begin{aligned}
    \delta n_A(t)&=n_A^{\rm eff}(t)-n_A^{\rm LAB}(t),\\
    \delta m_A(t)&=m_A^{\rm eff}(t)-m_A^{\rm LAB}(t),\\
    \delta P_A(t)&=P_A^{\rm eff}(t)-P_A^{\rm LAB}(t),
\end{aligned}
\label{eq:appC_observable_differences}
\end{equation}
where the last difference is evaluated only when both polarizations are numerically resolved.
Since \(P_A\) is expressed as a percentage, \(\delta P_A\) is measured in percentage points, rather than as a relative percentage error.
The stationary and finite-time electronic descriptions are compared with the same LAB calculation and the same injected state \(\rho_I\).
The benchmark results and convergence data are reported in Appendix~\ref{app:additional_results}; no agreement is assumed from the analytical reduction alone.

\section{Convergence and additional numerical results}
\label{app:additional_results}

This Appendix reports the convergence tests and supplementary calculations supporting Sec.~\ref{sec:results}.
The numerical construction and observables are defined in Appendix~\ref{app:numerics}.
Unless otherwise specified, the parameters are those of Table~\ref{tab:baseline_parameters}, the initial bridge contains four electrons, and the evolution ends at \(T_{\rm final}=1~\mathrm{ps}\).
The numerical calculations use \(\hbar=6.58\times10^{-4}~\mathrm{eV\,ps}\).
Polarization differences are expressed in percentage points, whereas occupation and bridge spin-imbalance differences are dimensionless.

\subsection{Reference convergence and projected bridge spin dynamics}
\label{appD:baseline_convergence}

\begin{figure}[htbp]
    \centering
    \includegraphics[width=0.86\linewidth]{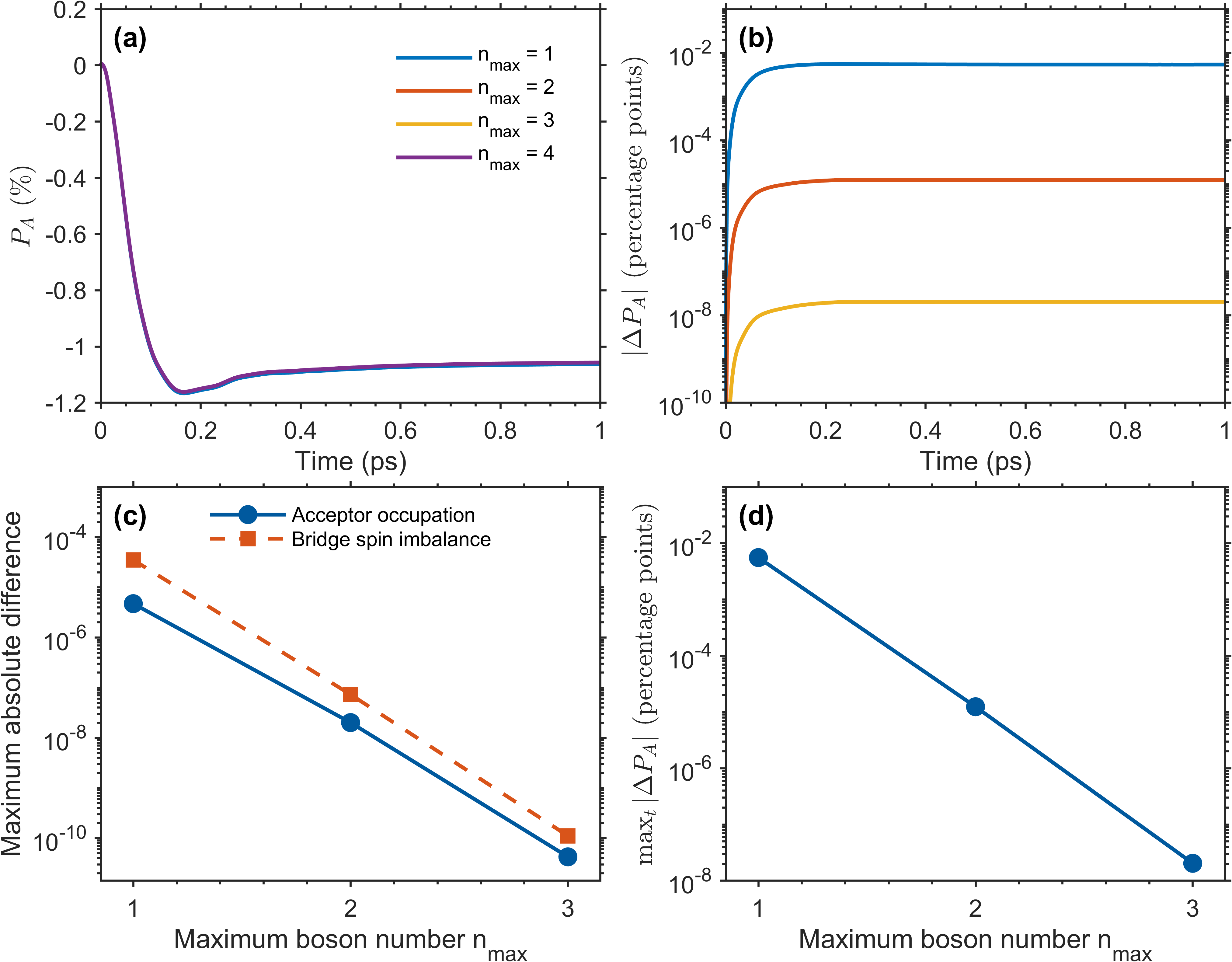}
    \caption{LAB vibrational-cutoff convergence at \(\Gamma=2~\mathrm{eV}\).
    (a) Acceptor polarization for \(n_{\max}=1,2,3,4\).
    Differences relative to \(n_{\max}=4\) are shown as (b) time-dependent absolute polarization differences, (c) maximum absolute differences in acceptor occupation and projected bridge spin imbalance, and (d) maximum absolute polarization differences.
    The bridge maximum includes all four sites and the common output times.
    Polarization differences are in percentage points.
    All calculations use the same injected electronic state and an initial vibrational vacuum.}
    \label{fig:appD_LAB_convergence}
\end{figure}

\begin{figure}[htbp]
    \centering
    \includegraphics[width=0.86\linewidth]{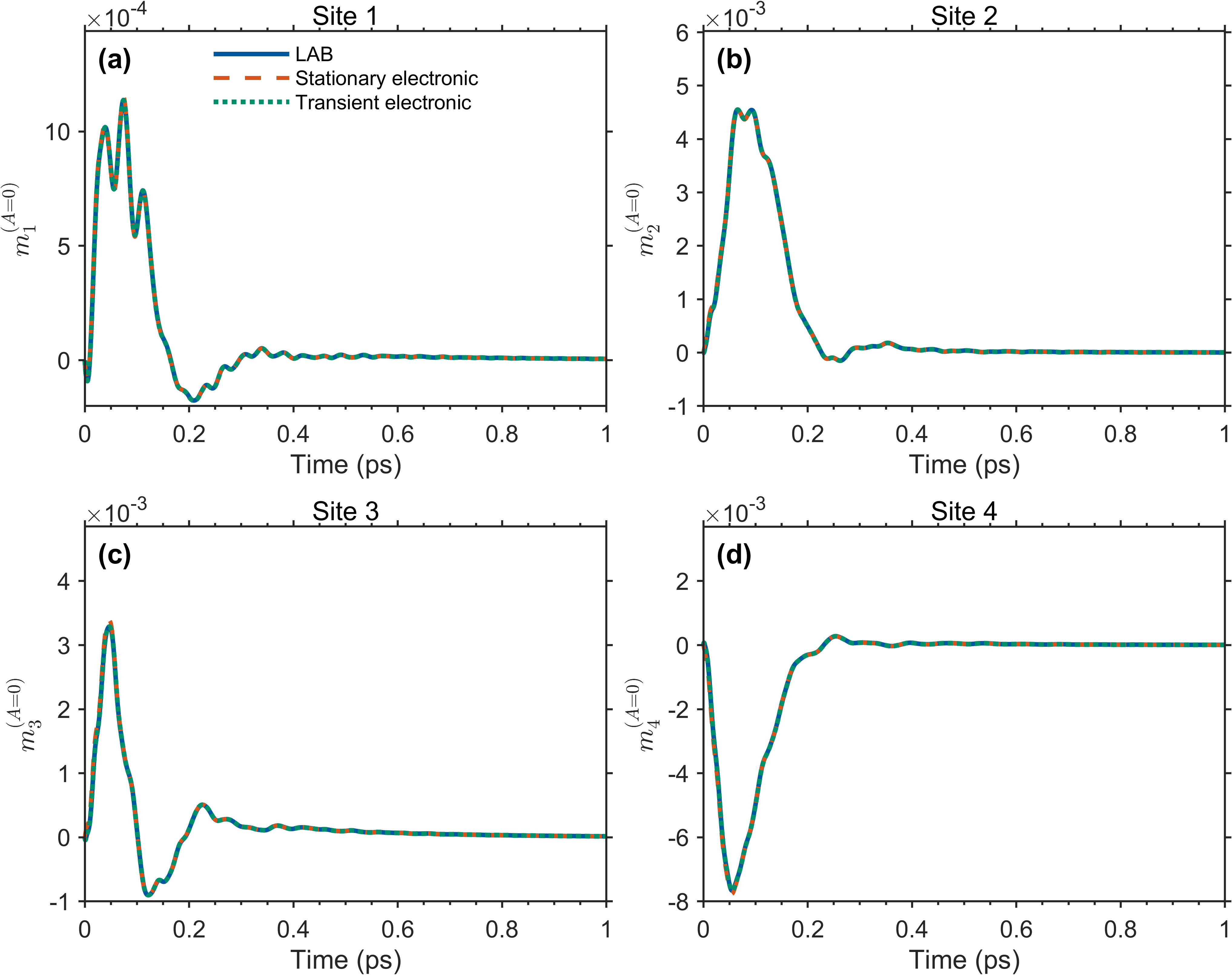}
    \caption{Projected local bridge spin imbalances for the reference parameters: (a) site \(1\), (b) site \(2\), (c) site \(3\), and (d) site \(4\).
    The LAB reference with \(n_{\max}=4\) is compared with the stationary and finite-time electronic responses; the latter is labeled ``Transient electronic'' in the legend.
    The quantities \(m_j^{(A=0)}\) include the probability weight of the empty-acceptor sector and are not percentages or separately normalized conditional polarizations.
    The scale factors differ between panels.}
    \label{fig:appD_bridge_spin}
\end{figure}

For the reference damping \(\Gamma=2~\mathrm{eV}\), the LAB calculation is repeated with oscillator states \(|0\rangle,\ldots,|n_{\max}\rangle\) and \(n_{\max}=1,2,3,4\).
Figure~\ref{fig:appD_LAB_convergence} compares the resulting polarization curves and their differences relative to \(n_{\max}=4\).
For an observable \(O\), the cutoff difference is
\begin{equation}
    \Delta O^{(n_{\max})}(t)=O_{\rm LAB}^{(n_{\max})}(t)-O_{\rm LAB}^{(4)}(t).
    \label{eq:appD_cutoff_difference}
\end{equation}
The polarization curves for \(n_{\max}=3\) and \(4\) are indistinguishable on the scale of panel (a), while the logarithmic plots resolve the remaining differences.
The acceptor occupation and projected bridge spin imbalances exhibit the same stabilization with increasing cutoff.
We use \(n_{\max}=4\), corresponding to five oscillator levels, as the reference in the LAB--electronic comparisons.
The observed cutoff dependence is substantially smaller than the discrepancies between the LAB and reduced electronic descriptions.
The additional local observables in Fig.~\ref{fig:appD_bridge_spin} are the projected imbalances \(m_j^{(A=0)}\) of Eq.~\eqref{eq:appC_projected_bridge_spin}.
The empty-acceptor block retains its statistical weight and is not normalized separately.
The LAB dynamics produces opposite-sign extrema at sites \(2\) and \(4\): approximately \(+4.55\times10^{-3}\) at \(0.066~\mathrm{ps}\) and \(-7.67\times10^{-3}\) at \(0.055~\mathrm{ps}\), respectively.
Site \(3\) reaches \(+3.29\times10^{-3}\) at \(0.047~\mathrm{ps}\) before changing sign and reaching approximately \(-9.03\times10^{-4}\) at \(0.122~\mathrm{ps}\).
These extrema refer to the sampled output grid.
The projected signals subsequently decrease as population leaves the empty-acceptor block, with residual oscillations and small nonzero values at the final time.
Table~\ref{tab:appD_baseline_errors} gives the maximum absolute discrepancies over the common output grid, supplementing the final acceptor values reported in the main text.
The bridge-spin discrepancy is maximized over both site and time.
The finite-time response gives the smaller discrepancy for both the projected bridge spins and acceptor polarization, while its maximum occupation discrepancy is slightly larger than that of the stationary response.
Thus the improvement is observable dependent, even in this reference benchmark.
\begin{table}[htbp]
    \caption{Maximum absolute differences from the LAB reference with \(n_{\max}=4\) at \(\Gamma=2~\mathrm{eV}\), over the common output grid up to \(T_{\rm final}=1~\mathrm{ps}\).
    Here \(\delta m_j^{(A=0)}=m_j^{(A=0),\rm eff}-m_j^{(A=0),\rm LAB}\).
    The polarization column is in percentage points; the other columns are dimensionless.}
    \label{tab:appD_baseline_errors}
    \begin{ruledtabular}
    \begin{tabular}{lccc}
        Electronic response & \(\max_t|\delta n_A|\) & \(\max_t|\delta P_A|\) & \(\max_{j,t}|\delta m_j^{(A=0)}|\) \\
        \hline
        Stationary & \(1.9753\times10^{-5}\) & \(0.032030\) & \(3.04\times10^{-4}\) \\
        Finite-time & \(2.4995\times10^{-5}\) & \(0.000616\) & \(5.70\times10^{-6}\) \\
    \end{tabular}
    \end{ruledtabular}
\end{table}
The electronic density matrices exhibit small negative eigenvalues, approximately \(-6\times10^{-7}\) for the stationary response and \(-10^{-8}\) for the finite-time response.

\subsection{Damping scan and limits of the reduced description}
\label{app:Gamma_scan}

\begin{figure}[htbp]
    \centering
    \includegraphics[width=0.78\linewidth]{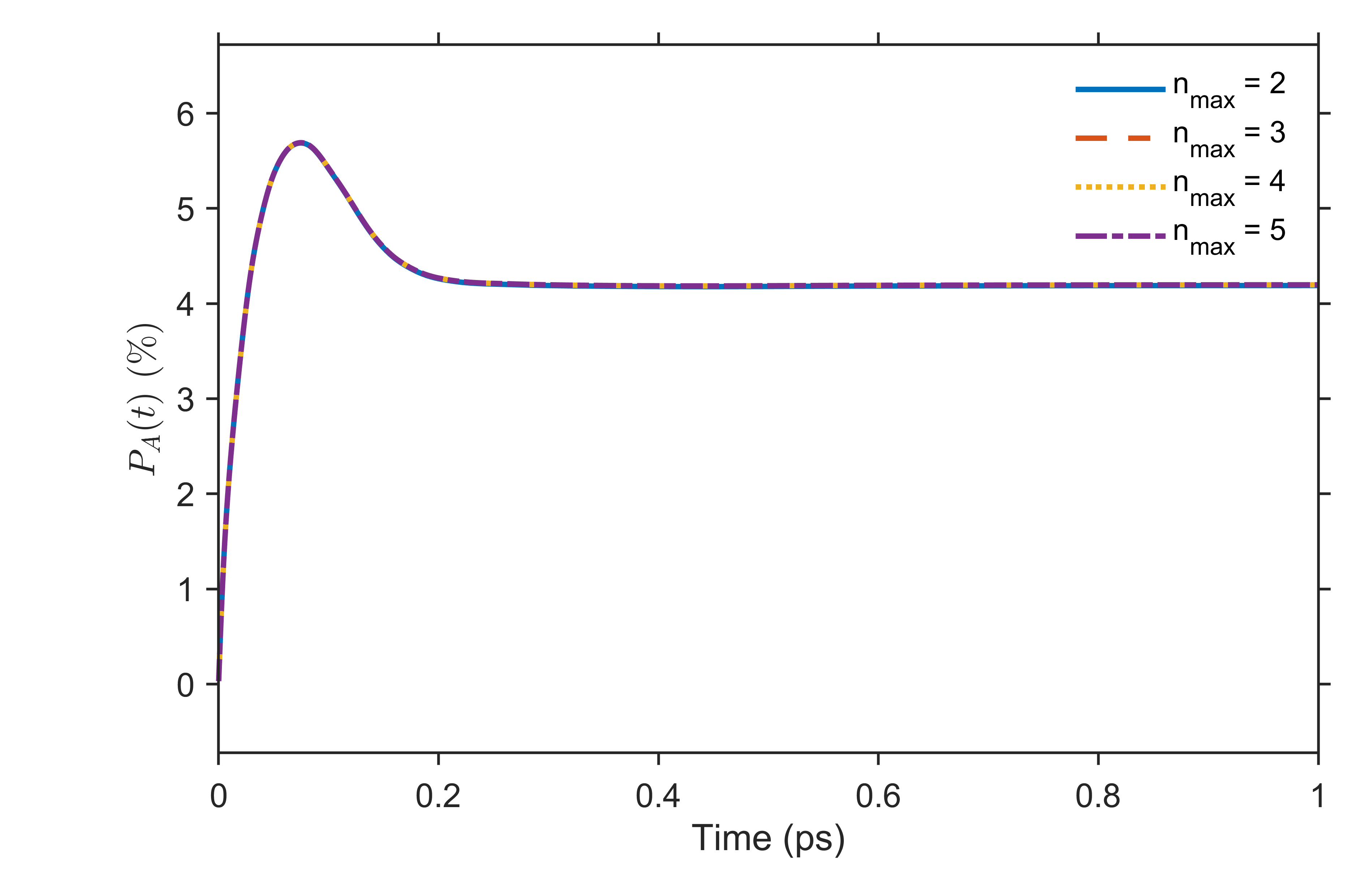}
    \caption{LAB acceptor polarization at the lowest scanned damping, \(\Gamma=0.2~\mathrm{eV}\), for \(n_{\max}=2,3,4,5\).
    The oscillator basis contains \(n_{\max}+1\) levels, and each calculation starts from the same electronic preparation and vibrational vacuum.
    The curves overlap at the plotted resolution; the cutoff comparison across the complete damping scan is quantified in the text.}
    \label{fig:appD_low_Gamma_convergence}
\end{figure}

Figure~\ref{fig:Gamma_scan_main} presents the complete damping scan discussed in Sec.~\ref{subsec:representative_dynamics}.
Here we report the corresponding low-damping values, vibrational-cutoff tests, positivity diagnostics, and control parameters of the electronic reduction.
The scan contains the ten values \(\Gamma=0.2,0.4,\ldots,2.0~\mathrm{eV}\), with all other parameters fixed and \(T_{\rm final}=1~\mathrm{ps}\).
At each point, the preparation is recomputed using \(R(\Gamma)\) and is shared by the LAB, stationary, and finite-time descriptions.\\
Table~\ref{tab:appD_low_damping} reports the two lowest-damping points, where the distinction between the stationary and finite-time responses is particularly relevant.
At \(\Gamma=0.4~\mathrm{eV}\), the stationary response predicts the opposite sign to LAB, whereas the finite-time response retains the LAB sign.
The absolute final-time discrepancies from LAB are approximately \(0.133\) percentage points for the stationary response and \(0.009\) percentage points for the finite-time response.
At \(\Gamma=0.2~\mathrm{eV}\), the corresponding discrepancies are approximately \(0.189\) and \(0.076\) percentage points.
These are endpoint discrepancies and should not be interpreted as maximum errors along the complete trajectories.
\begin{table}[htbp]
    \caption{Final acceptor polarization in percent at the two lowest damping values, for \(T_{\rm final}=1~\mathrm{ps}\).
    The LAB reference uses \(n_{\max}=4\).}
    \label{tab:appD_low_damping}
    \begin{ruledtabular}
    \begin{tabular}{cccc}
        \(\Gamma\) (eV) & LAB & Stationary electronic & Finite-time electronic \\
        \hline
        \(0.4\) & \(+0.05288\) & \(-0.08002\) & \(+0.06195\) \\
        \(0.2\) & \(+4.19698\) & \(+4.00824\) & \(+4.27335\) \\
    \end{tabular}
    \end{ruledtabular}
\end{table}
Vibrational-cutoff convergence is first examined at the lowest damping, \(\Gamma=0.2~\mathrm{eV}\), as shown in Fig.~\ref{fig:appD_low_Gamma_convergence}.
The common cutoff \(n_{\max}=4\) is then checked against \(n_{\max}=5\) at every point of the damping scan.
The largest polarization change over all sampled damping values and output times is below \(8.5\times10^{-6}\) percentage points.
The sign discrepancy of the stationary response at \(\Gamma=0.4~\mathrm{eV}\) is therefore much larger than the observed vibrational-cutoff variation.
Deviations from positivity also increase at the lowest damping.
At \(\Gamma=0.2~\mathrm{eV}\), the minimum eigenvalues of the propagated electronic density matrices are approximately \(-7.4\times10^{-4}\) for the stationary response and \(-1.6\times10^{-5}\) for the finite-time response.
The vibrational-cutoff convergence of the LAB calculation and the positivity properties of the reduced electronic evolution are separate diagnostics.\\
The lower end of the scan is determined by the working scales of the elimination, rather than by a universal threshold at which the acceptor polarization ceases to be reproduced.
For the reference couplings, the operator norms on the complete five-electron sector are
\begin{equation}
    \|B\|=0.05701~\mathrm{eV},
    \qquad
    \|J_A^\dagger J_A\|=2.
    \label{eq:appD_reference_norms}
\end{equation}
At \(\Gamma=0.2~\mathrm{eV}\), the response-envelope time is
\[
    \tau_{\rm vib}=6.58\times10^{-3}~\mathrm{ps},
\]
while the control parameters are
\[
    \xi_B\simeq0.57,
    \qquad
    \xi_A\simeq0.20.
\]
Although the response time remains short compared with the full \(1~\mathrm{ps}\) evolution, neither ratio defines a small expansion parameter.
For example, \(\xi_B^2\simeq0.33\) indicates that higher-order coupling corrections cannot be neglected solely on the basis of perturbative power counting; this quantity is not an estimate of the observable error.
The agreement of the finite-time response with the LAB polarization at this damping is therefore an observable- and preparation-specific numerical result, rather than a general validation of the reduced electronic dynamics for arbitrary states and observables.

\subsection{Computational cost}
\label{appD:computational_cost}
Let \(d_{\rm e}\) be the electronic Hilbert-space dimension and \(N_{\rm vib}=n_{\max}+1\) the number of retained oscillator levels.
The electronic and LAB density matrices contain, respectively,
\begin{equation}
    d_{\rm e}^2
    \qquad\text{and}\qquad
    N_{\rm vib}^2d_{\rm e}^2
    \label{eq:appD_matrix_storage}
\end{equation}
entries before exploiting sparsity or block structure.
For \(d_{\rm e}=196\) and \(n_{\max}=4\), eliminating the oscillator reduces this formal matrix size by a factor of \(25\).
If dense matrix--matrix products dominate a right-hand-side evaluation, their cubic dimension scaling gives a corresponding factor of \(N_{\rm vib}^3=125\).
This is a per-evaluation scaling guide, not an upper bound on the overall runtime or a prediction of the measured acceleration.
The actual cost also depends on matrix sparsity, block organization, response reconstruction, and the adaptive integration steps.\\
Table~\ref{tab:appD_computational_cost} reports measured wall-clock times for the \(\Gamma=0.2~\mathrm{eV}\) benchmark, using the same physical parameters, final time, and numerical tolerances in the three descriptions.
The stationary calculation constructs \(Q_\infty\) once, whereas the finite-time calculation reconstructs \(Q(t)\) during propagation.
The latter therefore incurs an additional cost while retaining a substantial reduction relative to LAB.
\begin{table}[htbp]
    \caption{Measured wall-clock times at \(\Gamma=0.2~\mathrm{eV}\) and \(T_{\rm final}=1~\mathrm{ps}\).
    These are implementation-dependent timings, not universal computational costs.
    The LAB calculation retains five oscillator levels.}
    \label{tab:appD_computational_cost}
    \begin{ruledtabular}
    \begin{tabular}{lrr}
        Description & Time (s) & Speed-up relative to LAB \\
        \hline
        LAB, \(n_{\max}=4\) & \(67.75\) & \(1.0\) \\
        Stationary electronic & \(5.06\) & \(13.4\) \\
        Finite-time electronic & \(7.87\) & \(8.6\) \\
    \end{tabular}
    \end{ruledtabular}
\end{table}
Across all ten damping values, the mean times are \(79.35~\mathrm{s}\) for LAB, \(5.73~\mathrm{s}\) for the stationary response, and \(8.46~\mathrm{s}\) for the finite-time response.
The ratios of these mean times are approximately \(13.8\) and \(9.4\), respectively.
These averages include the solver choices made at the individual scan points and depend on the computer, MATLAB version, and integration settings.
The timing comparison is separate from the accuracy and positivity tests above.

\subsection{Supplementary checks for the parameter scans and uniform-energy filling control}
\label{appD:scan_checks}
\begin{figure}[htbp]
    \centering
    \includegraphics[width=0.68\linewidth]{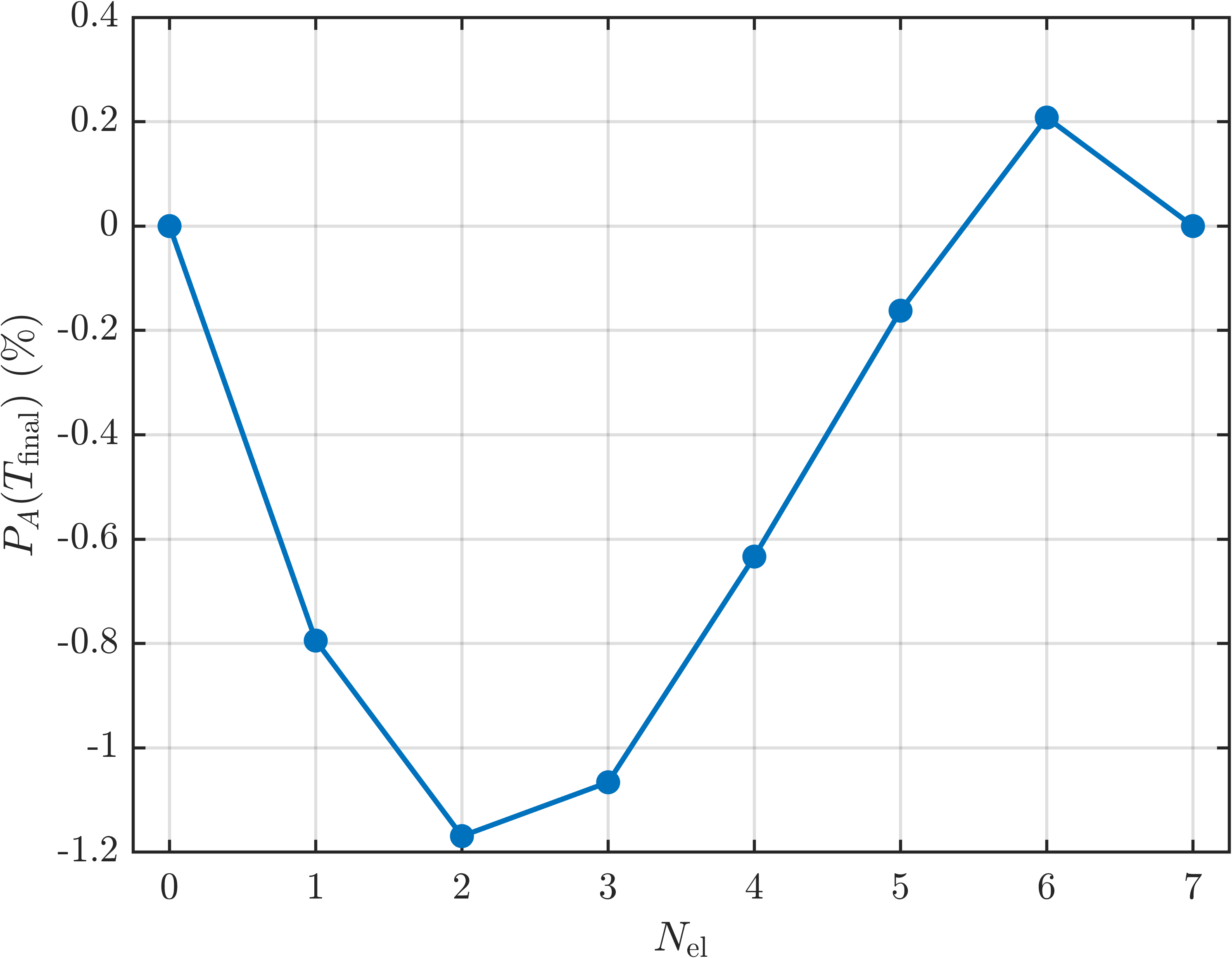}
    \caption{Uniform-energy filling control with \(\epsilon_j=0\) on the bridge and \(\epsilon_A=-1~\mathrm{eV}\).
    The finite-time electronic response is used at \(\Gamma=1~\mathrm{eV}\), \(E_0=100~\mathrm{meV}\), and \(T_{\rm final}=10~\mathrm{ps}\).
    The horizontal axis gives the number of bridge electrons before injection.
    The preparation is spin balanced for every filling and is recomputed for this site-energy profile.
    Lines connect the discrete filling values as a guide to the eye.}
    \label{fig:appD_uniform_filling}
\end{figure}

The coupling map, sign-reversal analysis, and vibrational-energy dependence are presented in Secs.~\ref{subsec:vibronic_symmetry} and \ref{subsec:physical_controls}.
Here we give additional control information without repeating those scans.
The norm entering \(\xi_B\) must refer to the complete electronic sector used for propagation.
In particular, \(\mathcal H_5\) contains bridge occupations \(N_{\rm bridge}=5\) and \(4\), corresponding to an empty and occupied acceptor.
The eigenvalues of the bilinear coupling in Eq.~\eqref{eq:app_q} give
\begin{equation}
    \|B\|_{N_{\rm bridge}=5}=\frac{2\phi+\phi^{-1}}{\sqrt{2}}\sqrt{g^2+\chi^2},
    \qquad
    \|B\|_{N_{\rm bridge}=4}=\sqrt{10}\sqrt{g^2+\chi^2}.
    \label{eq:appD_sector_norms}
\end{equation}
These follow by summing the largest single-particle coupling eigenvalues consistent with each fixed electron number.
Since \(B\) acts only on the bridge, its norm on \(\mathcal H_5\) is the larger of the two values.
At the largest couplings, \(g=\chi=20~\mathrm{meV}\),
\begin{equation}
    \|B\|_{N_{\rm bridge}=5}=0.0770820~\mathrm{eV},
    \qquad
    \|B\|_{\mathcal H_5}=0.0894427~\mathrm{eV}.
    \label{eq:appD_corner_norms}
\end{equation}
Thus, at \(\Gamma=1~\mathrm{eV}\), the full-sector control ratios at that corner are
\begin{equation}
    \xi_B=0.178885,
    \qquad \xi_A=0.040000.
    \label{eq:appD_corner_ratios}
\end{equation}
The smaller norm in Eq.~\eqref{eq:appD_corner_norms} applies only to the source block and is not the norm of the complete propagated space.
The ratios characterize the scale of the approximation; being below unity is not, by itself, a quantitative validation of every point in the map.\\
For the vibrational-energy scan, the maximum normalization error is \(1.443\times10^{-15}\), and the maximum deviation of the total electron number from five is \(7.105\times10^{-15}\).
These conservation checks supplement the occupation range and energy dependence reported in the main text; they do not bound the error in the polarization.\\
The filling control replaces the nonuniform bridge profile with \(\epsilon_j=0\) for all four sites while retaining \(\epsilon_A=-1~\mathrm{eV}\).
The remaining parameters, the spin-balanced preparation rule, and the finite-time response protocol are unchanged, with \(\Gamma=1~\mathrm{eV}\) and \(T_{\rm final}=10~\mathrm{ps}\).
The preparation Hamiltonian and its electronic state are rebuilt for the uniform profile at each filling.
Figure~\ref{fig:appD_uniform_filling} presents this control, and Table~\ref{tab:appD_filling_values} provides the complete numerical values for both energy profiles.
\begin{table}[htbp]
    \caption{Final polarization in percent for the filling scan at \(T_{\rm final}=10~\mathrm{ps}\).
    The reference profile is that of Table~\ref{tab:baseline_parameters}; the uniform profile has \(\epsilon_j=0\).
    Both use \(\Gamma=1~\mathrm{eV}\) and the finite-time electronic response.
    The small nonzero value at \(N_{\rm el}=0\) is retained as numerical output rather than replaced by zero.}
    \label{tab:appD_filling_values}
    \begin{ruledtabular}
    \begin{tabular}{crr}
        \(N_{\rm el}\) & Reference profile & Uniform profile \\
        \hline
        \(0\) & \(-0.00036\) & \(0\) \\
        \(1\) & \(+10.91558\) & \(-0.79439\) \\
        \(2\) & \(+0.94964\) & \(-1.16910\) \\
        \(3\) & \(-0.75543\) & \(-1.06593\) \\
        \(4\) & \(-1.51138\) & \(-0.63339\) \\
        \(5\) & \(-1.06953\) & \(-0.16217\) \\
        \(6\) & \(+0.26258\) & \(+0.20793\) \\
        \(7\) & \(0\) & \(0\) \\
    \end{tabular}
    \end{ruledtabular}
\end{table}
For the reference profile, \(n_A(T_{\rm final})\geq0.98866\), whereas for the uniform profile \(n_A(T_{\rm final})>0.99999\).
The residual empty-acceptor population in the reference profile is largest at \(N_{\rm el}=0\); the reported value there is a finite-time result, not an assumption of exactly complete transfer.
The two profiles are compared with the same preparation and injection rules, not with an identical initial density matrix.
The interpretation of the filling-dependent signs and extrema is discussed in Sec.~\ref{subsec:physical_controls}.
The coupling, energy, and filling scans use the finite-time electronic model; the LAB benchmarks above do not constitute a point-by-point validation of all these scans.

\subsection{Finite-temperature extension}
\label{app:finite_temperature}
\begin{figure}[htbp]
    \centering
    \includegraphics[width=0.64\linewidth]{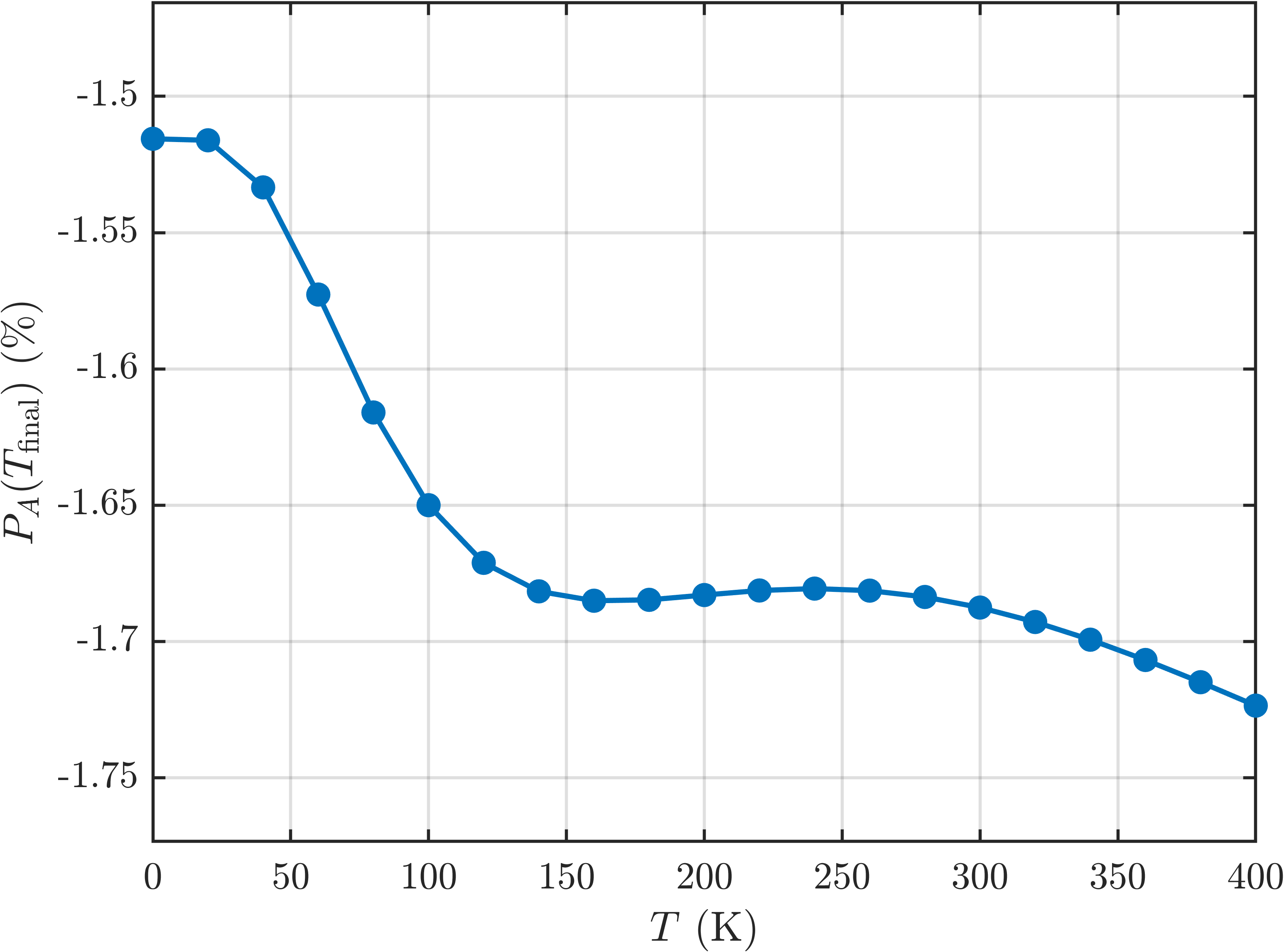}
    \caption{Finite-temperature electronic calculation at \(\Gamma=1~\mathrm{eV}\), \(E_0=100~\mathrm{meV}\), and \(T_{\rm final}=1~\mathrm{ps}\).
    The 21 points use the canonical electronic preparation and finite-time thermal response in Eqs.~\eqref{eq:appD_thermal_preparation}--\eqref{eq:appD_thermal_response_factors}.
    The same temperature enters both ingredients.
    The polarization remains negative, with a shallow nonmonotonic feature around \(160\)--\(240~\mathrm{K}\).
    This is a reduced electronic calculation, not a finite-temperature LAB comparison.}
    \label{fig:appD_temperature}
\end{figure}

The thermal scan is performed only in the finite-time electronic model, with \(g=10~\mathrm{meV}\), \(\chi=15~\mathrm{meV}\), \(E_0=100~\mathrm{meV}\), \(\Gamma=1~\mathrm{eV}\), and \(T_{\rm final}=1~\mathrm{ps}\).
The bare electronic parameters and transfer coefficient remain fixed.
Temperature enters both the initial electronic preparation and the vibrational response prescribed below.\\
At each \(T>0\), the four-electron state is the canonical density matrix of the restricted preparation Hamiltonian \(H_4\) defined in Eq.~\eqref{eq:appC_H4}:
\begin{equation}
    \rho_4(T)=\frac{e^{-H_4/(k_{\rm B}T)}}{\operatorname{Tr}_{\mathcal H_4}e^{-H_4/(k_{\rm B}T)}}.
    \label{eq:appD_thermal_preparation}
\end{equation}
At \(T=0\), the nondegenerate ground-state projector is used for this parameter set.
The injection maps of Eq.~\eqref{eq:appC_injection_maps} are applied to \(\rho_4(T)\), and the two spin contributions are summed before normalization as in Eq.~\eqref{eq:appC_rhoI}, giving \(\rho_I(T)\).
The thermal vibrational occupation is
\begin{equation}
    n_B(T)=\frac{1}{e^{E_0/(k_{\rm B}T)}-1},
    \qquad n_B(0)=0.
    \label{eq:appD_Bose_occupation}
\end{equation}
The electronic preparation and the vibrational response use the same temperature.\\
Let \(F^{E}=V^\dagger FV\) and \(\Delta_{mn}=E_m-E_n\) in the bare electronic energy basis of Eq.~\eqref{eq:appC_energy_basis}.
The thermal response used in the numerical calculation is
\begin{equation}
    [Q_T^{E}(t)]_{mn}=E_0^2F^{E}_{mn}\left[(n_B+1)f_+^{mn}(t)+n_B f_-^{mn}(t)\right],
    \qquad Q_T(t)=VQ_T^{E}(t)V^\dagger,
    \label{eq:appD_thermal_response}
\end{equation}
where
\begin{equation}
    f_\pm^{mn}(t)=\frac{1-e^{-[\gamma+i(\pm E_0+\Delta_{mn})]t/\hbar}}{\gamma+i(\pm E_0+\Delta_{mn})},
    \qquad \gamma=\frac{\Gamma}{2}.
    \label{eq:appD_thermal_response_factors}
\end{equation}
The weights \(n_B+1\) and \(n_B\) correspond to emission and absorption contributions, respectively.
The response satisfies \(Q_T(0)=0\) for every temperature and reduces to Eq.~\eqref{eq:appC_Q_finite_spectral} when \(T\to0\).
No stationary replacement or commutator cutoff is introduced.
Equations~\eqref{eq:appD_thermal_preparation}--\eqref{eq:appD_thermal_response_factors} specify the thermal extension used here; the vacuum-based derivation in Appendix~\ref{app:adiabatic_elimination} concerns the zero-temperature reduction.\\
For this scan, the code propagates the unnormalized empty-acceptor block \(\rho_0(t)\), corresponding to Eq.~\eqref{eq:appC_empty_acceptor_block}, and accumulates the acceptor populations from the transfer fluxes.
With all electronic operators in the following equation restricted to the source block, the evolution is
\begin{equation}
    \hbar\dot\rho_0=-i[H_{\rm el},\rho_0]-\frac{\Gamma_A}{2}\{J_A^\dagger J_A,\rho_0\}
    -[F,Q_T(t)\rho_0-\rho_0Q_T^\dagger(t)],
    \qquad \rho_0(0)=\rho_I(T).
    \label{eq:appD_thermal_source}
\end{equation}
The spin-resolved acceptor populations obey
\begin{equation}
    \dot n_{A\sigma}(t)=\frac{\Gamma_A}{\hbar}\operatorname{Tr}\left[J_{A\sigma}^\dagger J_{A\sigma}\rho_0(t)\right],
    \qquad n_{A\sigma}(0)=0.
    \label{eq:appD_thermal_fluxes}
\end{equation}
The separate fluxes resolve the populations generated by the single transfer operator \(J_A=J_{A\uparrow}+J_{A\downarrow}\); they do not replace it with two independent transfer dissipators.
The final polarization is obtained from these accumulated populations using Eq.~\eqref{eq:appC_acceptor_polarization}.\\
Figure~\ref{fig:appD_temperature} shows the results at \(T=0,20,\ldots,400~\mathrm{K}\).
The zero-temperature point reproduces \(P_A(1~\mathrm{ps})=-1.515582\%\).
The polarization remains negative and changes only weakly between \(0\) and \(20~\mathrm{K}\).
It reaches \(-1.68504\%\) at \(160~\mathrm{K}\), becomes slightly less negative at \(240~\mathrm{K}\), where it is \(-1.68059\%\), and then decreases to \(-1.72352\%\) at \(400~\mathrm{K}\).
The magnitude therefore grows by approximately \(0.208\) percentage points between the endpoints, with a shallow nonmonotonic feature at intermediate temperatures.
The Bose occupation reaches \(n_B=0.05816\) at \(400~\mathrm{K}\).
Both the canonical preparation and the emission--absorption weights change in this calculation; their individual contributions to the temperature dependence are not isolated by this scan.
The weak temperature dependence follows from the structure of the thermal response.
Equation~\eqref{eq:appD_thermal_response} can be written as \(Q_T=Q_++n_B(Q_++Q_-)\), where \(Q_\pm\) are the emission and absorption responses built from \(f_\pm^{mn}\), so the vibrational thermal correction has a relative size of order \(2n_B\).
For \(E_0=100~\mathrm{meV}\), this is below \(1.5\times10^{-3}\) up to \(160~\mathrm{K}\) and reaches only \(0.12\) at \(400~\mathrm{K}\).
The change of \(P_A\) between \(0\) and \(160~\mathrm{K}\) must therefore originate mainly from the canonical electronic preparation rather than from vibrational absorption.
Moreover, for \(\gamma\gg E_0\) the response distinguishes energy-releasing from energy-absorbing electronic transitions only weakly already at zero temperature: the Lorentzian weight \(\gamma/[\gamma^2+(E_0+\Delta_{mn})^2]\) differs between the two by a relative amount of order \(4E_0|\Delta_{mn}|/\gamma^2\), corresponding to an effective temperature \(k_{\rm B}T_{\rm eff}\approx\gamma^2/(4E_0)\simeq0.6~\mathrm{eV}\) at \(\Gamma=1~\mathrm{eV}\).
Thermal vibrational occupation therefore adds little to an asymmetry that is already weak in this strongly damped regime.
The bare working ratios remain \(\xi_B=0.11402\) and \(\xi_A=0.04000\), since \(B\), \(J_A\), \(\Gamma\), and \(\Gamma_A\) are fixed.
Their constancy does not remove the temperature dependence of the response amplitudes in Eq.~\eqref{eq:appD_thermal_response}.
The conservation checks do not establish positivity or quantify the discrepancy from a finite-temperature LAB calculation; no such LAB benchmark is included here.

\twocolumngrid

\end{document}